\documentclass[11pt]{article}

\usepackage[margin=1in]{geometry}
\RequirePackage{amsthm,amsmath,amsfonts,amssymb}
\usepackage{newtxtext,newtxmath}
\usepackage{mathtools}
\usepackage{cutwin}
\usepackage[sort&compress,square,numbers]{natbib}
\usepackage{bbm}
\usepackage{algorithmic}
\usepackage{graphicx}
\usepackage{textcomp}
\usepackage{sidecap}
\usepackage{enumerate}
\usepackage{xcolor}
\usepackage{csquotes}
\usepackage{chngcntr}
\usepackage{apptools}
\usepackage{tikz}
\usepackage{upgreek}
\usepackage[caption=false,font=footnotesize]{subfig}
\usepackage{cases}
\usepackage{wrapfig}
\usepackage[hidelinks]{hyperref}
\usepackage{cleveref}

\newcommand{\Pm}{\mathcal{P}}

\newcommand{\Q}{\mathcal{Q}}
\newcommand{\X}{\mathcal{X}}

\newcommand{\B}{\mathcal{B}}

\newcommand{\Y}{\mathcal{Y}}
\newcommand{\F}{\mathcal{F}}

\DeclareMathOperator*{\esssup}{ess\,sup}
\DeclareMathOperator*{\essinf}{ess\,inf}

\DeclareMathOperator{\dd}{d\!}

\makeatletter
\newcounter{labelcnt}
\renewcommand{\thelabelcnt}{(\alph{labelcnt})}
\newcommand{\setlabel}[1]{%
	\refstepcounter{labelcnt}\ltx@label{lbl:#1}%
	{\text{\upshape\thelabelcnt}}%
}

\makeatother

\newcounter{relctr} %
\everydisplay\expandafter{\the\everydisplay\setcounter{relctr}{0}} %

\AtBeginDocument{}

\theoremstyle{plain}

\newtheorem{theorem}{Theorem}[section]
\newtheorem{corollary}{Corollary}[section]
\newtheorem{lemma}[theorem]{Lemma}
\newtheorem{proposition}{Proposition}
\theoremstyle{remark}
\newtheorem{definition}[theorem]{Definition}
\newtheorem{remark}{Remark}
\newtheorem{example}{Example}

\begin{document}

\title{Tight Bounds for Linear and Non-Linear Contraction of Divergences via Duality}

\author{Amedeo Roberto Esposito\thanks{Amedeo Roberto Esposito was with the Institute of Science and Technology Austria (ISTA), Klosterneuburg, Austria and is now with the Okinawa Institute of Science and Technology (OIST), Okinawa, Japan. Email: amedeo.esposito@oist.jp.}, Marco Mondelli\thanks{Marco Mondelli is with the Institute of Science and Technology Austria (ISTA), Klosterneuburg, Austria. Email: marco.mondelli@ist.ac.at.}
}        %

\maketitle

\begin{abstract}
We develop a novel framework for bounding the contraction of information divergences, using duality and associated norms in Orlicz spaces. By working in the dual space, we obtain a principled approach to bounding both distribution-dependent strong data-processing inequality (SDPI) constants and \(F_\varphi\)-curves of divergences. Our bounds are either available in closed form or reducible to one-dimensional convex optimisation problems, in contrast to the infinite-dimensional optimisation problems that characterise SDPIs. These bounds depend on the densities of the reverse kernels with respect to a reference measure. To the best of our knowledge, they are the first universal closed-form bounds on distribution-dependent SDPI constants. We establish tightness for the \(\chi^2\)-divergence on several important channel classes, including full-rank binary kernels.

We apply our results to several settings. In particular, we derive bounds on the mixing times of Markov chains, including chains with heavy-tailed stationary distributions; obtain improved bounds on burn-in periods for Markov chain Monte Carlo; and strengthen concentration-of-measure bounds for dependent random variables.
\end{abstract}

\smallskip

\smallskip


\section{Introduction}
Strong data-processing inequalities (SDPIs) quantify the loss of information induced by a Markov kernel. While the ordinary data-processing inequality guarantees that a divergence cannot increase under the action of a kernel, an SDPI seeks a strict and quantitative form of this contraction. Such inequalities play a fundamental role in information theory and statistics, and they provide a natural tool for studying Markov processes~\cite{sdpiRaginsky, contractionInputConstraint}. Indeed, when the relevant contraction is strict, its iteration yields quantitative convergence to the stationary distribution and corresponding bounds on the mixing time.

The usual SDPI constant provides a linear description of this phenomenon through a single worst-case contraction factor. A more informative description is given by the $F_\varphi$-curve, introduced in~\cite{contractionInputConstraint}, which records how the divergence after applying the kernel depends on its value before applying the kernel. Controlling this curve amounts to establishing a nonlinear data-processing inequality and can yield substantially sharper information than a global linear coefficient. In particular, iterating a nonlinear contraction bound can provide refined convergence estimates even in regimes in which the corresponding linear bound is loose or vacuous.

These questions have direct practical implications for Markov chain Monte Carlo (MCMC) methods. MCMC approximates integrals with respect to a target measure $\pi$, when direct sampling from $\pi$ is unavailable or computationally inefficient, by constructing a Markov chain whose stationary distribution is $\pi$~\cite{mcmc92,gilks1995markov,mcmcThesis}. Quantitative contraction estimates determine how rapidly the chain approaches stationarity and therefore provide bounds on the mixing time, the burn-in period, and the resulting estimation error~\cite{mcmc92,gilks1995markov,mcmcThesis}.

A closely related and extensively studied problem is that of bounding the contraction coefficients of Markov operators on function spaces, particularly on $L_2$; see~\cite{Roberts_2004} for a survey. Many existing estimates rely on spectral-gap arguments and therefore apply most naturally in $L_2$~\cite{mcmcThesis,dependentViaStatAndChange}. Their extension to other $L_p$ spaces typically proceeds through Riesz--Thorin interpolation, which can produce loose or even vacuous bounds. Other approaches rely on strong uniform assumptions, such as Doeblin's minorisation condition or ultra-mixing~\cite{contractionLedoux,hiddenMarkovModels,sdpiRaginsky}. These limitations make it difficult to obtain effective contraction estimates for kernels acting on more general spaces or for distributions with irregular or heavy-tailed behaviour.

In this work, we develop a universal framework for bounding both linear and nonlinear information contraction through the action of Markov operators on Orlicz spaces. Orlicz spaces form a broad generalisation of the classical $L_p$ scale and allow the underlying function space to be adapted to the integrability and tail behaviour of the distributions under consideration. We first show that Markov kernels are contractions on these spaces: their operator contraction coefficients are always bounded above by $1$. Our main result, \Cref{thm:psiNormsContraction}, goes beyond this qualitative statement and gives an explicit upper bound on the contraction coefficient in terms of appropriate nested norms of the transition densities. For norm-like divergences, \Cref{cor:chiSquareBound} translates this operator estimate into a closed-form, distribution-dependent upper bound on the corresponding $H_\alpha$-SDPI constants, and in particular on the $\chi^2$-SDPI constant. On finite alphabets, the bound can be evaluated directly from $\mu$ and $K$. For $\chi^2$, it is tight for centred rank-one channels, including binary-input and binary-output kernels, while \Cref{ex:graph} shows that it can improve on existing estimates beyond these tight cases.

The main technical ingredient is a duality argument. The convergence of the Markov process governed by a kernel $K$ is controlled by the contraction properties of its dual $K^\star$, which we bound using nested Orlicz norms of the densities of $K^\star$ with respect to the reference measure. Unlike spectral methods, this approach does not require the existence of a spectral gap and applies directly in $L_p$ and, more generally, in Orlicz spaces. In the $L_p$ setting, it avoids the loss incurred by interpolation and improves on the corresponding Riesz--Thorin bounds. The flexibility of the Orlicz framework also makes it possible to treat heavy-tailed distributions that are difficult to accommodate using classical techniques~\cite{polynomialMC,heavyTailedMCMC}; see \Cref{ex:heavyTailOrlicz}.

A central feature of our approach is that the resulting bounds are universal with respect to the choice of divergence. They control not only the associated linear SDPI constant but also the full $F_\varphi$-curve, thereby providing nonlinear contraction bounds for arbitrary divergences. The extension is particularly direct for norm-like divergences, for which the operator estimates can be translated immediately into information-contraction bounds. General divergences need not admit such a norm representation and require additional functional machinery, which we develop through the polar gauges of divergence balls and their Amemiya-type dual representations.

For the squared Hellinger distance, \Cref{thm:hellingerNonlinearContraction} gives a closed-form upper bound on the full $F_\varphi$-curve in terms of the $L_2$-norms and essential ranges of the reverse-kernel densities, while \Cref{cor:hellingerSDPI} yields the corresponding linear SDPI bound. The improvement over standard distribution-independent estimates is illustrated in \Cref{ex:ternaryHellingerImprovement}. For the Kullback--Leibler divergence, \Cref{thm:klNonlinearContraction} first expresses the nonlinear contraction bound through the polar gauges of KL balls, whose dual representation reduces their estimation to one-dimensional convex optimisation; \Cref{cor:klSDPIPolarGauge} then gives the corresponding linear SDPI estimate. We next derive a hierarchy of increasingly explicit relaxations. \Cref{thm:klBennettContraction} uses Bennett's inequality to obtain bounds that are explicit up to the Lambert $W$ function, while \Cref{cor:klBernsteinClosedFormSDPI} gives a closed-form Bernstein estimate of the SDPI constant. The complementary sub-Gaussian route in \Cref{thm:subGaussianKLContraction} controls both the nonlinear curve and its linear counterpart through sub-Gaussian norms, and \Cref{cor:subGaussianKLRelaxations} replaces these norms by successively simpler Kearns--Saul and Hoeffding expressions. Thus, the bounds trade potential sharpness for increasing ease of evaluation, from one-dimensional convex optimisation to formulas involving only norms, moments, or ranges of the reverse-kernel densities. \Cref{ex:generalBinaryChannel,ex:biasedRefreshmentKL} compare this hierarchy, show that no single relaxation is uniformly preferable, and exhibit improvements over existing bounds. To the best of our knowledge, these are the first general bounds of this kind that are either available in closed form or reducible to one-dimensional convex optimisation.

Our framework also recovers well-known connections between contraction and ergodicity (\Cref{thm:alphaInfinityCase}), as well as the classical implications of ultra-mixing and Doeblin's minorisation condition (\Cref{thm:ultraMixing})~\cite{contractionLedoux,hiddenMarkovModels,sdpiRaginsky}.

Finally, our results yield several probabilistic and statistical applications. \Cref{thm:psiMixingAndContraction} relates Orlicz-space contraction to mixing times, while \Cref{thm:probBoundPsiNorm} and \Cref{thm:probBoundAlphaNorms} translate convergence in these norms into bounds on probabilities of events away from stationarity. Leverage generalised Orlicz spaces allows us to treat heavy-tailed stationary distributions for which no useful $L_p$ bound with $p>1$ is available, as shown in \Cref{ex:heavyTailOrlicz}. The same tools provide stronger convergence guarantees for MCMC methods (\Cref{sec:MCMC}), leading to more refined estimates of the burn-in period~\cite{mcmcThesis,dependentViaStatAndChange}; \Cref{ex:twoSpinGibbs} gives a concrete Gibbs-sampling comparison. Building on the strategy of~\cite{esposito2024concentration}, \Cref{thm:sdpiConcentrationBound} also gives concentration inequalities for Markov-dependent random variables, whose improvement over existing results is illustrated in \Cref{ex:nonStationaryTimeDependentConcentration}. These applications show how universal bounds on both linear and nonlinear information contraction translate into explicit convergence and finite-sample guarantees.

 \section{Preliminaries}\label{sec:preliminaries}

	We %
 adopt a measure-theoretic framework.
	Given a measurable space $(\X,\mathcal{F})$ and two measures $\mu,\nu$ making it a measure space, if $\nu$ is absolutely continuous w.r.t.\ $\mu$ (%
 $\nu\ll\mu$), %
 we represent with $\frac{d\nu}{d\mu}$ the Radon-Nikodym derivative of $\nu$ w.r.t.\ $\mu$. Given a (measurable) function $f:\X \to \mathbb{R}$ and a measure $\mu$, we denote with $\mu(E)$ the measure of an event $E$ and with $\mu(f) = \int f d\mu$ the 
Lebesgue integral of $f$ with respect to the measure $\mu$. %
    \subsection{Markov kernels and Markov chains}\label{sec:markovPreliminaries}
    \begin{definition}[Markov kernel]
    Let $(\Omega,\mathcal{F})$ be a measurable space. A \emph{Markov kernel} $K:\mathcal{F}\times \Omega \to [0,1]$ is a mapping such that \begin{enumerate}[(i)]
        \item $\forall \, x\in \Omega$, the mapping $E\in\mathcal{F}\to K(E|x)$ is a probability measure on $(\Omega,\mathcal{F})$;
        \item $\forall\, E\in\mathcal{F}$, the mapping $x \in \Omega \to K(E|x)$ is an $\mathcal{F}$-measurable real-valued function.    
    \end{enumerate}
    
    \end{definition}
    A Markov kernel acts %
    on measures ``from the right'', \textit{i.e.}, given a measure $\mu$ on $(\Omega, \mathcal{F})$, 
    $$\mu K(E) = \mu(K(E|\cdot)) = \int  K(E|x) \mu(\mathrm{d}x),
    $$
    and on functions ``from the left'',  \textit{i.e.}, given a function $f:\Omega \to \mathbb{R}$,
    $$
        (K f)(x) = \int f(y) K(\mathrm{d}y|x).
    $$
     Hence, given a function $f\in L^p(\mu K)$, one can see $K$ as a mapping $K:L_p(\mu K)\to L_p(\mu)$.
     Indeed, an application of Jensen's inequality gives that
$         \mu((Kf)^p)\leq \mu(K(f^p)) = \mu K(f^p)$,

     Given a Markov kernel $K$ and a measure $\mu$, let %
     $K^\star_\mu$ denote the dual operator. By definition, the dual kernel can be seen as acting on $(L^p(\mu ))^\star \cong L^q(\mu )$ with $q = p/(p-1)$ and returning a function in $(L^p(\mu K))^\star \cong L^q(\mu K)$. In particular, given a probability measure $\mu$ and two measurable functions $f,g$, one can define the inner-product $\langle f,g \rangle_\mu = \int (f\cdot g) d\mu,$
     and $K^\star_\mu$ is %
     the operator such that 
         $\langle Kf, g \rangle_\mu = \langle f, K^\star_\mu g \rangle_{\mu K}$, %
     for all $f$ and $g$. %
     In discrete settings, the dual operator can be explicitly characterised via $K$ and $\mu$ as $           K_\mu^\star(x|y) = \frac{K(y|x)\mu(x)}{\mu K (y)}$, see~\cite[Eq. (1.2)]{sdpiRaginsky}. 
   
    Given a sequence of random variables $(X_t)_{t\in\mathbb{N}}$, one says that it represents a \emph{Markov chain} if, given $i\geq 1$,  there exists a Markov kernel $K_i$ such that for every measurable event $E$: 
    \begin{equation}
        \mathbb{P}(X_i \in E | X_1,\ldots,X_{i-1}) = \mathbb{P}(X_i \in E | X_{i-1}) = K_i(E|X_{i-1})\qquad  \text{ almost surely}.
    \end{equation}
    If for every $i\geq 1$, $K_i = K$ for some Markov kernel $K$, then the Markov chain is time-homogeneous, and we suppress the index from $K$. %
    The kernel $K$ of a Markov chain is the probability of getting from $x$ to $E$ in one step, \textit{i.e.}, for every $i\geq 1$, 
    $K(E|x) = \mathbb{P}(X_i \in E| X_{i-1}=x)$. One can then define (inductively) the $t$-step %
    kernel $K^t$ as
     $   K^t(E|x) = \int K^{t-1}(E|y)dK(y|x)$.
    Note that $K^t$ is also a Markov kernel, and it represents the probability of getting from $x$ to $E$ in $t$ steps: $K^t(E|x)= \mathbb{P}(X_{t+1}\in E|X_{1}=x)$.
    If $(X_t)_{t\in\mathbb{N}}$ is the Markov chain associated to the kernel $K$ and $X_0 \sim \mu$, then $\mu K^m$ denotes the measure of $X_{m}$ for every $m\in \mathbb{N}$. Furthermore, a probability measure $\pi$ is a stationary measure for $K$ if $\pi K(E) = \pi(E)$ for every measurable event $E$. When the state space is discrete, $K$ can be represented via a stochastic matrix.

Given a Markov kernel and a measure, one can define the corresponding dual~\citep{contractionLedoux} and show the following result, whose proof is deferred to %
Appendix \ref{app:kStarRND}.
    
\begin{lemma}\label{lem:kStarRND}
    Let $\mu$ be a positive measure, $\nu$ a positive measure s.t.\ $\nu\ll\mu$, and $f$ the Radon-Nikodym derivative   $\frac{d\nu}{d\mu}$. Let $K$ be a Markov kernel and $g=\frac{d\nu K}{d\mu K}$. Assume a reverse Kernel $K_\mu^\star$ satisfying the adjoint relation below exists. Then, %
  \begin{equation}
        g = K_\mu^\star f\hspace{2em} \mu K\text{-a.e.},
    \end{equation}
    where $K_\mu^\star$ is the operator such that, given two admissible functions $h,u$, one has that
    $\langle Kh, u \rangle_\mu = \langle h, K_\mu^\star u\rangle_{\mu K}.$
    \end{lemma}
   The adjoint $K_\mu^\star$ is quasi-Markovian: it is positivity
preserving and satisfies
\[
    K_\mu^\star \mathbbm{1}=\mathbbm{1},
    \qquad \mu K\text{-a.e.}
\]
On standard Borel spaces, $K_\mu^\star$ admits a representative induced by a Markov kernel, unique up to $\mu K$-almost-everywhere equality~\cite[Section 2, Lemma 2.2]{contractionLedoux}.
Whenever this reverse-kernel representative is absolutely continuous with respect to $\mu$, we use, throughout the paper, the notation
\begin{equation}
    g_y
    =
    \frac{\mathrm dK_\mu^\star(\cdot\mid y)}{\mathrm d\mu},
    \qquad
    m_y
    =
    \essinf_{\mu} g_y,
    \qquad
    M_y
    =
    \esssup_{\mu} g_y.
    \label{eq:reverseKernelDensityEndpoints}
\end{equation}
The notation $g_Y,m_Y,M_Y$ denotes the corresponding random quantities when $Y\sim\mu K$. Since $g_y$ is a probability density, $0\leq m_y\leq1\leq M_y\leq+\infty$ for $\mu K$-almost every $y$.
\subsection{$\varphi$-Divergences}

    \begin{definition}[$\varphi$-divergences]\label{def:fDiv}
		Let $(\Omega,\F,\Pm),(\Omega,\F,\Q)$ be two probability spaces. Let $\varphi:\mathbb{R}^+\to \mathbb{R}$ be a convex function such that $\varphi(1)=0$. Consider a measure $\mu$ such that $\Pm\ll\mu$ and $\Q\ll\mu$. Denoting with $p,q$ the densities of the measures with respect to $\mu$, the \emph{$\varphi$-divergence of $\Pm$ from $\Q$} is defined as %
		\begin{align}
			D_\varphi(\Pm\|\Q):=\int q \varphi\left(\frac{p}{q}\right) \dd\mu.
		\end{align}
	\end{definition} 
    $\varphi$-divergences quantify in different ways how far two probability measures are~\cite{fInformativity}. %
    When a Markov kernel $K$ is applied to both measures, divergences contract. This is known in information theory as a Data-Processing Inequality (DPI): given two measures $\mu,\nu$ and a Markov kernel $K$, one has that, for every convex $\varphi$,
    \begin{equation}
        D_\varphi(\nu K\| \mu K) \leq D_\varphi(\nu\|\mu).\label{eq:DPI}
    \end{equation}
    This property holds as well for R\'enyi's $\alpha$-divergences, despite them not being a $\varphi$-divergence~\cite[Theorem 9]{RenyiKLDiv}.
    Similarly to how one wants to compute the coefficient of a Markovian kernel in any norm, one can define a similar notion quantifying how much a kernel $K$ is contracting a specific divergence $D_\varphi$. 
 This is known in the literature as a ``strong Data-Processing Inequality''~\cite[Definition 3.1]{sdpiRaginsky}.  
    
\begin{definition}[Strong Data-Processing Inequality]\label{def:sdpi}
    Given a probability measure $\mu$, a Markov kernel $K$ and  a convex function $\varphi$, we say that $K$ satisfies a $\varphi$-type \emph{Strong Data-Processing Inequality (SDPI)} at $\mu$ with constant $c\in[0,1]$ if
    \begin{equation}
    D_\varphi(\nu K\|\mu K) \leq c\cdot D_\varphi(\nu\|\mu),
    \end{equation}
    for all $\nu\ll\mu$. The sharpest such constant $c$ is denoted by
\begin{align*}
\eta_\varphi(\mu,K) &= \sup_{\substack{\nu\ll\mu:\\
0<D_\varphi(\nu\|\mu)<\infty}}
\frac{D_\varphi(\nu K\|\mu K) }{D_\varphi(\nu\|\mu)}, \\
\eta_\varphi(K) &= \sup_{\mu} \eta_\varphi(\mu,K).
\end{align*}
\end{definition}

 \begin{example}[SDPI for the KL and the BSC] 
     Let $\mu=\text{Ber}(1/2)$ and $K=\text{BSC}(\lambda)$ with $\lambda<\frac12$. One has that $\eta_{x\log x}(\mu,K)=(1-2\lambda)^2$~\cite{spreadingOfSetsHypercontractivity}, which implies that  $\eta_{x\log x}(\mu,K) < 1$ for all $\lambda>0$.
 \end{example} 
 Computing the SDPI constant has implications in various areas of information theory~\cite{contractionInputConstraint, sdpiRaginsky}, Markov chain theory~\cite{sdpiRaginsky,mcmcThesis, spreadingOfSetsHypercontractivity, contractionLedoux}, probability theory~\cite{concentrationMeasureII, esposito2024concentration} and statistics~\cite{bayesRiskRaginsky, esposito2023lower}.
 While $\eta_\varphi$ can be a difficult object to compute even for simple kernels, some universal upper and lower bounds are known~\cite[Theorems 3.1, 3.3]{sdpiRaginsky}:
 \begin{equation}
 \eta_\varphi(K) \leq \sup_{x,\hat{x}} \left\lVert K(\cdot|x)-K(\cdot|\hat{x})\right\rVert_{TV} = \eta_{|x-1|}(K)=\eta_{TV}(K), \label{eq:etaTV}
 \end{equation}
  \begin{align}
 \eta_\varphi(\mu,K) \geq  \eta_{\chi^2}(\mu,K), \label{eq:universalLower} \end{align}
 where $\eta_{\chi^2}(\mu,K)$ denotes the SDPI constant induced by the choice of $\varphi(x)=x^2-1$ \textit{i.e.}, the $\chi^2$-divergence.
 We remark that these bounds hold for functions $\varphi$ such that $\varphi(1)=0$ or, equivalently, when the divergence $D_\varphi(\nu\|\mu)$ is defined to be $\mu\left(\varphi\left(\frac{d\nu}{d\mu}\right)\right)-\varphi(1)$. For general convex functions $\varphi$, as well as for R\'enyi's divergences, the DPI holds and SDPI constants are still defined analogously, however one cannot use common techniques to bound said quantities~\cite[Example 2]{esposito2024concentration}.
\subsection{$L_p$ and Orlicz spaces} \label{sec:lpAndOrliczSpaces}
Markov kernels are well-known to be contracting in $L_p$-spaces. This is formalised in the following result.
\begin{lemma}
    Let $p\in [1,+\infty]$ and assume $K$ is a Markov kernel with stationary distribution $\pi$, then for every function $f\in L_p(\pi)$
    \begin{equation}
        \left\lVert Kf \right\rVert_{L_p(\pi)} \leq \left\lVert f \right\rVert_{L_p(\pi)}.
    \end{equation}
    Moreover, \begin{equation}
        \left\lVert K \right\rVert_{L_p(\pi)\to L_p(\pi)} = 1,
    \end{equation}
    where $ \left\lVert K \right\rVert_{L_p(\pi)\to L_p(\pi)} = \sup_{f: \left\lVert f\right\rVert_{L_p(\pi)}\leq 1} \left\lVert K f \right\rVert_{L_p(\pi)}.$
\end{lemma} 
While the contraction coefficient of Markovian kernels in $L_p$ spaces is always $1$, focusing on a subspace of $L_p$ leads to a more interesting object. In particular we can define %
\begin{equation}
    L_p^0(\pi) = \left\{f\in L_p(\pi): \pi(f)=0\right\},
\end{equation}
and %
the corresponding contraction coefficient $\left\lVert K\right\rVert_{L_p^0(\pi)\to L_p^0(\pi)} $ which, differently from the $L_p(\pi)$ counterpart, will typically be strictly less than $1$. In fact, this quantity is pivotal in understanding the speed of convergence of $\mu K^n$ to the stationary $\pi$ for increasing $n\in \mathbb{N}$. Take for instance $p=2$,
    if $\left\lVert K \right\rVert_{L_2^0\to L_2^0}=\gamma<1$, then $K$ %
    admits an $L_2$-spectral gap $(1-\gamma)$, which implies 
``$L_2$-geometric ergodicity'', as stated below \cite[Proposition 3.12]{mcmcThesis}.
    
    \begin{proposition}\label{prof:l2GeometricErgodicity}
        Let $K$ be a Markov kernel with stationary distribution $\pi$ and spectral gap $1-\gamma>0$. For every probability measure $\nu$ on the same space s.t.\ $\left\lVert d\nu/d\pi \right\rVert_{L_2(\pi)}<\infty$, we have %
          \begin{equation}
        \left\lVert \frac{d(\nu K^t)}{d\pi}-1 \right\rVert_{L_2(\pi)} \leq \gamma^t \left\lVert \frac{d\nu}{d\pi}-1\right\rVert_{L_2(\pi)}.
    \end{equation}
    \end{proposition} 
    In fact, in the specific setting of $L_2$, geometric ergodicity and %
    existence of a spectral gap are equivalent notions~\cite[Proposition 3.13]{mcmcThesis}.
    The first connection between divergence contraction and operator norms appears here and links the spectral gap to the $\chi^2$-divergence. Indeed, given a Kernel $K$ with spectral gap $(1-\gamma)$ and stationary distribution $\pi$, one has that 
    $
    \eta_{\chi^2}(\pi, K) = \gamma^2.$
    For a stationary reversible kernel, this is the square of its nontrivial absolute spectral radius and is therefore directly related to the spectral gap. The results developed in this work extend this operator perspective beyond the Hilbert-space setting. The preceding $L_2$ example shows that the contraction of the $\chi^2$-divergence can be studied through the action of the reverse operator on centred functions. This connection is not limited to the quadratic setting. Generalising the $L_2$-norm to $L_p$-norms leads to other norm-generated divergences and provides a first extension beyond $\chi^2$. Orlicz spaces further generalise this perspective by replacing power functions with more general convex functions. We introduce the required definitions next.

\begin{definition}[Young Function, {\cite[Chapter 3]{theoryOrliczSpaces}}]
    Let $\psi:\mathbb{R}^+\to\mathbb{R}^+$ be a convex non-decreasing function s.t.\ $\psi(0)=0$ and $\psi(x)\xrightarrow[x\to\infty]{} \infty$. Then, $\psi$ is called a Young function.
\end{definition}
\begin{definition}[{\cite[Definition 5]{theoryOrliczSpaces}}]
    Let $\psi$ be a Young function. Let $(\Omega,\mathcal{F}, \mu)$ be a measure space and denote with $L_0(\mu)$ the space of all the $\mathcal{F}$-measurable and real-valued functions on $\Omega$. An Orlicz space can be defined as the following set:
    \begin{equation}
        L_\psi(\mu) = \left\{f\in L_0(\mu) : \int_\Omega \psi(|\lambda f|)d\mu < +\infty\text{ for some }\lambda>0\right\}. \label{eq:orliczSpace}
    \end{equation}
\end{definition}
    Given a Young function $\psi: [0,+\infty) \to \overline{\mathbb{R}}^+$, the complementary function to $\psi$ in the sense of Young, denoted as  $\psi^\star: \mathbb{R} \to  \overline{\mathbb{R}}^+ $, is defined as follows \cite[Section 1.3]{theoryOrliczSpaces}: %
    \begin{align} \label{eq:youngComplement}
        \psi^\star(x) = \sup\left\{\lambda|x| - \psi(\lambda) : \lambda\geq 0 \right\}.
    \end{align}
An Orlicz space can be endowed with several norms that render it a Banach space~\cite{orliczAmemiyaNorm}: the Luxemburg norm, the Orlicz norm, and the Amemiya norm. We will henceforth ignore the Orlicz norm, as it %
is equivalent to the Amemiya norm~\cite{orliczAmemiyaNorm}, and we define the Luxemburg norm $L_\psi^L$ and the Amemiya norm $L_\psi^A$ as
\begin{equation}
    \lVert f\rVert_{L^L_\psi(\mu)} =\inf \left\{ \sigma > 0 : \mu\left(\psi\left(|f|/\sigma\right)\right)\leq 1\right\} ,\qquad
 \lVert f\rVert_{L^A_\psi(\mu)} = \inf_{t>0}  \frac{\mu\left(\psi(t|f|)\right)+1}{t}.\label{eq:amemiyaNorm}
\end{equation}
Orlicz spaces and the corresponding 
norms 
recover well-known objects (\emph{e.g.}, $L_p$-norms) and characterise random variables according to their ``tail''. For more details, see Appendix \ref{app:orliczSpaces}. %

    \section{Contraction of Divergences}\label{sec:contractionDivergences}
 
\subsection{Norm-Like Divergences}\label{sec:sdpiBound}

The tools and results commonly utilised to study contraction of Markovian operators in $L_p$ spaces can be leveraged and adapted to provide bounds on the SDPI constant of various divergences. Indeed, some of the most well-known divergences are either norms or can be related to them. Examples include \emph{(i)} the Total Variation distance -- corresponding to the $L_1$-norm, \emph{(ii)} the $\chi^2$-divergence -- corresponding to the $L_2$-norm; and \emph{(iii)} the Hellinger $\alpha$-divergence -- which can be related to the general $L_\alpha$-norm. %
Indeed, take $\alpha>1$ and $\varphi_\alpha(x)= (x^\alpha-1)/(\alpha-1)$, then one has that $D_{\varphi_\alpha}(\nu\|\mu) = \mathcal{H}_\alpha(\nu\|\mu)$ and for $1<\alpha \leq 2$ one has that
\begin{equation}
    (\alpha-1)\mathcal{H}_\alpha(\nu\|\mu) \leq \left\lVert \frac{d\nu}{d\mu}-1\right\rVert_{L_\alpha(\mu)}^\alpha \leq (\alpha-1)\mathcal{H}_\alpha(\nu\|\mu)+2.
\end{equation}
Similar bounds can be found for $\alpha>2$.
Given that the Hellinger divergence can be upper and lower bounded by the $L_\alpha$-norm and given that $\left\lVert \frac{d\nu}{d\mu}-1 \right\rVert_{L_\alpha(\mu)}^\alpha = H_\alpha(\nu\|\mu)$ can be seen itself to be a divergence stemming from $\tilde{\varphi}_\alpha(x)= |x-1|^\alpha$, we will focus on bounding the SDPI constant of $H_\alpha$ denoted as $\eta_{\alpha}$. This also allows us to provide results on both the Total Variation and $\chi^2$-Divergence. 
Indeed, for $\alpha=1$ and $\alpha=2$, one has that
\begin{align}
H_1(\nu\|\mu) &= 2TV(\nu\|\mu) = \left\lVert \frac{d\nu}{d\mu}-1\right\rVert_{L_1(\mu)}, \quad %
    H_2(\nu\|\mu) = \chi^2(\nu\|\mu) = \left\lVert \frac{d\nu}{d\mu}-1\right\rVert_{L_2(\mu)}^2.
\end{align}
For these objects, a closed-form bound on the contraction coefficient in $L_\alpha$-spaces would automatically yield a closed-form bound on the corresponding SDPI coefficient. In order to deliver such a theorem we need to state a few intermediate results on contraction of norms. We will state the first result in its full generality.

\begin{theorem}\label{thm:psiNormsContraction}
        Let $\psi$ and $\varphi$ be Young functionals and $\mu$ %
        a probability measure. Let $K$ be a Markov kernel, %
        $K^\star=K^\star_\mu$ its dual, and $N\in \{A,L\}$. Assume the jointly measurable Radon--Nikodym versions below exist for $\mu$-a.e.\ $x$ and $\mu K$-a.e.\ $y$, and denote them by $g_x = \frac{dK(\cdot|x)}{d\mu K}$ and $g_y =\frac{dK^\star(\cdot|y)}{d\mu}$. Then, for every $f\in L_\varphi^{0}(\mu K)$ and $h \in L_\varphi^0(\mu)$,
    \begin{align}
       \left\lVert Kf \right\rVert_{L_\psi^N(\mu)}
       &\leq
       \left\lVert f \right\rVert_{L_\varphi^N(\mu K)}
       \left\lVert
       \left\lVert g_X-1\right\rVert_{L_{\varphi^\star}^{N^\star}(\mu K)}
       \right\rVert_{L_\psi^N(\mu )},
       \label{eq:forwardMixedOrlicz}\\
       \left\lVert K^\star h \right\rVert_{L_\psi^N(\mu K)}
       &\leq
       \left\lVert h \right\rVert_{L_\varphi^N(\mu )}
       \left\lVert
       \left\lVert g_Y-1\right\rVert_{L_{\varphi^\star}^{N^\star}(\mu )}
       \right\rVert_{L_\psi^N(\mu K)}.
       \label{eq:reverseMixedOrlicz}
    \end{align}
    \end{theorem}
      The proof of~\Cref{thm:psiNormsContraction} can be found in~\Cref{app:proofPsiNormsContraction}.
\begin{remark}
           Let $\psi(x)=|x|^p$ and $\varphi(x)=|x|^q$ for $p,q>1$, and denote with $q^\star= q/(q-1)$. Then~\Cref{thm:psiNormsContraction} gives
    \begin{equation}
          \left\lVert K\right\rVert_{L_q^0(\mu K)\to L_p(\mu)}
          \leq \left\lVert \left\lVert g_X -1  \right\rVert_{L_{q^\star}(\mu K)} \right\rVert_{L_p(\mu )},
          \qquad
          \left\lVert K^\star\right\rVert_{L_q^0(\mu)\to L_p(\mu K)}
          \leq\left\lVert \left\lVert g_Y-1\right\rVert_{L_{q^\star}(\mu )}\right\rVert_{L_p(\mu K)}.
          \label{thm:alphaNormsHyperContraction}
    \end{equation}
    If $q<p$ then~\Cref{thm:alphaNormsHyperContraction} represents a bound on the hypercontractivity constant of $K$ and $K^\star$, see~\cite[Section 2]{logSobMarkovChains}. 
    Moreover, selecting $q=p$ one recovers the following bound on the contraction coefficient of $K$ and $K^\star$:
      \begin{equation}
          \left\lVert K\right\rVert_{L_p^0(\mu K)\to L_p(\mu)}
          \leq \left\lVert \left\lVert g_X -1  \right\rVert_{L_{p^\star}(\mu K)} \right\rVert_{L_p(\mu )},
          \qquad
          \left\lVert K^\star\right\rVert_{L_p^0(\mu)\to L_p(\mu K)}
          \leq\left\lVert \left\lVert g_Y-1\right\rVert_{L_{p^\star}(\mu )}\right\rVert_{L_p(\mu K)},
          \label{thm:alphaNormsContraction}
    \end{equation}
    where $p^\star = p/(p-1)$.
        \end{remark}
        
When working with divergences, leveraging the dual kernel $K^\star$ is pivotal in linking the SDPI constant to the contraction coefficient of a related norm. Indeed, divergences act as functionals of the Radon-Nikodym derivatives and, relating $D_\varphi(\nu K\|\mu K)$ (which acts on $d\nu K/d\mu K$) to $D_\varphi(\nu\|\mu)$ (which acts on $d\nu/d\mu$) heavily relies on the identity $d\nu K/d\mu K = K^\star (d\nu/d\mu)$, see~Lemma~\ref{lem:kStarRND}. This observation, along with~\Cref{thm:psiNormsContraction},  yields the following result.
\begin{corollary}\label{cor:chiSquareBound}
    Let $\mu$ be a probability measure, $K$ a Markov kernel and $\alpha>1$. Then,
    \begin{equation}
        \eta_{\alpha}(\mu, K ) \leq \min\left\{
1,\,
\left\lVert
\left\lVert g_Y-1\right\rVert_{L_\beta(\mu)}
\right\rVert_{L_\alpha(\mu K)}^\alpha\right\} \label{eq:alphaNormBound}
    \end{equation}
    where $\beta=\alpha/(\alpha-1)$. %
\end{corollary}
The proof of~\Cref{cor:chiSquareBound} can be found in~\Cref{app:proofChiSquareBound}.
On finite alphabets~\eqref{eq:alphaNormBound} reduces to the following
\begin{equation}
\left\lVert
\left\lVert g_Y-1\right\rVert_{L_\beta(\mu)}
\right\rVert_{L_\alpha(\mu K)}^\alpha =\sum_{y:\mu K(y)>0}\mu K(y)
\left[
\sum_x\mu(x)
\left|
\frac{K^\star(x|y)}{\mu(x)}-1
\right|^\beta
\right]^{\alpha/\beta}.
\label{eq:finiteBalpha}
\end{equation}
It requires only $\mu$, $K$, and $\mu K$, leaves no optimisation over
$\nu$, and is evaluated in $O(|\X||\Y|)$ operations. 
Note that, to the best of our knowledge, specialising \Cref{cor:chiSquareBound} to $\alpha=2$ provides the first closed-form upper bound on $\eta_{\chi^2}(\mu, K)$. Said object has gained importance over the years due to its connections with the maximal correlation and Poincaré constant~\cite{sdpiRaginsky}, as well as from the fact that it represents a universal lower bound for the SDPI constant of both $\varphi$ and R\'enyi-divergences~\cite{sdpiRaginsky, jin2024properties}.

We now discuss a variety of settings in which the bound of~\Cref{cor:chiSquareBound} is tight.
For every $\alpha>1$, the bound is sharp for the binary symmetric channel with the uniform input distribution. To understand the sharpness it is useful to look at a potential refinement of~\Cref{cor:chiSquareBound}.
In particular, the reverse-kernel identity can be generalised as follows:
\begin{equation}
    K_\mu^\star f(y) = \int f(x) (g_y(x)-c) \mu(\mathrm{d}x),
\end{equation}
for every constant $c$. 
Hence, the corresponding (albeit, not closed-form bound) sharpening of~\Cref{cor:chiSquareBound} is the following
\begin{equation}
    \eta_\alpha(\mu, K) \leq \sum_y \mu K(y) \left[ \inf_{c\in \mathbb{R}}\left\lVert g_y(x)-c\right\rVert_{L_\beta(\mu)} \right]^\alpha. \label{eq:genCenteredBound}
\end{equation}
\Cref{eq:genCenteredBound} is sharp for every channel such that $K_\mu^\star - 1_\mu$ is rank-one. This is equivalent to saying that the matrix $K_\mu = (K(y|x))_{x\in \text{supp}\mu , y \in \text{supp} \mu K}$ has rank at most two. Important examples of centred rank-one channels include all binary-input and binary-output channels (the BSC, BEC and Z-channel, binary-input AWGN channel and binary-input Poisson channel), or, more generally, channels $K: \mathcal{X} \mapsto \mathcal{Y}$ such that $K=WV$ with $W: \mathcal{X} \mapsto \{0,1\}$ and $V: \{0,1\} \mapsto \mathcal{Y}$ \textit{i.e.}, channels that can be decomposed as $X \rightarrow Z \rightarrow Y$ with $Z$ binary. Said channels have been studied, for instance, in the context of Principal Inertia Components (PIC)~\cite{Calmon2017PrincipalInertia} and low-nonnegative-rank approximation of Markov chains~\cite{Duan2020LowNonnegativeRank}.
For $\alpha=2$, the minimising constant is $1$ for every $y$, so the refinement leaves the bound unchanged, and the optimality for the same class of channels of~\Cref{cor:chiSquareBound} holds without the additional minimisation over $c$. 

A notable example that shows how~\Cref{cor:chiSquareBound}, despite not being sharp, can still yield better results compared to the state-of-the-art is the following.
  
\begin{example}[Random walk on a graph]\label{ex:graph}
    Consider a simple connected and undirected graph $G = (V,E)$, where $V$ is the set of vertices and $E\subseteq V\times V$ is the set of edges. Define $\mathrm{deg}(x)=|\left\{ y : (x,y) \in E \right\}|$. Consider the following probability distribution over $V$:
    \begin{equation}
        \pi(x) = \frac{\mathrm{deg}(x)}{2|E|}, \qquad x \in V.
    \end{equation}
    Given a parameter $\lambda \in (0,1)$  and $\bar{\lambda}=1-\lambda$, consider the following kernel:
 \begin{equation}
K_\lambda(y\mid x)
=
(1-\lambda)\,\mathbbm{1}_{\{y=x\}}
+
\frac{\lambda}{\deg(x)}\,\mathbbm{1}_{\{y\sim x\}}.
\end{equation}
    It is easy to see that $\pi K_\lambda = \pi$, hence $\pi$ is the invariant distribution with respect to $K_\lambda$. Consequently, $g_y(x) =  \frac{K^\star_\lambda(x|y)}{\pi(x)} = \frac{K_\lambda(y|x)}{\pi K_\lambda(y)} = \frac{K_\lambda(y|x)}{\pi(y)}$, and we proceed to bound $\eta_{\chi^2}(K_\lambda, \pi)$. For $y \in V$, we have
    \begin{align}
        \sum_{x} (g_y(x)-1)^2 \pi(x)   &=
        \left(\frac{2|E| \bar{\lambda}}{\mathrm{deg}(y)}\right)^2 \pi(y) + \sum_{x: \{x,y\}\in E} \left( \frac{2|E|\lambda}{\mathrm{deg}(x)\mathrm{deg}(y) } \right)^2 \pi(x) - 1 \\
        &= \bar{\lambda}^2 \frac{2|E|}{\mathrm{deg}(y)} +  \lambda^2 \frac{2|E|}{\mathrm{deg}(y)^2}  \sum_{x: \{x,y\}\in E} \frac{1}{\mathrm{deg}(x)} - 1 \\
        &= \bar{\lambda}^2 \frac{2|E|}{\mathrm{deg}(y)} +  \lambda^2 \frac{2|E|}{\mathrm{deg}(y)^2}h(y) - 1,
    \end{align}
    where we have set $h(y)=\sum_{x: \{x,y\}\in E} \frac{1}{\mathrm{deg}(x)}$.
    Consequently,
    \begin{align}
        \sum_y \left(\sum_{x} (g_y(x)-1)^2 \pi(x)\right) \pi(y) &= \sum_y \pi(y) \left(\bar{\lambda}^2 \frac{2|E|}{\mathrm{deg}(y)} +  \lambda^2 \frac{2|E|}{\mathrm{deg}(y)^2}h(y) - 1\right) \\
        &= |V|(1-\lambda)^2+ \lambda^2 \sum_{y} \frac{1}{\mathrm{deg}(y)}h(y) - 1. \label{eq:boundEtaChiGraph}
    \end{align}
    Let us compute~\Cref{eq:boundEtaChiGraph} in specific settings of interest. 
    In case the graph is complete, then $\sum_y \frac{1}{\mathrm{deg}(y)}h(y) = \frac{|V|}{|V|-1}$ and~\Cref{eq:boundEtaChiGraph} yields:
    \begin{equation}
        \eta_{\chi^2}(K_\lambda,\pi) \leq |V|(1-\lambda)^2+ \lambda^2 \frac{|V|}{|V|-1} - 1 \label{eq:boundEtaChiGraphComplete}.
    \end{equation}
    If $V=\{0,1\}$, then $K_\lambda$ corresponds to the $\text{BSC}(\lambda)$ and~\Cref{eq:boundEtaChiGraph} is tight.  For general $V$, one has that~\Cref{eq:boundEtaChiGraphComplete} is non trivial (smaller than $1$) if 
    \begin{equation}
        \frac{|V|-1-\sqrt{|V|-1}}{|V|}<\lambda <\frac{|V|-1+\sqrt{|V|-1}}{|V|}. \label{eq:rangeValidityCompleteGraph}
    \end{equation}
    To provide a concrete comparison, let us consider a setting studied in~\cite[Example 3.4]{sdpiRaginsky}: the path graph on the ternary vertex set $V=\{0,1,2\}$, \textit{i.e.}, $E=\{\{0,1\},\{1,2\}\}.$
    In this case, the two nonconstant eigenvalues are
    $1-\lambda$ and $1-2\lambda$.  Therefore
    $\eta_{\chi^2}(\pi,K_\lambda)
    =
    \max\{(1-\lambda)^2,(1-2\lambda)^2\}$, while \Cref{eq:boundEtaChiGraph} yields:
    \begin{equation}
        \eta_{\chi^2}(K_\lambda,\pi) \leq (1-\lambda)^2+(1-2\lambda)^2 = 5\lambda^2-6\lambda+2. \label{eq:boundEtaChiGraphPath}
    \end{equation}
    \Cref{fig:comparisonChi2VsKL} shows that our approach significantly improves upon~\cite[Eq. (3.41)]{sdpiRaginsky}, and it leads to a bound which is non-trivial for a larger range of values of $\lambda$: \Cref{eq:boundEtaChiGraphPath} is non-trivial for $0.200<\lambda<1$ while~\cite[Eq. (3.41)]{sdpiRaginsky} is non-trivial for $ (8-2\sqrt{3})/13<\lambda<1/2$. %
    Furthermore, one can find distributions $\nu$ over $V$ and choices of $\lambda$ such that $\chi^2(\nu K_\lambda\|\pi)/\chi^2(\nu\|\pi)= 5\lambda^2 -6\lambda + 2$, \textit{e.g.}, $\nu=(p,1/2,1/2-p)$ with $0\leq p<1/2$ and $\lambda=1/2$, while~\cite[Eq. (3.41)]{sdpiRaginsky} is strictly larger than $\chi^2(\nu K_\lambda\|\pi)/\chi^2(\nu\|\pi)$ for every choice of $\lambda$ and $\nu$.
    \begin{figure}
        \centering
        \includegraphics[width=0.70\textwidth]{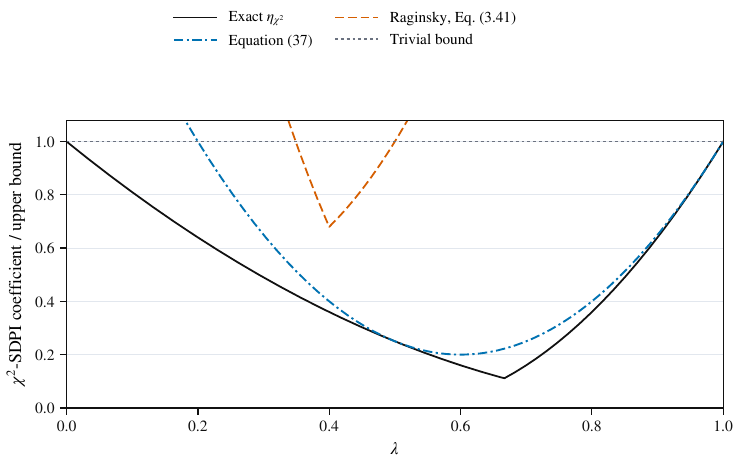}
        \caption{Behaviour of~\cite[Eq. (3.41)]{sdpiRaginsky} and~\Cref{eq:boundEtaChiGraphPath}, as a function of $\lambda$. 
        }
        \label{fig:comparisonChi2VsKL}
    \end{figure}
\end{example}

To conclude this section, focusing on the norm itself, rather than on $H_\alpha$, the case of $\alpha=\infty$ represents another %
     setting in which we can prove the tightness of~\Cref{thm:alphaNormsContraction}, as formalised below.
         \begin{corollary}~\label{thm:alphaInfinityCase}
    Let $K$ be a Markov kernel with stationary distribution $\pi$. Then, the following holds:
  \begin{equation}
        \left\lVert K^t-1_\pi \right\rVert_{L_\infty \to L_\infty} \leq \left\lVert \left\lVert g^t_X -1  \right\rVert_{L_1(\pi)} \right\rVert_{L_\infty(\pi)} = \esssup_{\pi} 2\left\lVert K^t(\cdot|X)-\pi \right\rVert_{TV},~\label{eq:alphaInfinityCase}
    \end{equation}
    where $g_x^t(y)=\frac{dK^t(y|x)}{d\pi K^t(y)} = \frac{dK^t(y|x)}{d\pi(y)}$.
\end{corollary}
The proof of~\Cref{thm:alphaInfinityCase} can be found in~\Cref{app:proofAlphaInfinityCase}.
In fact, $\left\lVert K^t-1_\pi \right\rVert_{L_\infty \to L_\infty}=\esssup_{x} 2\left\lVert K^t(\cdot|x)-\pi \right\rVert_{TV}$ by Proposition 3.23 in~\cite{mcmcThesis}. 
In the Markov chain literature, it is quite common to show that %
the right-hand side of~\Cref{eq:alphaInfinityCase} is bounded. This usually goes under the name of uniform (or strong) ergodicity. Hence,~\Cref{thm:alphaInfinityCase} leads to another proof that asking for uniform ergodicity implies $L_\infty$-exponential convergence. 

\Cref{thm:psiNormsContraction} also %
recovers %
other well-known results. %
Indeed, \cite{contractionLedoux} shows that, if ultra-mixing holds (which, in turn, implies Doeblin's minorisation~\cite[Section 4.3.3]{hiddenMarkovModels}), then one can bound the contraction coefficient of Markovian kernels in arbitrary norms. %
\begin{corollary}\label{thm:ultraMixing}
        Given a Markov kernel $K$ and probability measures $\mu,\nu$ with $\nu\ll\mu$, assume there exists $0<\epsilon<1$ s.t.\ the kernel satisfies the ultra-mixing condition,  \textit{i.e.}, for all $x,y$,
    \begin{equation}
        \frac{dK(\cdot|x)}{dK(\cdot|y)}\geq \epsilon,  \hspace{2em} K(\cdot|y)-a.e. \label{eq:ultraMixing}
    \end{equation}
    Then, for every Young function $\psi$ and $N\in\{A,L\}$, we have
     \begin{equation}
        \left\lVert \frac{d\nu K}{d\mu K} -1\right\rVert_{L^N_\psi(\mu K)} \leq (1-\epsilon) \left\lVert \frac{d\nu}{d\mu } -1 \right\rVert_{L^N_\psi(\mu)}.
    \end{equation}
    \end{corollary}
\Cref{thm:ultraMixing} is indeed a consequence of~\Cref{thm:psiNormsContraction}, and
its proof 
can be found in~\Cref{app:proofUltraMixing}.

\subsection{Kullback-Leibler and beyond}
    Similar tools can be leveraged to provide a bound on the KL divergence $\eta_{\mathrm{KL}}$. To do so, we draw inspiration from the mathematics underlying the Amemiya norm, generalising it to include objects like the Kullback-Leibler divergence, that are not norms.
    \subsubsection{Non-linear contraction and polar gauges}
   Let us first define the nonlinear contraction profile, or F-curve~\cite{contractionInputConstraint}.
   Let $\varphi$ be a convex functional, %
   $\mu$ a reference measure, and define the corresponding $\varphi$-Divergence as a functional acting on measures: $D_\varphi(\cdot\|\mu) = \psi_\mu^\varphi(\cdot)$. Define the divergence ball centred around $\mu$ as follows
   \begin{equation}
       \mathcal{B}_\varphi(\mu,c) = \{\nu\in \mathcal{P}(\X) : \nu\ll\mu , D_\varphi(\nu\|\mu)\leq c\} \label{eq:phiDivBall}.
   \end{equation}
   Given a Markov kernel $K$, the nonlinear contraction profile can be defined as follows
   \begin{equation}
       F_\varphi(\mu,K; c) = \sup_{\nu \in \B_\varphi(\mu,c) } D_\varphi(\nu K\|\mu K). \label{eq:F_phiCurve}
   \end{equation}
   One then has that:
   \begin{equation}
       D_\varphi(\nu K\|\mu K) \leq F_\varphi(\mu,K; D_\varphi(\nu\|\mu)) \overset{\mu,K,\varphi \text{ fixed}}{\equiv} F(D_\varphi(\nu\|\mu)).\label{phiNonLinearContraction}
   \end{equation}
   To obtain the linear contraction that characterises SDPI then, one needs the following:
   \begin{equation}
       \eta_\varphi(\mu,K) = \sup_{c>0} \frac{F_\varphi(\mu,K;c)}{c}.
   \end{equation}
   Some fundamental properties of~\Cref{eq:F_phiCurve} are the following
   \begin{equation}
       F_\varphi(\mu, K; 0)=0, \hspace{3em} F_\varphi(\mu,K; c)\leq c,
   \end{equation}
   where the second inequality follows from data processing. Moreover, one can have that $\eta_\varphi(\mu, K)=1$ but $F(c)<c$~\cite{contractionInputConstraint}.
    The following are useful observations:
    \begin{align}
        F_\varphi(\mu, K;c) &= \sup_{\nu \in \B_\varphi(\mu,c)} \int \varphi\left(1+\mu \left[\left(\frac{d\nu}{d\mu}-1\right)\left(g_Y-1\right)\right]\right) \mu K(\mathrm{d}y) \\
        &\leq \int \esssup_{\nu \in \B_\varphi(\mu,c)}\varphi\left(1+\mu \left[\left(\frac{d\nu}{d\mu}-1\right)\left(g_Y-1\right)\right]\right) \mu K(\mathrm{d}y).\label{eq:gaugeInFphi}
    \end{align}
    This motivates the following definition of the polar gauge induced by $D_\varphi$, defined for functions $f$ such that $\mu(f)=0$:
   \begin{equation}
       \rho_{\mu,\varphi}^c(f) = \sup_{\nu \in \B_\varphi(\mu,c)} \mu\left[\left(\frac{d\nu}{d\mu}-1\right)f\right] = \sup_{\nu \in \B_\varphi(\mu,c)} \nu(f) = \sup_{\nu \in \B_\varphi(\mu,c)} \langle \nu, f \rangle  . \label{eq:polarGauge}
   \end{equation}
   \begin{remark}
       The functional defined in~\Cref{eq:polarGauge} satisfies the following properties:
        \begin{enumerate}
        \item is non-negative and null at 0; \label{it:nonNeg}
        \item invariant under constant displacement, \textit{i.e.}, $\rho^c_{\mu,\varphi}(f+a)= \rho^c_{\mu,\varphi}(f)$ for $a \in \mathbb{R}$;
        \item positive homogeneous, \textit{i.e.}, given $\lambda\geq 0$ one has that $\rho^c_{\mu,\varphi}(\lambda f) = \lambda \rho^c_{\mu,\varphi}(f);$\label{it:posHom}
      
        \item need not be symmetric, \textit{i.e.}, $\rho^c_{\mu,\varphi}(f)$ need not equal $\rho^c_{\mu,\varphi}(-f)$;
        \item convex. \label{it:convex}
    \end{enumerate}
   \end{remark}
   A bound on the polar gauge (also known as the support function of $\B_\varphi(\mu,c)$) defined in~\Cref{eq:polarGauge} will naturally yield a bound on the corresponding $F_\varphi$-curve via \Cref{eq:gaugeInFphi}, given that $g_Y-1$ has mean zero with respect to $\mu$.
   Moreover, leveraging the variational representations of $\varphi$-divergences, it admits an Amemiya-norm-like representation which reduces the problem to a one-dimensional convex optimisation problem. 
   More precisely, given a reference measure $\mu$, by convexity of $\varphi$, $D_\varphi(\cdot\|\mu)$ is a convex functional and it admits a Legendre-Fenchel dual $D_\varphi(\cdot\|\mu)^\star(\cdot)$ which acts on functions. For technical details of the spaces over which these objects are defined, the reader is referred to~\cite{minimizationMeasures,generalisedTPC}.
   From these observations, along with~\Cref{eq:polarGauge} and Fenchel-Young duality one recovers the following bound for every $\nu \in \B_\varphi(\mu,c)$, $f$ such that $\mu(f)=0$ and $t>0$
   \begin{align}
        \langle \nu, f\rangle = \nu(f) &\overset{t\neq 0}{=} \frac{\nu(tf)}{t} \overset{t>0}{\leq} \frac{D_\varphi(\nu\|\mu) + D_\varphi(\cdot\|\mu)^\star(tf)}{t}\overset{\nu \in \B_\varphi(\mu,c)}{\leq} \frac{c + D_\varphi(\cdot\|\mu)^\star(tf)}{t}.
   \end{align}
   Consequently,
   \begin{equation}
        \rho_{\mu,\varphi}^c(f) = \sup_{\nu \in \B_\varphi(\mu,c)} \nu(f) \leq \inf_{\substack{t>0}} \frac{c + D_\varphi(\cdot\|\mu)^\star(tf)}{t}. \label{eq:amemiyaLikePhiFunctional}
   \end{equation}
   \begin{remark}[Equality in the dual representation]
Define the probability-constrained dual functional by
\begin{equation}
\Lambda_{\varphi,\mu}(f)
\coloneqq
\sup_{\substack{\nu\in\mathcal P(\mathcal X)\\ \nu\ll\mu}}
\left\{
\nu(f)-D_\varphi(\nu\Vert\mu)
\right\}.
\end{equation}
For every centred function $f$, weak duality gives
\begin{equation}
\rho_{\mu,\varphi}^{c}(f)
\leq
\inf_{t>0}
\frac{c+\Lambda_{\varphi,\mu}(tf)}{t}.
\end{equation}
If $c>0$ and strong duality holds for the divergence-ball
optimisation, then the preceding inequality is an equality. This is
the case, in particular, on finite alphabets for the usual closed
convex divergence generators. Indeed, $\mu$ is strictly feasible
because $D_\varphi(\mu\Vert\mu)=0<c$.

The restriction of the dual problem to probability measures is
essential. If the conjugate is instead computed over arbitrary
nonnegative measures, one generally obtains only an upper bound on
the polar gauge. The probability constraint can equivalently be
incorporated through the normalisation parameter
\begin{equation}
\Lambda_{\varphi,\mu}(f)
=
\inf_{a\in\mathbb R}
\left\{
a+\mu\!\left[\varphi^\star(f-a)\right]
\right\}.
\end{equation}
\end{remark}
   We therefore refer to the right-hand side of~\eqref{eq:amemiyaLikePhiFunctional} (cf.~\Cref{eq:amemiyaNorm}) as an Amemiya-type dual functional. This terminology concerns its variational structure only: unlike an Amemiya norm, the functional is generally one-sided, depends on the divergence radius \(c\), and need not be symmetric.
   \begin{remark}
       The one-sided polar gauge can be symmetrised by defining
\begin{equation}
\|f\|_{\mu,\varphi,c}^{\circ}
\coloneqq
\max\left\{
\rho_{\mu,\varphi}^{c}(f),
\rho_{\mu,\varphi}^{c}(-f)
\right\}
=
\sup_{\nu\in\B_\varphi(\mu,c)}
\left|\nu(f)-\mu(f)\right|.
\end{equation}
This functional is absolutely homogeneous and subadditive. Since it
vanishes on constant functions, it is generally an extended seminorm,
and it induces a norm on functions modulo constants whenever it is
finite and
\[
\|f\|_{\mu,\varphi,c}^{\circ}=0
\quad\Longrightarrow\quad
f \text{ is constant } \mu\text{-a.s.}
\]
Applying~\Cref{eq:amemiyaLikePhiFunctional} to both $f$ and $-f$
provides an Amemiya-type upper bound on this symmetric polar
seminorm. Whenever equality holds in the dual representation, this
becomes an Amemiya-type representation of the corresponding norm on
functions modulo constants.
   \end{remark}
   
\subsubsection{Hellinger distance}
In order to illustrate the applicability of the preceding framework, let us consider the squared Hellinger distance \textit{i.e.}, the divergence induced by $\varphi_{\mathrm H}(x)=\frac{1}{2}(\sqrt{x}-1)^2$. First we provide a closed-form bound on the polar gauge and then we translate it into a bound on the corresponding $F_{\varphi_{\mathrm H}}$-curve.
   \begin{proposition}[Polar gauge of a squared Hellinger ball]\label{prop:hellingerPolarGauge}
    Let $\mu$ be a probability measure and
    \(
        \varphi_{\mathrm H}(u)=\frac{1}{2}(\sqrt{u}-1)^2,
        u\geq 0,
    \)
    and let
    \( 
        H^2(\nu,\mu)
        =
        D_{\varphi_{\mathrm H}}(\nu\Vert\mu)
        =
        \frac{1}{2}
        \mu\left[
        \left(
        \sqrt{\frac{\mathrm d\nu}{\mathrm d\mu}}-1
        \right)^2
        \right].
    \) Let $f\in L^2(\mu)\cap L^\infty(\mu)$ satisfy $\mu(f)=0$.
    One has that
    \begin{align} \rho_{\mu,\varphi_{\mathrm H}}^c(f)
        &\leq
        2\sqrt{2c}\,
        \lVert f\rVert_{L^2(\mu)}
        +
        2c\operatorname*{ess\,sup}_{\mu}f.
        \label{eq:hellingerPolarExplicit}\\
  \rho_{\mu,\varphi_{\mathrm H}}^c(-f)
        &\leq
        2\sqrt{2c}\,
        \lVert f\rVert_{L^2(\mu)}
        -
        2c\operatorname*{ess\,inf}_{\mu}f.
        \label{eq:hellingerPolarExplicitNegative}
    \end{align}
\end{proposition}

The proof of~\Cref{prop:hellingerPolarGauge} can be found in~\Cref{app:proofHellingerPolarGauge}.

For the remainder of this subsection, let $\mu$ be a probability measure on a standard Borel space $\mathcal X$, let $K$ be a Markov kernel from $\mathcal X$ to a standard Borel space $\mathcal Y$, and let $K_\mu^\star$ be its reverse Markov kernel. Assume that, for $\mu K$-almost every $y$, $K_\mu^\star(\cdot\mid y)\ll\mu$, that the reverse-kernel density $g_y$ in~\Cref{eq:reverseKernelDensityEndpoints} admits a jointly measurable version, and that $g_y\in L^2(\mu)\cap L^\infty(\mu)$.
Since $\mu(g_y)=1$, applying
\Cref{eq:hellingerPolarExplicit,eq:hellingerPolarExplicitNegative}
to $f=g_y-1$ gives
\begin{align}
    \rho_{\mu,\varphi_{\mathrm H}}^c(g_y-1)
    &\leq
    \min\Bigl\{
        M_y-1,\,
        2\sqrt{2c}\,
        \lVert g_y-1\rVert_{L^2(\mu)}
        +
        2c\left(
        M_y-1
        \right)
    \Bigr\},
    \label{eq:hellingerPositiveGaugeChannelClipped}\\
    \rho_{\mu,\varphi_{\mathrm H}}^c(1-g_y)
    &\leq
    \min\Bigl\{
        1-m_y,\,
        2\sqrt{2c}\,
        \lVert g_y-1\rVert_{L^2(\mu)}
        +
        2c\left(
        1-m_y
        \right)
    \Bigr\}.
    \label{eq:hellingerNegativeGaugeChannelClipped}
\end{align}
Consequently, one has the following bound on the corresponding $F_{\varphi_{\mathrm H}}$-curve.
\begin{theorem}[Nonlinear contraction of the squared Hellinger divergence]
\label{thm:hellingerNonlinearContraction}

Under the standing assumptions on $\mu$, $K$, and $K_\mu^\star$ above, for every $c>0$,
\begin{align}
F_{\varphi_{\mathrm H}}(\mu,K;c)
\leq
\min\Bigg\{
c,\,
\frac{1}{2}
\int
\max\Bigg\{
&\left[
1-
\sqrt{
1-
\min\Bigl\{
1-m_y,\,
2\sqrt{2c}\,
\lVert g_y-1\rVert_{L^2(\mu)}
+
2c\left(
1-m_y
\right)
\Bigr\}
}
\right]^2,
\nonumber\\
&
\left[
\sqrt{
1+
\min\Bigl\{
M_y-1,\,
2\sqrt{2c}\,
\lVert g_y-1\rVert_{L^2(\mu)}
+
2c\left(
M_y-1
\right)
\Bigr\}
}
-1
\right]^2
\Bigg\}
\,\mu K(\mathrm dy)
\Bigg\}.
\label{eq:hellingerNonlinearContraction}
\end{align}
\end{theorem}

The proof of~\Cref{thm:hellingerNonlinearContraction} can be found in~\Cref{app:proofHellingerNonlinearContraction}.
To conclude, we can present a bound on the linear SDPI coefficient of the Hellinger distance.

\begin{corollary}[Hellinger SDPI bound]
\label{cor:hellingerSDPI}

Under the assumptions of
\Cref{thm:hellingerNonlinearContraction}, the contraction coefficient of the Hellinger distance with respect to the measure $\mu$ and the kernel $K$ satisfies
\begin{align}
\eta_{\varphi_{\mathrm H}}(\mu,K)
\leq
\min\Bigg\{
1,\,
\int
\max\Bigg\{
&
\left(
\frac{
\lVert g_y-1\rVert_{L^2(\mu)}
+
\sqrt{
\lVert g_y-1\rVert_{L^2(\mu)}^2
+
(1-m_y)^2
}
}{
1+\sqrt{m_y}
}
\right)^2,
\nonumber\\
&
\frac{1}{4}
\left(
\lVert g_y-1\rVert_{L^2(\mu)}
+
\sqrt{
\lVert g_y-1\rVert_{L^2(\mu)}^2
+
(M_y-1)^2
}
\right)^2
\Bigg\}
\,\mu K(\mathrm dy)
\Bigg\}.
\label{eq:hellingerSDPIClosedFormBound}
\end{align}
\end{corollary}

The proof of~\Cref{cor:hellingerSDPI} can be found in~\Cref{app:proofHellingerSDPI}.

Given that $\varphi_{\mathrm H}$ is operator convex, one has that~\cite{sdpiRaginsky}
\begin{equation}
    \eta_{\chi^2}(\mu ,K) \leq \eta_{\varphi_{\mathrm H}}(\mu,K) \leq \eta_{\chi^2}(K) \leq \eta_{TV}(K). \label{eq:trivialBoundsHellinger}
\end{equation}
Below we compare~\Cref{eq:hellingerNonlinearContraction} and~\Cref{eq:hellingerSDPIClosedFormBound} to~\Cref{eq:trivialBoundsHellinger} in a concrete ternary example.
\begin{example}
\label{ex:ternaryHellingerImprovement}

Let $\mathcal X=\mathcal Y=\{1,2,3\}$, and consider
\begin{equation}
    \mu
    =
    \begin{pmatrix}
        0.050 & 0.850 & 0.100
    \end{pmatrix},
    \qquad
    K
    =
    \begin{pmatrix}
        0.550 & 0.150 & 0.300\\
        0.350 & 0.250 & 0.400\\
        0.200 & 0.350 & 0.450
    \end{pmatrix},
\end{equation}
where the rows of $K$ are indexed by $x\in\mathcal X$. Since
$\det(K)=-0.005$, the channel has full rank.
The output distribution is
\begin{equation}
    \mu K
    =
    \begin{pmatrix}
        0.345 & 0.255 & 0.400
    \end{pmatrix}.
\end{equation}

For every $y\in\mathcal Y$, the reverse-kernel density in~\Cref{eq:reverseKernelDensityEndpoints} in this example yields the following matrix
\begin{equation}
    \begin{pmatrix}
        g_1(1) & g_2(1) & g_3(1)\\
        g_1(2) & g_2(2) & g_3(2)\\
        g_1(3) & g_2(3) & g_3(3)
    \end{pmatrix}
    =
    \begin{pmatrix}
        \dfrac{110}{69} & \dfrac{10}{17} & \dfrac{3}{4}\\[1mm]
        \dfrac{70}{69}  & \dfrac{50}{51} & 1\\[1mm]
        \dfrac{40}{69}  & \dfrac{70}{51} & \dfrac{9}{8}
    \end{pmatrix}.
\end{equation}
Consequently, the endpoint quantities defined in~\Cref{eq:reverseKernelDensityEndpoints} and the corresponding $L^2(\mu)$ norms are
\begin{align}
    (m_1,m_2,m_3)
    &=
    \left(
        \frac{40}{69},
        \frac{10}{17},
        \frac{3}{4}
    \right),\\
    (M_1,M_2,M_3)
    &=
    \left(
        \frac{110}{69},
        \frac{70}{51},
        \frac{9}{8}
    \right),\\
    (\lVert g_1-1\rVert_{L^2(\mu)},\lVert g_2-1\rVert_{L^2(\mu)},\lVert g_3-1\rVert_{L^2(\mu)})
    &\simeq
    (0.188,0.151,0.068).
\end{align}

The closed-form Hellinger SDPI bound from
\Cref{cor:hellingerSDPI} is
obtained by writing $s_y=\lVert g_y-1\rVert_{L^2(\mu)}$:
\begin{align}
    \eta_{H^2}(\mu,K)
    \leq
    \min\Bigg\{
    1,\,
    \sum_y \mu K(y)
    \max\Bigg\{&
        \left(
        \frac{
            s_y+\sqrt{s_y^2+(1-m_y)^2}
        }{
            1+\sqrt{m_y}
        }
        \right)^2,
        \frac{1}{4}
        \left(
            s_y+\sqrt{s_y^2+(M_y-1)^2}
        \right)^2
    \Bigg\}
    \Bigg\} \simeq
    0.098    .
\end{align}
Let $\overline F_{\varphi_{\mathrm H}}(\mu,K;c)$ denote the right-hand side of~\Cref{eq:hellingerNonlinearContraction}. It follows from \Cref{thm:hellingerNonlinearContraction} that
\begin{equation}
    \eta_{\varphi_{\mathrm H}}(\mu,K)
    \leq
    \sup_{0<c\leq1}
    \frac{
        \overline F_{\varphi_{\mathrm H}}(\mu,K;c)
    }{c} \simeq
    0.082.
\end{equation}

For comparison, the distribution independent contraction coefficient of $\chi^2$ is
\begin{align}
    \eta_{\chi^2}(K)
    =
    \max_{x\neq x'}
    \sup_{0<\lambda<1}
    \lambda(1-\lambda)
    \sum_y
    \frac{
        \left(
            K(y\mid x)-K(y\mid x')
        \right)^2
    }{
        \lambda K(y\mid x)
        +(1-\lambda)K(y\mid x')
    }
    \simeq
    0.137.
\end{align}
The Dobrushin contraction coefficient is
\begin{equation}
    \eta_{\mathrm{TV}}(K)
    =
    \frac{1}{2}
    \max_{x\neq x'}
    \sum_y
    \left|
        K(y\mid x)-K(y\mid x')
    \right|
    =
    0.350.
\end{equation}
Consequently,
\begin{equation}
    0.039 \simeq \eta_{\varphi_{\mathrm H}}(\mu,K)
    \leq
    0.082
    <
    0.098
    <
    \eta_{\chi^2}(K)
    \simeq
    0.137
    <
    \eta_{\mathrm{TV}}(K)
    =
    0.350,
\end{equation}
where the contraction coefficient of the Hellinger distance has been computed numerically. 
This example demonstrates that the distribution-dependent Hellinger
bounds can improve on standard distribution independent bounds.
\end{example}
\subsubsection{Kullback-Leibler Divergence} 
We will now provide bounds on the contraction of KL using similar tools as above. First we derive a general bound and then, in order to reach computable expressions, we will provide a sequence of approximations that render it possible.  First we prove a bound on the $F_{\varphi_{KL}}$-curve.
\begin{theorem}
\label{thm:klNonlinearContraction}
Let $\mu$ be a probability measure and $K$ a Markov kernel. Assume that, for $\mu K$-almost every $y$, the centred exponential
moment $\mu(e^{t(g_y-1)})$ is finite for $t$ in a neighbourhood of the
origin. 
Let $\varphi_{\mathrm{KL}}(u)
    =
    u\log u-u+1, u\geq0,$ with the convention $\varphi_{\mathrm{KL}}(0)=1$. 
    The following holds true \begin{align}
F_{\mathrm{KL}}(\mu,K;c)
\leq
\min\Bigg\{
c,\,
\int
\max\Bigg\{&
\varphi_{\mathrm{KL}}
\left(
1-\rho_{\mu,\mathrm{KL}}^c(1-g_y)
\right),
\varphi_{\mathrm{KL}}
\left(
1+\rho_{\mu,\mathrm{KL}}^c(g_y-1)
\right)
\Bigg\}
\,\mu K(\mathrm dy)
\Bigg\}.
\label{eq:klNonlinearContraction}
\end{align}

\end{theorem}

The proof of~\Cref{thm:klNonlinearContraction} can be found in~\Cref{app:proofKLNonlinearContraction}.

As for the corresponding bound on the SDPI constant, one retrieves the following.
\begin{corollary}\label{cor:klSDPIPolarGauge}
    \begin{align}
\eta_{\mathrm{KL}}(\mu,K) =  
    \sup_{c>0}
    \frac{
        F_{\mathrm{KL}}(\mu,K;c)
    }{c} 
\leq
\sup_{c>0}
\frac{1}{c}
\min\Bigg\{
c,\,
\int
\max\Bigg\{&
\varphi_{\mathrm{KL}}
\left(
1-\rho_{\mu,\mathrm{KL}}^c(1-g_y)
\right),
\nonumber\\
&
\varphi_{\mathrm{KL}}
\left(
1+\rho_{\mu,\mathrm{KL}}^c(g_y-1)
\right)
\Bigg\}
\,\mu K(\mathrm dy)
\Bigg\}.
\label{eq:klSDPIExactMGF}
\end{align}
\end{corollary}
The difficulty in computing the bound and solving the optimisation over $c$ comes from two sources: $\varphi_{KL}$ and the infimal representation of $\rho_{\mu,KL}^c$. We can thus provide looser estimates by approximating both quantities into expressions that are more amenable to analysis. 
Starting with $\rho_{\mu,KL}^c$, we will provide closed-form quantities approximating $\log \mu(\exp(\pm t(g_y-1)))$. The first result stems from a quadratic approximation known in the literature as Bennett's inequality.
\begin{theorem}
\label{thm:klBennettContraction}
Assume that $M_y<\infty$ for $\mu K$-almost every $y$. Then, for
every $c>0$,
\begin{align}
F_{\varphi_{\mathrm{KL}}}(\mu,K;c)
\leq
\min\left\{
\begin{aligned}
&c,\\
&\int
\max\left\{
\begin{aligned}
&\varphi_{\mathrm{KL}}\left(
1-
\min\left\{
\begin{aligned}
&1-m_y,\\
&\begin{aligned}
&\frac{\lVert g_y-1\rVert_{L^2(\mu)}^2}{1-m_y}\\[-1mm]
&\quad{}\times
\left[\varphi_{\mathrm{KL}}(1+\cdot)\right]^{-1}
\left(
\frac{c(1-m_y)^2}{\lVert g_y-1\rVert_{L^2(\mu)}^2}
\right)
\end{aligned}
\end{aligned}
\right\}
\right),\\
&\varphi_{\mathrm{KL}}\left(
1+
\min\left\{
\begin{aligned}
&M_y-1,\\
&\begin{aligned}
&\frac{\lVert g_y-1\rVert_{L^2(\mu)}^2}{M_y-1}\\[-1mm]
&\quad{}\times
\left[\varphi_{\mathrm{KL}}(1+\cdot)\right]^{-1}
\left(
\frac{c(M_y-1)^2}{\lVert g_y-1\rVert_{L^2(\mu)}^2}
\right)
\end{aligned}
\end{aligned}
\right\}
\right)
\end{aligned}
\right\}
\,\mu K(\mathrm dy)
\end{aligned}
\right\}.
\label{eq:klBennettContractionCurve}
\end{align}

Consequently,
\begin{align}
\eta_{\mathrm{KL}}(\mu,K)
\leq
\sup_{c>0}\frac{1}{c}
\min\left\{
\begin{aligned}
&c,\\
&\int
\max\left\{
\begin{aligned}
&\varphi_{\mathrm{KL}}\left(
1-
\min\left\{
\begin{aligned}
&1-m_y,\\
&\begin{aligned}
&\frac{\lVert g_y-1\rVert_{L^2(\mu)}^2}{1-m_y}\\[-1mm]
&\quad{}\times
\left[\varphi_{\mathrm{KL}}(1+\cdot)\right]^{-1}
\left(
\frac{c(1-m_y)^2}{\lVert g_y-1\rVert_{L^2(\mu)}^2}
\right)
\end{aligned}
\end{aligned}
\right\}
\right),\\
&\varphi_{\mathrm{KL}}\left(
1+
\min\left\{
\begin{aligned}
&M_y-1,\\
&\begin{aligned}
&\frac{\lVert g_y-1\rVert_{L^2(\mu)}^2}{M_y-1}\\[-1mm]
&\quad{}\times
\left[\varphi_{\mathrm{KL}}(1+\cdot)\right]^{-1}
\left(
\frac{c(M_y-1)^2}{\lVert g_y-1\rVert_{L^2(\mu)}^2}
\right)
\end{aligned}
\end{aligned}
\right\}
\right)
\end{aligned}
\right\}
\,\mu K(\mathrm dy)
\end{aligned}
\right\}.
\label{eq:klBennettSDPI}
\end{align}
Whenever $g_y=1$ $\mu$-almost surely, the corresponding integrand is
understood to be zero. Here $
    \left[
        \varphi_{\mathrm{KL}}(1+\cdot)
    \right]^{-1}$ denotes the inverse of the strictly increasing function
$u\mapsto\varphi_{\mathrm{KL}}(1+u)$ on $[0,\infty)$.
\end{theorem}

The proof of~\Cref{thm:klBennettContraction} can be found in~\Cref{app:proofKLBennettContraction}.
\begin{remark}
\label{rem:inverseKLGenerator}

The inverse appearing in
\Cref{thm:klBennettContraction} can be written explicitly using the
principal branch $W_0$ of the Lambert $W$ function. In particular,
\begin{equation}
    \left[
        \varphi_{\mathrm{KL}}(1+\cdot)
    \right]^{-1}(r)
    =
    \frac{
        r-1
    }{
        W_0\left((r-1)/e\right)
    }
    -1,
    \qquad r\geq0.
    \label{eq:inverseKLGeneratorLambertW}
\end{equation}
At $r=1$, the right-hand side is understood by continuity and equals
$e-1$. At $r=0$, it equals zero. Hence, the Bennett bounds in
\Cref{thm:klBennettContraction} are explicit up to evaluation of the
standard Lambert $W$ function and require no additional optimisation
over the MGF parameter.
\end{remark}
An elementary relaxation of Bennett's approximation of the CGF is known as Bernstein's inequality and leads to the following bound.
\begin{corollary}[Closed-form Bernstein bound on the KL contraction coefficient]
\label{cor:klBernsteinClosedFormSDPI}

Under the assumptions of
\Cref{thm:klBennettContraction} one has that
\begin{align}
\eta_{\mathrm{KL}}(\mu,K)
\leq
\min\Bigg\{
1,\,
\int
\max\Bigg\{
&
\frac{
    \varphi_{\mathrm{KL}}(m_y)
}{
    2(1-m_y)^2
}
\left(
    \lVert g_y-1\rVert_{L^2(\mu)}
    +
    \sqrt{
        \lVert g_y-1\rVert_{L^2(\mu)}^2
        +
        \frac{2}{3}(1-m_y)^2
    }
\right)^2,
\nonumber\\
&
\frac{1}{4}
\left(
    \lVert g_y-1\rVert_{L^2(\mu)}
    +
    \sqrt{
        \lVert g_y-1\rVert_{L^2(\mu)}^2
        +
        \frac{2}{3}(M_y-1)^2
    }
\right)^2
\Bigg\}
\,\mu K(\mathrm dy)
\Bigg\}.
\label{eq:klBernsteinClosedFormSDPI}
\end{align}
If $g_y=1$ $\mu$-almost surely, the corresponding integrand is
understood to be zero.
\end{corollary}

The proof of~\Cref{cor:klBernsteinClosedFormSDPI} can be found in~\Cref{app:proofKLBernsteinClosedFormSDPI}.
A different approximation of the CGF of $g_Y-1$ comes from Hoeffding's inequality and the sub-Gaussian norm~\cite{vershynin_2018}. 
\begin{definition}
Let 
$ \Psi_2(x)\coloneqq e^{x^2}-1, x\geq0.$
The centred sub-Gaussian Orlicz norm of a measurable function $f$ is
\begin{align}
\lVert f\rVert_{L^L_{\Psi_2}(\mu)}
&\coloneqq
\lVert f-\mu(f)\rVert_{L^L_{\Psi_2}(\mu)} =
\inf\left\{
\lambda>0:
\mu\left(
\Psi_2\left(
\frac{|f-\mu(f)|}{\lambda}
\right)
\right)
\leq1
\right\}.
\label{eq:subGaussianOrliczNorm}
\end{align}
This is the Luxemburg norm associated with $\Psi_2$, applied after
centring. It is therefore a seminorm on functions and a norm on
functions modulo constants.
\end{definition}

A closely connected quantity is the sub-Gaussian norm used
in~\cite{sdpiRaginsky}:
\begin{equation}
\lVert f\rVert_{\mathrm{SG}(\mu)}
\coloneqq
\inf\left\{
\sigma\geq0:
\log\mu\left(
e^{t(f-\mu(f))}
\right)
\leq
\frac{\sigma^2t^2}{2}
\quad\text{for every }t\in\mathbb R
\right\}.
\label{eq:subGaussianNorm}
\end{equation}
The Orlicz norm
$\lVert f\rVert_{L_{\Psi_2}^{L}(\mu)}$ characterises
sub-Gaussian functions through exponential-square integrability,
whereas~\eqref{eq:subGaussianNorm} gives the smallest quadratic
upper bound on the centred log moment-generating function.
The two norms are equivalent up to universal constants, but are not
in general numerically equal.  The latter is particularly convenient
for the KL dual representation, since, by definition,
\begin{equation}
\log\mu\left(e^{t(f-\mu(f))}\right)
\leq
\frac{t^2}{2}
\lVert f\rVert_{\mathrm{SG}(\mu)}^2,
\qquad t\in\mathbb R.
\label{eq:subGaussianMGFBound}
\end{equation}
\begin{theorem}[Sub-Gaussian contraction of relative entropy]
\label{thm:subGaussianKLContraction}
Let $Y\sim\mu K$, with $m_Y$ and $M_Y$ as defined in~\Cref{eq:reverseKernelDensityEndpoints}.
Then, for every $c>0$,
\begin{align}
F_{\varphi_{\mathrm{KL}}}(\mu,K;c)
\leq
\min\Bigg\{
c,\,
\mu K\Bigg(
\max\Bigg\{
&
\varphi_{\mathrm{KL}}\left(
1-
\min\left\{
1-m_Y,\,
\sqrt{2c}\,
\lVert g_Y-1\rVert_{\mathrm{SG}(\mu)}
\right\}
\right),
\nonumber\\
&
\varphi_{\mathrm{KL}}\left(
1+
\min\left\{
M_Y-1,\,
\sqrt{2c}\,
\lVert g_Y-1\rVert_{\mathrm{SG}(\mu)}
\right\}
\right)
\Bigg\}
\Bigg)
\Bigg\}.
\label{eq:subGaussianKLNonlinearContraction}
\end{align}
Consequently,
\begin{equation}
\eta_{\mathrm{KL}}(\mu,K)
\leq
\min\left\{
1,\,
2\mu K\left(
\kappa(m_Y)
\lVert g_Y-1\rVert_{\mathrm{SG}(\mu)}^2
\right)
\right\},
\label{eq:subGaussianKLSDPI}
\end{equation}
where \(\kappa(a)\coloneqq \varphi_{\mathrm{KL}}(a)/(1-a)^2\) for \(0\leq a<1\), with \(\kappa(1)\coloneqq 1/2\).
\end{theorem}

The proof of~\Cref{thm:subGaussianKLContraction} can be found in~\Cref{app:proofSubGaussianKLContraction}.

The last relaxations we will present recover results established in the literature (cf.~\cite{sdpiRaginsky}) along with some closed-form expression upper-bounds stemming from Kearns--Saul's and Hoeffding's classical approximation of the CGF~\cite{kearnsSaul1998, berendKontorovich2013}. 
\begin{corollary}
\label{cor:subGaussianKLRelaxations}
Suppose that $m_Y$ and $M_Y$ are finite $\mu K$-almost surely.
Then
\begin{equation}
\eta_{\mathrm{KL}}(\mu,K)
\leq
\min\left\{
1,\,
2\mu K\left(
\lVert g_Y-1\rVert_{\mathrm{SG}(\mu)}^2
\right)
\right\}.
\label{eq:raginskySubGaussianBound}
\end{equation}
Moreover, the sub-Gaussian norm admits the closed-form upper bounds
\begin{align}
\lVert g_Y-1\rVert_{\mathrm{SG}(\mu)}^2
&\leq
\frac{
(M_Y-m_Y)(M_Y+m_Y-2)
}{
2\log\left(
\dfrac{M_Y-1}{1-m_Y}
\right)
}
\label{eq:kearnsSaulSubGaussianNorm}\\
&\leq
\frac{(M_Y-m_Y)^2}{4}.
\label{eq:hoeffdingSubGaussianNorm}
\end{align}
Consequently,
\begin{align}
\eta_{\mathrm{KL}}(\mu,K)
&\leq
\min\Bigg\{
1,\,
\mu K\left(
\frac{
(M_Y-m_Y)(M_Y+m_Y-2)
}{
\log\left(
\dfrac{M_Y-1}{1-m_Y}
\right)
}
\right)
\Bigg\}
\label{eq:kearnsSaulKLBound}\\
&\leq
\min\left\{
1,\,
\frac12
\mu K\left(
(M_Y-m_Y)^2
\right)
\right\}.
\label{eq:hoeffdingKLBound}
\end{align}
When $M_Y+m_Y=2$, the quotient in
\eqref{eq:kearnsSaulSubGaussianNorm} and~\eqref{eq:kearnsSaulKLBound} is defined as
$\frac{(M_Y-m_Y)^2}{4}.$
\end{corollary}

The proof of~\Cref{cor:subGaussianKLRelaxations} can be found in~\Cref{app:proofSubGaussianKLRelaxations}.
   
   We will now showcase the results above in a variety of examples, showing how each approach is not uniformly better than any other and can yield meaningful and, most importantly, improved with respect to the state-of-the-art bounds on the SDPI constant of KL. 

\begin{example}[General binary channel]\label{ex:generalBinaryChannel}
Let $\mathcal X=\mathcal Y=\{0,1\}$, let
$\mu=\operatorname{Bern}(p)$, and consider
\begin{equation}
K=
\begin{pmatrix}
1-a & a\\
b & 1-b
\end{pmatrix}.
\end{equation}
Then
\begin{equation}
\mu K(0)=(1-p)(1-a)+pb,
\qquad
\mu K(1)=(1-p)a+p(1-b),
\end{equation}
and
\begin{equation}
g_1(x)-1
=
\frac{1-a-b}{\mu K(1)}(x-p),
\qquad
g_0(x)-1
=
-\frac{1-a-b}{\mu K(0)}(x-p).
\label{eq:generalBinaryReverseDensities}
\end{equation}
The endpoint values are
\begin{align}
m_0&=\frac{\min\{1-a,b\}}{\mu K(0)},
&
M_0&=\frac{\max\{1-a,b\}}{\mu K(0)},
\nonumber\\
m_1&=\frac{\min\{a,1-b\}}{\mu K(1)},
&
M_1&=\frac{\max\{a,1-b\}}{\mu K(1)}.
\label{eq:generalBinaryEndpoints}
\end{align}

Every probability distribution on the input alphabet can be written as
$\nu=\operatorname{Bern}(r)$, and
\[
\nu K
=
\operatorname{Bern}\bigl((1-r)a+r(1-b)\bigr).
\]
Consequently,
\begin{equation}
\eta_{\mathrm{KL}}(\mu,K)
=
\sup_{\substack{0\leq r\leq1\\r\neq p}}
\frac{
d_{\mathrm{KL}}\!\left(
(1-r)a+r(1-b)
\,\middle\|\,
\mu K(1)
\right)
}{
d_{\mathrm{KL}}(r\|p)
}.
\label{eq:generalBinaryExactKL}
\end{equation}

The optimal sub-Gaussian variance proxy of a centred Bernoulli random
variable gives
\begin{align}
\left\lVert g_0-1\right\rVert_{\mathrm{SG}(\mu)}^2
&=
\frac{(1-a-b)^2}{\mu K(0)^2}
\frac{1-2p}{
2\log\!\left(\frac{1-p}{p}\right)},
\nonumber\\
\left\lVert g_1-1\right\rVert_{\mathrm{SG}(\mu)}^2
&=
\frac{(1-a-b)^2}{\mu K(1)^2}
\frac{1-2p}{
2\log\!\left(\frac{1-p}{p}\right)}.
\label{eq:generalBinarySubGaussianNorms}
\end{align}
At $p=1/2$, the quotient in
\eqref{eq:generalBinarySubGaussianNorms} has continuous value $1/4$.

Substitution into the nonlinear sub-Gaussian contraction bound yields
\begin{small}
\begin{align}F_{\varphi_{\mathrm{KL}}}(\mu,K;c)\leq
\min\left\{
\begin{aligned}
&c,\\
&\begin{aligned}
&\sum_{y\in\{0,1\}}\mu K(y)
\max\left\{
\begin{aligned}
&\varphi_{\mathrm{KL}}\!\left(
1-
\min\left\{
\begin{aligned}
&1-\frac{\min_{x\in\{0,1\}}K(y\mid x)}{\mu K(y)},\\
&\frac{|1-a-b|}{\mu K(y)}
\sqrt{\frac{c(1-2p)}{\log((1-p)/p)}}
\end{aligned}
\right\}
\right),\\
&\varphi_{\mathrm{KL}}\!\left(
1+
\min\left\{
\begin{aligned}
&\frac{\max_{x\in\{0,1\}}K(y\mid x)}{\mu K(y)}-1,\\
&\frac{|1-a-b|}{\mu K(y)}
\sqrt{\frac{c(1-2p)}{\log((1-p)/p)}}
\end{aligned}
\right\}
\right)
\end{aligned}
\right\}
\end{aligned}
\end{aligned}
\right\}.
\label{eq:generalBinaryNonlinearSubGaussianCurve}
\end{align}
\end{small}
The corresponding closed-form refinement is
\begin{align}
\eta_{\mathrm{KL}}(\mu,K)
\leq
\min\Bigg\{1,\,
&(1-a-b)^2
\frac{1-2p}{\log((1-p)/p)}
\Bigg[
\frac{
\kappa\!\left(
\frac{\min\{1-a,b\}}{\mu K(0)}
\right)
}{\mu K(0)}
+
\frac{
\kappa\!\left(
\frac{\min\{a,1-b\}}{\mu K(1)}
\right)
}{\mu K(1)}
\Bigg]
\Bigg\}.
\label{eq:generalBinaryRefinedSubGaussian}
\end{align}

Raginsky's binary-input sub-Gaussian bound
\cite[Theorem~3.7 and Example~3.3]{sdpiRaginsky} gives
\begin{equation}
\eta_{\mathrm{KL}}(\mu,K)
\leq
\min\left\{
1,\,
\frac{
(1-a-b)^2(1-2p)
}{
\mu K(0)\mu K(1)\log((1-p)/p)
}
\right\}.
\label{eq:generalBinaryRaginskyBound}
\end{equation}
The range-Hoeffding relaxation gives
\begin{equation}
\eta_{\mathrm{KL}}(\mu,K)
\leq
\min\left\{
1,\,
\frac{(1-a-b)^2}
{2\mu K(0)\mu K(1)}
\right\}.
\label{eq:generalBinaryHoeffdingBound}
\end{equation}

For a numerical illustration, take
\begin{equation}
p=\frac{1}{10},
\qquad
K=
\begin{pmatrix}
\frac{7}{10} & \frac{3}{10}\\[1mm]
\frac{2}{5} & \frac{3}{5}
\end{pmatrix}.
\label{eq:generalBinaryNumericalChannel}
\end{equation}
Then
\begin{equation}
\mu K(0)=\frac{67}{100},
\qquad
\mu K(1)=\frac{33}{100},
\end{equation}
and
\begin{equation}
g_0-1=
\left(
\frac{3}{67},-\frac{27}{67}
\right),
\qquad
g_1-1=
\left(
-\frac{1}{11},\frac{9}{11}
\right).
\label{eq:generalBinaryNumericalReverseDensities}
\end{equation}
Numerical maximisation of~\eqref{eq:generalBinaryExactKL} gives
 $\eta_{\mathrm{KL}}(\mu,K)
\simeq 0.069$.
The nonlinear curve in
\eqref{eq:generalBinaryNonlinearSubGaussianCurve} gives
\begin{equation}
\eta_{\mathrm{KL}}(\mu,K)
\lesssim  
0.076,
\end{equation}
whereas~\eqref{eq:generalBinaryRefinedSubGaussian} gives
\begin{equation}
\eta_{\mathrm{KL}}(\mu,K)
\lesssim
0.080.
\end{equation}

For Bennett's inequality, direct substitution into the nonlinear
contraction bound gives
\begin{align}
F_{\varphi_{\mathrm{KL}}}(\mu,K;c)
\leq
\min\Bigg\{c,\,
&
\frac{33}{100}
\varphi_{\mathrm{KL}}\!\left(
1+
\frac{9}{11}
\min\left\{
1,\,
\frac{
[\varphi_{\mathrm{KL}}(1+\cdot)]^{-1}(9c)
}{9}
\right\}
\right)
\nonumber\\
&+
\frac{67}{100}
\varphi_{\mathrm{KL}}\!\left(
1-
\frac{27}{67}
\min\left\{
1,\,
\frac{
[\varphi_{\mathrm{KL}}(1+\cdot)]^{-1}(9c)
}{9}
\right\}
\right)
\Bigg\}.
\label{eq:generalBinaryBennettCurve}
\end{align}
To obtain a linear SDPI bound from the nonlinear Bennett estimate, we maximise its ratio to $c$. Consequently, said optimisation yields  
\begin{equation}
\eta_{\mathrm{KL}}(\mu,K)
\leq
\frac{
d_{\mathrm{KL}}\!\left(
\frac35\,\middle\|\,\frac{33}{100}
\right)
}{
(10\log 10-9)/9
}
\simeq
0.098.
\label{eq:generalBinaryBennettValue}
\end{equation}

Similarly, Bernstein's inequality gives
\begin{align}
F_{\varphi_{\mathrm{KL}}}(\mu,K;c)
\leq
\min\Bigg\{c,\,
&
\frac{33}{100}
\varphi_{\mathrm{KL}}\!\left(
1+
\frac{9}{11}
\min\left\{
1,\,
\frac{\sqrt{2c}+c}{3}
\right\}
\right)+
\frac{67}{100}
\varphi_{\mathrm{KL}}\!\left(
1-
\frac{27}{67}
\min\left\{
1,\,
\frac{\sqrt{2c}+c}{3}
\right\}
\right)
\Bigg\}.
\label{eq:generalBinaryBernsteinCurve}
\end{align}
Maximising the ratio of the Bernstein bound to $c$ over $c>0$ yields
\begin{equation}
\eta_{\mathrm{KL}}(\mu,K)
\leq
\frac{
d_{\mathrm{KL}}\!\left(
\frac35\,\middle\|\,\frac{33}{100}
\right)
}{
4-\sqrt7
}
\simeq
0.113.
\label{eq:generalBinaryBernsteinValue}
\end{equation}

\begin{figure}[tb] 
   \centering
   \includegraphics[width=1\textwidth]{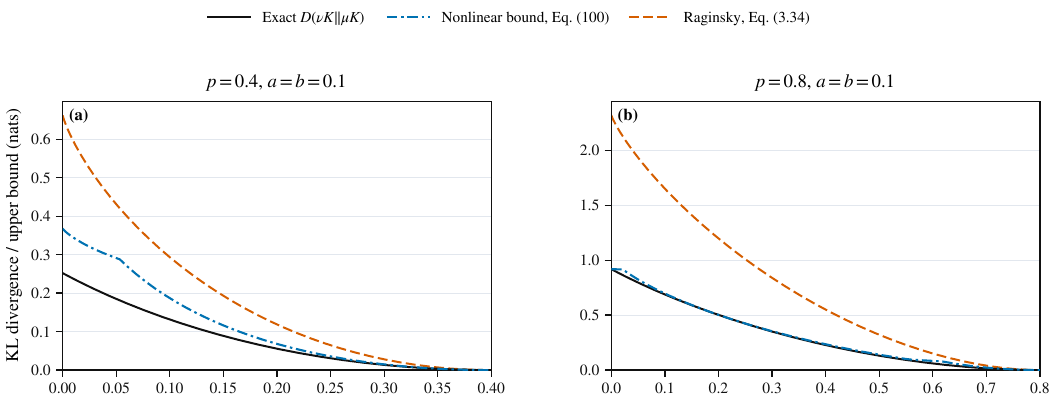}
 
    \caption{Comparison of the nonlinear KL contraction bounds in~\Cref{eq:generalBinaryNonlinearSubGaussianCurve} and~\cite[Eq. (3.34)]{sdpiRaginsky} for the channel in~\Cref{ex:generalBinaryChannel}, as functions of $0\leq r\leq p$, with $a=b=0.1$.}  \label{fig:comparisonKLSimple}
\end{figure}

For comparison,
\begin{equation}
\eta_{\chi^2}(K)
=
\left(
\sqrt{(1-a)(1-b)}-\sqrt{ab}
\right)^2
\simeq
0.091,
\end{equation}
while
\begin{equation}
\eta_{\mathrm{TV}}(K)
=
|1-a-b|
=
0.300.
\end{equation}
\Cref{tab:generalBinaryKLBounds} and \Cref{fig:comparisonKLSimple} summarise the comparison between the various bounds derived in this work and the corresponding bounds available in the existing literature.
\begin{table}[t]
\centering
\caption{Bounds for the general binary channel
in~\eqref{eq:generalBinaryNumericalChannel}.}
\label{tab:generalBinaryKLBounds}
\begin{tabular}{l c}
\hline
Quantity & Value\\
\hline
$\eta_{\mathrm{KL}}(\mu, K)$
& $0.069$\\
\Cref{eq:generalBinaryNonlinearSubGaussianCurve} (nonlinear sub-Gaussian certificate)
& $0.076$\\
\Cref{eq:generalBinaryRefinedSubGaussian} (closed refined sub-Gaussian)
& $0.080$\\
$\eta_{\chi^2}(K)$
& $0.091$\\
\Cref{eq:generalBinaryBennettValue} (Bennett approximation)
& $0.098$\\
\Cref{eq:generalBinaryBernsteinValue} (Bernstein approximation)
& $0.113$\\
~\cite[Thm 3.7, Ex 3.3]{sdpiRaginsky}
& $0.148$\\
\Cref{eq:generalBinaryHoeffdingBound} (Classical Hoeffding)
& $0.204$\\
$\eta_{\mathrm{TV}}(K)$
& $0.300$\\
\hline
\end{tabular}
\end{table}
\end{example}

\begin{example}[Random walk on a weighted three-vertex graph]
\label{ex:biasedRefreshmentKL}
Let $V=\{0,1,2\}$ and consider the weighted undirected graph with
conductance matrix
\begin{equation}
C
=
\frac{1}{1000}
\begin{pmatrix}
656&72&72\\
72&19&9\\
72&9&19
\end{pmatrix}.
\label{eq:biasedRefreshmentConductances}
\end{equation}
The weighted degrees are
\begin{equation}
\sum_{y\in V}C(x,y)
=
\pi(x),
\qquad
\pi
=
\left(
\frac45,\frac1{10},\frac1{10}
\right).
\end{equation}
The transition kernel of the corresponding random walk is
\begin{equation}
K(y\mid x)
=
\frac{C(x,y)}{\pi(x)}
=
\frac{1}{10}\mathbbm{1}_{\{y=x\}}
+
\frac{9}{10}\pi(y).
\label{eq:biasedRefreshmentKernel}
\end{equation}
Equivalently,
\begin{equation}
K
=
\begin{pmatrix}
0.820&0.090&0.090\\
0.720&0.190&0.090\\
0.720&0.090&0.190
\end{pmatrix}.
\end{equation}
Since $C$ is symmetric, the random walk is reversible with respect to
$\pi$. The diagonal entries of $C$ correspond to self-loops.

We consider the non-uniform reference distribution
\begin{equation}
\mu
=
\left(
\frac{3}{100},
\frac{94}{100},
\frac{3}{100}
\right).
\label{eq:biasedRefreshmentPrior}
\end{equation}
Its output distribution is
\begin{equation}
\mu K
=
\left(
\frac{723}{1000},
\frac{184}{1000},
\frac{93}{1000}
\right).
\label{eq:biasedRefreshmentOutput}
\end{equation}
In this example one has that  for every $y\in V$,
\begin{equation}
g_y(x)-1
=
\frac{
\mathbbm{1}_{\{x=y\}}-\mu(y)
}{
10\,\mu K(y)
}.
\label{eq:biasedRefreshmentCentredDensity}
\end{equation}
Thus, every $g_y-1$ is a centred two-point random variable under
$\mu$.

The endpoints and second moments required below are
\begin{center}
\small
\setlength{\tabcolsep}{4pt}
\renewcommand{\arraystretch}{1.15}
\begin{tabular}{@{}cccc@{}}
\hline
$y$
&
$m_y$
&
$M_y$
&
{\scriptsize$\left\lVert g_y-1\right\rVert_{L^2(\mu)}^2$}
\\
\hline
$0$
&
$\frac{240}{241}$
&
$\frac{820}{723}$
&
$0.001$
\\[1mm]
$1$
&
$\frac{45}{92}$
&
$\frac{95}{92}$
&
$0.017$
\\[1mm]
$2$
&
$\frac{30}{31}$
&
$\frac{190}{93}$
&
$0.034$
\\
\hline
\end{tabular}
\end{center}

\paragraph*{Bound via moment-generating-function}
From~\Cref{eq:biasedRefreshmentCentredDensity}, one has
\begin{align}
\log\mu\left(e^{t(g_y-1)}\right)
=
\log\Bigg[
&
(1-\mu(y))
\exp\left(
-\frac{t\mu(y)}{10\mu K(y)}
\right)+
\mu(y)
\exp\left(
\frac{t(1-\mu(y))}{10\mu K(y)}
\right)
\Bigg].
\label{eq:biasedRefreshmentExactMGF}
\end{align}
Consequently,
\begin{align}
\rho_{\mu,\varphi_{\mathrm{KL}}}^{c}(\pm (g_y-1))
&=
\inf_{t>0}
\frac{
c+\log\mu\left(e^{\pm t(g_y-1)}\right)
}{t}.
\label{eq:biasedRefreshmentExactGauges}
\end{align}
The sharpest nonlinear contraction theorem gives
\begin{align}
F_{\varphi_{\mathrm{KL}}}(\mu,K;c)
\leq
\min\Bigg\{
c,\,
\sum_{y\in V}
\mu K(y)
\max\Bigg\{
&
\varphi_{\mathrm{KL}}\left(
1-
\rho_{\mu,\varphi_{\mathrm{KL}}}^{c}(1-g_y)
\right),
\varphi_{\mathrm{KL}}\left(
1+
\rho_{\mu,\varphi_{\mathrm{KL}}}^{c}(g_y-1)
\right)
\Bigg\}
\Bigg\}.
\label{eq:biasedRefreshmentExactMGFContraction}
\end{align}
In this example, the infima in \Cref{eq:biasedRefreshmentExactGauges} reduce to one-dimensional inversions of the binary KL divergence and can be computed exactly. Thus, maximising the right-hand side of
\Cref{eq:biasedRefreshmentExactMGFContraction} divided by $c$ over $0<c\leq
\log\left(\frac{100}{3}\right)$ 
gives
\begin{equation}
\eta_{\mathrm{KL}}(\mu,K)
\leq
0.024.
\label{eq:biasedRefreshmentExactMGFSDPI}
\end{equation}

\paragraph*{Sub-Gaussian bounds}
Since the random variables in
\Cref{eq:biasedRefreshmentCentredDensity} are two-valued, their optimal
sub-Gaussian variance proxies are available in closed form:
\begin{equation}
\left\lVert g_y-1\right\rVert_{\mathrm{SG}(\mu)}^2
=
\frac{1}{100\left(\mu K(y)\right)^2}
\frac{
1-2\mu(y)
}{
2\log\left(
\dfrac{1-\mu(y)}{\mu(y)}
\right)
}.
\label{eq:biasedRefreshmentSubGaussianProxy}
\end{equation}
Numerically,
\begin{align}
\left\lVert g_0-1\right\rVert_{\mathrm{SG}(\mu)}^2
=
0.003,
\left\lVert g_1-1\right\rVert_{\mathrm{SG}(\mu)}^2
=
0.047,
\left\lVert g_2-1\right\rVert_{\mathrm{SG}(\mu)}^2
=
0.156.
\label{eq:biasedRefreshmentSubGaussianValues}
\end{align}
The nonlinear sub-Gaussian contraction bound becomes
\begin{align}
F_{\varphi_{\mathrm{KL}}}(\mu,K;c)
\leq
\min\Bigg\{
c,\,
\sum_{y\in V}
\mu K(y)
\max\Bigg\{
&
\varphi_{\mathrm{KL}}\left(
1-
\min\left\{
1-m_y,\,
\sqrt{2c}
\left\lVert g_y-1\right\rVert_{\mathrm{SG}(\mu)}
\right\}
\right),
\nonumber\\
&
\varphi_{\mathrm{KL}}\left(
1+
\min\left\{
M_y-1,\,
\sqrt{2c}
\left\lVert g_y-1\right\rVert_{\mathrm{SG}(\mu)}
\right\}
\right)
\Bigg\}
\Bigg\}.
\label{eq:biasedRefreshmentNonlinearSubGaussian}
\end{align}
Maximising the right-hand side of
\Cref{eq:biasedRefreshmentNonlinearSubGaussian} divided by $c$ gives
\begin{equation}
\eta_{\mathrm{KL}}(\mu,K)
\leq
0.025.
\label{eq:biasedRefreshmentNonlinearSubGaussianSDPI}
\end{equation}

Recall that
\begin{equation}
\kappa_{\mathrm{KL}}(u)
=
\begin{cases}
\displaystyle
\frac{\varphi_{\mathrm{KL}}(u)}{(1-u)^2},
&u\neq1,
\\[2mm]
\displaystyle\frac12,
&u=1.
\end{cases}
\end{equation}
The corresponding closed-form linear estimate is
\begin{align}
\eta_{\mathrm{KL}}(\mu,K)
&\leq
2
\sum_{y\in V}
\mu K(y)
\kappa_{\mathrm{KL}}(m_y)
\left\lVert g_y-1\right\rVert_{\mathrm{SG}(\mu)}^2 \simeq
0.027.
\label{eq:biasedRefreshmentRefinedSubGaussian}
\end{align}

By contrast, applying Raginsky's standard sub-Gaussian estimate gives
\begin{align}
\eta_{\mathrm{KL}}(\mu,K)
&\leq
2
\sum_{y\in V}
\mu K(y)
\left\lVert g_y-1\right\rVert_{\mathrm{SG}(\mu)}^2=
0.050.
\label{eq:biasedRefreshmentRaginsky}
\end{align}
Therefore, the improvement in
\Cref{eq:biasedRefreshmentRefinedSubGaussian} is produced by retaining
the nonlinear behaviour of the KL generator, rather than merely by
using optimal sub-Gaussian proxies.

For comparison, Hoeffding's range estimate gives
\begin{equation}
\left\lVert g_y-1\right\rVert_{\mathrm{SG}(\mu)}^2
\leq
\frac{(M_y-m_y)^2}{4}
=
\frac{1}{400\left(\mu K(y)\right)^2}.
\end{equation}
Substitution into the standard sub-Gaussian estimate yields
\begin{align}
\eta_{\mathrm{KL}}(\mu,K)
&\leq
\frac{1}{200}
\sum_{y\in V}
\frac{1}{\mu K(y)}
\simeq
0.088.
\label{eq:biasedRefreshmentRangeHoeffding}
\end{align}

\paragraph*{Bennett bounds}
Let
\begin{equation}
h(u)
\coloneqq
\varphi_{\mathrm{KL}}(1+u),
\qquad u\geq0.
\end{equation}
Bennett's moment-generating-function inequality gives
\begin{align}
\rho_{\mu,\varphi_{\mathrm{KL}}}^{c}(g_y-1)
\leq
\min\Bigg\{
&M_y-1,\,
\frac{
\left\lVert g_y-1\right\rVert_{L^2(\mu)}^2
}{
M_y-1
}
h^{-1}\left(
\frac{
c(M_y-1)^2
}{
\left\lVert g_y-1\right\rVert_{L^2(\mu)}^2
}
\right)
\Bigg\},
\label{eq:biasedRefreshmentBennettPositiveGauge}
\\
\rho_{\mu,\varphi_{\mathrm{KL}}}^{c}(1-g_y)
\leq
\min\Bigg\{
&1-m_y,\,
\frac{
\left\lVert g_y-1\right\rVert_{L^2(\mu)}^2
}{
1-m_y
}
h^{-1}\left(
\frac{
c(1-m_y)^2
}{
\left\lVert g_y-1\right\rVert_{L^2(\mu)}^2
}
\right)
\Bigg\}.
\label{eq:biasedRefreshmentBennettNegativeGauge}
\end{align}
Substituting
\Cref{eq:biasedRefreshmentBennettPositiveGauge,eq:biasedRefreshmentBennettNegativeGauge}
into the coordinatewise contraction bound and maximising the resulting
expression divided by $c$ gives
\begin{equation}
\eta_{\mathrm{KL}}(\mu,K)
\lesssim 0.030.
\label{eq:biasedRefreshmentNonlinearBennettSDPI}
\end{equation}

A completely closed-form Bennett estimate follows by maximising each
coordinate separately:
\begin{align}
\eta_{\mathrm{KL}}(\mu,K)
&\leq
\sum_{y\in V}
\mu K(y)
\max\Bigg\{
\frac{
(M_y-1)^2\varphi_{\mathrm{KL}}(M_y)
}{
\left\lVert g_y-1\right\rVert_{L^2(\mu)}^2
h\left(
\dfrac{(M_y-1)^2}{
\left\lVert g_y-1\right\rVert_{L^2(\mu)}^2
}
\right)
},
\frac{
(1-m_y)^2\varphi_{\mathrm{KL}}(m_y)
}{
\left\lVert g_y-1\right\rVert_{L^2(\mu)}^2
h\left(
\dfrac{(1-m_y)^2}{
\left\lVert g_y-1\right\rVert_{L^2(\mu)}^2
}
\right)
}
\Bigg\}
\nonumber\\
&\simeq
0.032.
\label{eq:biasedRefreshmentClosedBennett}
\end{align}
The difference between
\Cref{eq:biasedRefreshmentNonlinearBennettSDPI} and
\Cref{eq:biasedRefreshmentClosedBennett} is caused by moving the
maximisation over $c$ through the sum over $y\in V$.

\paragraph*{Distribution-independent comparisons}
By the binary-input reduction for channel-wide contraction
coefficients~\cite{ordentlichPolyanskiyBinaryInputs}, it is sufficient
to inspect the three two-input subchannels. The largest coefficient is
attained by restricting the input to $\{1,2\}$. For these inputs, the
output equals $0$ with probability $0.720$, independently of the input.
Conditional on the complementary event, the remaining channel is a
binary symmetric channel with crossover probability $9/28$.
Consequently,
\begin{align}
\eta_{\chi^2}(K)
&=
0.280
\left(
1-\frac{18}{28}
\right)^2 = 
\frac{1}{28}
\simeq 
0.036.
\label{eq:biasedRefreshmentChannelChiSquare}
\end{align}

The Dobrushin coefficient is
\begin{equation}
\eta_{\mathrm{TV}}(K)
=
0.100.
\label{eq:biasedRefreshmentTV}
\end{equation}

Combining the preceding estimates gives
\begin{align}
\eta_{\mathrm{KL}}(\mu,K)
&\lesssim 
0.024
&&\text{sharpest MGF estimate},
\nonumber\\
&\lesssim
0.025
&&\text{nonlinear sub-Gaussian},
\nonumber\\
&\lesssim
0.027
&&\text{closed refined sub-Gaussian},
\nonumber\\
&\lesssim
0.030
&&\text{nonlinear Bennett},
\nonumber\\
&\lesssim
0.032
&&\text{closed Bennett},
\nonumber\\
&<
0.036
\simeq 
\eta_{\chi^2}(K).
\label{eq:biasedRefreshmentComparisonChain}
\end{align}
The inequalities in
\Cref{eq:biasedRefreshmentComparisonChain} are intended as a comparison
of numerical values: each displayed quantity is separately an upper
bound on $\eta_{\mathrm{KL}}(\mu,K)$. The comparison is summarised in
\Cref{tab:biasedRefreshmentComparison}. 
\end{example}

\begin{table}[t]
\centering
\caption{KL-SDPI comparison for the weighted three-vertex random walk.
The first row is the numerically computed SDPI coefficient; the remaining rows
are upper bounds. }
\label{tab:biasedRefreshmentComparison}
\begin{tabular}{lc}
\hline
Quantity & Value\\
\hline
$\eta_{\mathrm{KL}}(\mu, K)$
& $0.020$
\\
sharpest MGF estimate,
\Cref{eq:biasedRefreshmentExactMGFSDPI}
& $0.024$
\\
Nonlinear sub-Gaussian,
\Cref{eq:biasedRefreshmentNonlinearSubGaussianSDPI}
& $0.025$
\\
Closed refined sub-Gaussian,
\Cref{eq:biasedRefreshmentRefinedSubGaussian}
& $0.027$
\\
Nonlinear Bennett,
\Cref{eq:biasedRefreshmentNonlinearBennettSDPI}
& $0.030$
\\
Closed Bennett,
\Cref{eq:biasedRefreshmentClosedBennett}
& $0.032$
\\$\eta_{\chi^2}(K)$
& $0.036$
\\
Raginsky's sub-Gaussian,
\Cref{eq:biasedRefreshmentRaginsky}
& $0.050$
\\
Range-Hoeffding,
\Cref{eq:biasedRefreshmentRangeHoeffding}
& $0.088$
\\
$\eta_{\mathrm{TV}}(K)$
& $0.100$
\\
\hline
\end{tabular}
\end{table}

\section{Applications}
 
 \subsection{Markov chains mixing times}\label{sec:mixingTimes}
    \begin{definition}\label{def:alphaMixing}
       Given a Markov kernel $K$ with stationary distribution $\pi$, $N\in\{A, L\}$, and a Young functional $\psi$, the $\psi$-mixing time of $K$ is the function $\tau_\psi(K,\cdot):\mathbb{R}^+\to\mathbb{N}$ such that
      \begin{equation}
           \tau_{\psi}(K,\epsilon) = \min \left\{t \in \mathbb{N} : \sup_{\nu\ll\pi} \left\lVert \frac{d\nu K^t}{d\pi} -1 \right\rVert_{L_\psi^{N}(\pi)} \leq \epsilon \right\}.\label{eq:mixingTimeDef}
       \end{equation}
   \end{definition}
    The first application of our framework is stated below and follows from~\Cref{thm:psiNormsContraction} and Eq.~\eqref{eq:mixingTimeDef}.
    \begin{corollary}\label{thm:psiMixingAndContraction}
        Let $K$ be a Markov operator with stationary distribution $\pi$, $K^\star = K^\star_\pi$ its dual, $N\in\{A, L\}$, and $\psi$ a Young functional. Then, the following holds:
       \begin{equation}
            \tau_\psi(K,\epsilon) \leq \left\lceil\max\left\{0,\frac{\log\left(\left(\sup_{\nu\ll\pi}\left\lVert  \frac{d\nu}{d\pi}-1\right\rVert_{L_\psi^{N}(\pi)}\right)/\epsilon\right)}{-\log\left(\left\lVert K^\star\right\rVert_{L_\psi^{N,0}\to L_\psi^{N,0}}\right)}\right\}\right\rceil.\label{eq:mixingTimeBoundPsiNorm}
        \end{equation} 
    \end{corollary}
    \begin{remark}\label{rmk:maxNormsDiscrete}
        For discrete spaces and a discrete-valued Markov kernel, by leveraging the convexity of the norms and the fact that convex functions on a compact convex set attain their maximum at an extreme point, we can upper bound $\left\lVert  \frac{d\nu}{d\pi}-1\right\rVert_{L_\psi^{N}(\pi)}$ with $\max_x \left\lVert  \frac{d\delta_x}{d\pi}-1\right\rVert_{L_\psi^{N}(\pi)}$, which admits a closed-form expression. Furthermore, the following closed-form expressions for $\max_x \left\lVert  \frac{d\delta_x}{d\pi}-1\right\rVert_{L_\psi^{N}(\pi)}$ hold:
           \begin{equation}\label{eq:Lcs}
            \max_x \left\lVert  \frac{d\delta_x}{d\pi}-1\right\rVert_{L_\psi^{L}(\pi)} \leq \max_x \begin{cases}
                1/{\psi}^{-1}\left(\frac{1}{2\pi(\Omega\setminus\{x\})}\right) &\text{ if } \pi(\{x\})\geq 1/2,\\ 
                \frac{\pi(\Omega\setminus\{x\})}{\pi(\{x\})}\cdot 1/{\psi}^{-1}\left(\frac{1}{2\pi(\{x\})}\right) &\text{ else },
            \end{cases}
        \end{equation}   
       \begin{equation}
            \max_x \left\lVert  \frac{d\delta_x}{d\pi}-1\right\rVert_{L_\psi^{A}(\pi)} \leq 2\max_x \pi(\Omega\setminus\{x\})\begin{cases}
                {\psi^\star}^{-1}\left(\frac{1}{2\pi(\Omega\setminus\{x\})}\right) &\text{ if } \pi(\{x\})\geq 1/2,\\ 
                {\psi^\star}^{-1}\left(\frac{1}{2\pi(\{x\})}\right) &\text{ else },
            \end{cases}
        \end{equation}
         where $\Omega$ is the space over which $\nu$ and $\pi$ are defined and $\psi^\star$ the Young conjugate, see~\Cref{eq:youngComplement}.
   We start by proving \Cref{eq:Lcs}. Let $x$ be fixed. Then, for a given $\psi$ and $\sigma>0$, 
    \begin{equation}
        \pi\left(\psi\left(\frac{\left|\frac{d\delta_x}{d\pi}-1\right|}{\sigma}\right)\right) = \pi(\{x\})\psi\left(\frac{\pi(\Omega\setminus\{x\})}{ \pi(\{x\})}\frac1\sigma\right)+\pi(\Omega\setminus\{x\})\psi\left(\frac1\sigma\right).
    \end{equation}
    We need to consider two cases. First assume that $\pi(\Omega\setminus\{x\})/\pi(\{x\}) \leq 1$, which is equivalent to $\pi(\{x\})\geq 1/2$. Then, since $\psi$ is convex and $\psi(0)=0$, one has that
    \begin{align}
        \pi(\{x\})\psi\left(\frac{\pi(\Omega\setminus\{x\})}{ \pi(\{x\})}\frac1\sigma\right)+\pi(\Omega\setminus\{x\})\psi\left(\frac1\sigma\right) \leq 2\pi(\Omega\setminus\{x\})\psi\left(\frac1\sigma\right).\label{eq:maxLuxNormDiscrete1}
    \end{align}
    Thus, in order to have the right-hand side of~\Cref{eq:maxLuxNormDiscrete1} less than one, one needs $$\sigma \geq \frac{1}{\psi^{-1}\left(\frac{1}{2\pi(\Omega\setminus\{x\})}\right)}.$$ If instead $\pi(\Omega\setminus\{x\})/\pi(\{x\}) \geq 1$, then  %
     \begin{align}
        \pi(\{x\})\psi\left(\frac{\pi(\Omega\setminus\{x\})}{ \pi(\{x\})}\frac1\sigma\right)+\pi(\Omega\setminus\{x\})\psi\left(\frac1\sigma\right) \leq 2\pi(\{x\})\psi\left(\frac{\pi(\Omega\setminus\{x\})}{ \pi(\{x\})}\frac1\sigma\right)\label{eq:maxLuxNormDiscrete2}.
    \end{align}
    Consequently, one needs $$\sigma \geq \frac{\pi(\Omega\setminus\{x\})}{ \pi(\{x\})\psi^{-1}\left(\frac{1}{2\pi(\{x\})}\right)}.$$
   
    Analogous considerations allow us to derive a bound on the Amemiya norm. Indeed, if $\pi(\Omega\setminus\{x\})/\pi(\{x\}) \leq 1$, then
    \begin{align}
        \inf_{t>0} \frac{1+\pi\left(\psi\left(t\left|\frac{d\delta_x}{d\pi}-1\right|\right)\right)}{t} &\leq   \inf_{t>0} \frac{1+2\pi(\Omega\setminus\{x\}) \psi(t)}{t} = 2\pi(\Omega\setminus\{x\}) {\psi^{\star}}^{-1}\left(\frac{1}{2\pi(\Omega\setminus\{x\})}\right).
    \end{align}
    One can then also similarly bound the Amemiya norm when $\pi(\Omega\setminus\{x\})/\pi(\{x\}) \geq 1$.

    \end{remark}
   \Cref{thm:psiMixingAndContraction} relates the contraction coefficient of the dual operator to the mixing time of the corresponding Markov chain, thus motivating our study of said contraction coefficient. 
    Moreover, as %
    $\psi$ is arbitrary, a natural question concerns the choice of the norm. %
    To address it, first we %
    derive a general result providing intuition over what it means to be close in a specific norm. In particular, we relate the probability of any event under $\mu K^t$ with the probability of the same event under the stationary measure $\pi$ and an $L^{N}_\psi$-norm. We then focus on the well-known family of $L_p$-norms (particularly relevant for exponentially decaying probabilities under $\pi$), and show that asking for a bounded norm with larger $p$ guarantees a faster decay of the probability of the same event under $\mu K^t$. Finally, we design a Young function tailored to the challenging (and less understood) setting in which $\pi$ behaves like a power law \cite{polynomialMC}. %
    \begin{theorem}\label{thm:probBoundPsiNorm}
     Let $\mu$ be a probability measure, $K$ a Markov kernel with stationary distribution $\pi$, $t\in\mathbb{N}$, and $E$ a measurable event. Assume that $\mu K^t\ll\pi$. Then,
    \begin{equation}
            \mu K^t(E) \leq \min \begin{cases}
               \inf_{\psi} 1/{\psi^\star}^{-1}(1/\pi(E))\left(\left\lVert \frac{d\mu K^t}{d\pi}-1\right\rVert_{L_\psi^A(\pi)}+{\psi^\star}^{-1}(1)\right), \\
                   \inf_{\psi}\pi(E)\psi^{-1}(1/\pi(E))\left(\left\lVert \frac{d\mu K^t}{d\pi}-1\right\rVert_{L_\psi^L(\pi)}+1/\psi^{-1}(1)\right),
            \end{cases}
        \end{equation}
        where the infimum is over all Young functionals $\psi$.
    \end{theorem}
    The proof of~\Cref{thm:probBoundPsiNorm} can be found in~\Cref{app:proofProbBoundPsiNorm}.
    We now apply~\Cref{thm:probBoundPsiNorm} to various $\psi$, and start by
demonstrating the advantage of considering $L_p$-norms with $p\neq 2$. We will now prove the following result: 
\begin{corollary}\label{thm:probBoundAlphaNorms}
    Let $\mu$ be a probability measure, $K$ a Markov kernel with stationary distribution $\pi$, and assume that $\mu K^t\ll\pi$. Let $\varphi_p(x)=|x|^p/p$ with $p>1$, $q=p/(p-1)$, $t\in\mathbb{N}$, and $E$ a measurable event. Then,
    \begin{align}
        \mu K^t(E) &\leq \pi(E)^{\frac1q}\left(\pi(E)^\frac1p + \left\lVert \frac{d\mu K^t}{d \pi}-1\right\rVert_{L_p(\pi)}\right).\label{eq:newResultHalpha}
    \end{align}
\end{corollary} 
The proof of~\Cref{thm:probBoundAlphaNorms} can be found in~\Cref{app:proofProbBoundAlphaNorms}.
As an immediate consequence, if 
$\left\lVert \frac{d\mu K^t}{d\pi}-1\right \rVert_{L_p(\pi)}\leq  \epsilon,$
then for every %
$E$
 \begin{align}\label{eq:ubLp}
    \mu K^t(E) &\leq \pi(E)^\frac1q\left(\pi(E)^\frac1p + \epsilon \right) = \pi(E) + \pi(E)^\frac1q \epsilon. 
\end{align}
The following setting is especially interesting for the applicability of~\Cref{thm:probBoundAlphaNorms}. Consider a function $f:\mathcal{X}^n\to\mathbb{R}$ and $E=\{(x_1, \ldots, x_n): |f(x_1, \ldots, x_n)-\pi(f)|\geq \eta \}$ for some fixed $\eta>0$, with $\pi$ a product law. If $f(x_1, \ldots, x_n)=\frac1n \sum_{i=1}^n f_i(x_i)$ for $f_1,\ldots,f_n$ with $0\leq f_i(x) \leq C$, by Hoeffding's inequality one has:%
 \begin{equation}
    \pi(E) \leq 2\exp\left(-\frac{2\eta^2 n}{ C^2 }\right),
\end{equation}
where the constant $C^2$ at the exponent can be improved under additional assumptions, see \emph{e.g.}~\cite{concentrationMeasureII}.
Then, given $t\geq \tau_{\varphi_p}(K,\epsilon)$ and $q=\frac{p}{p-1}$, we have
 \begin{equation}
    \mu K^t(E) \leq 2\exp\left(-\frac{2n\eta^2 }{C^2 }\right) + \epsilon 2^\frac{1}{q}\exp\left(-\frac{2n\eta^2}{q C^2 }\right). \label{eq:probBoundMixing}
\end{equation}
Starting from~\Cref{eq:probBoundMixing}, one can now select $\epsilon$ such that the probability is exponentially decaying in the dimension of the ambient space $n$. %
In fact, a bound that decays exponentially in the dimension $n$ is guaranteed whenever, for a given $\eta$, $f$, and finite $p$, 
\begin{equation}
     \epsilon < \exp\left(\gamma\frac{ 2 n \eta^2}{q C^2 } \right), \text{ with } \gamma<1.
\end{equation} 
For example, $\epsilon = \exp\left(n\eta^2/q C^2 \right)$ guarantees exponential convergence in~\Cref{eq:probBoundMixing} and plugging this choice in~\Cref{eq:mixingTimeBoundPsiNorm} gives the bound below on the mixing time:
   \begin{equation}
       \tau_p(K,\epsilon) \leq \left\lceil\max\left\{0,\frac{\log\left(L_p^\star(\pi)\right)-\frac{n\eta^2}{qC^2}}{-\log\left(\left\lVert K^\star \right\rVert_{L_p^0\to L_p^0} \right)}\right\}\right\rceil,\label{eq:mixingTimeExponentialConvergence}
   \end{equation}
   where $L_p^\star(\pi)=\sup_{\nu\ll\pi} \left\lVert \frac{d\nu}{d\pi}-1\right\rVert_{L^p(\pi)}$. We note that, besides $\epsilon$, both $L_p^\star(\pi)$ and $\left\lVert K^\star \right\rVert_{L_p^0\to L_p^0}$ depend on the dimension $n$. 
To maximise the rate of exponential decay in~\Cref{eq:probBoundMixing}, one has to take $p\to\infty$, which implies $q\to 1$. The price to pay is an increase of the mixing time and, %
to quantify such increase, it is crucial to provide tight bounds on the contraction coefficient of Markovian operators, beyond the usual $L_2$ space. 
\Cref{thm:psiNormsContraction} can then be leveraged to compute said bounds, and we now compare it with the standard approach in the literature, \textit{i.e.}, the Riesz-Thorin interpolation theorem between norms, which is stated below for convenience. 

\begin{figure}
    \centering
    
    \begin{minipage}{0.47\textwidth}
        \centering
        \includegraphics[width=1\textwidth]{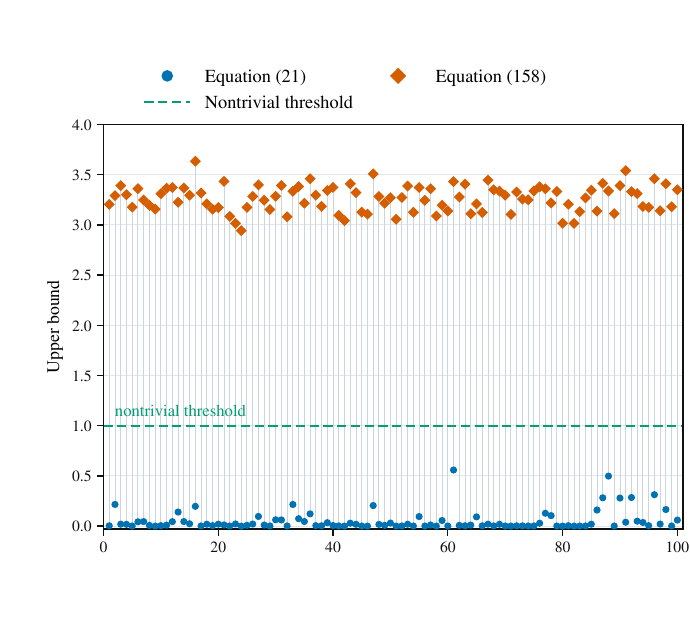}
        \caption{Bounds on the contraction coefficient given by~\Cref{thm:alphaNormsContraction,eq:contractionMarkovChainLalpha} in $100$ randomly generated $5\times 5$ stochastic matrices, for $p=100$ and $t=10$. %
        }
        \label{fig:comparisonOursVsInterpolation}
    \end{minipage}\hfill
    \begin{minipage}{0.47\textwidth} 
        \centering
        \includegraphics[width=1\textwidth]{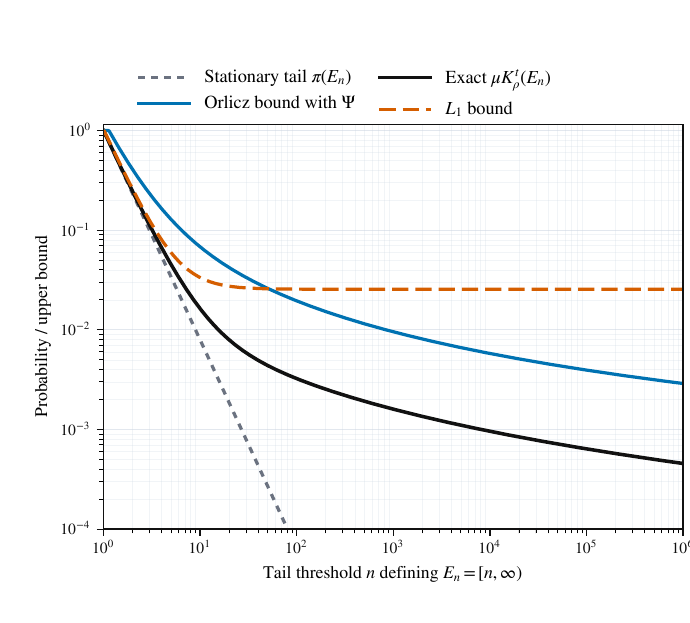}
      
        \caption{ Exact tail probability and comparison with the \(L_1\) and Orlicz bounds in Example 6, as functions of the tail threshold \(n\), where \(E_n=[n,\infty)\). We set \(\rho^t=0.1\). Both axes use logarithmic scales.
        }
        \label{fig:comparisonPowerLaw}
    \end{minipage}
\end{figure}

\begin{proposition}\label{prop:L2toLalphaConvergence}
        Let $p \in (1,+\infty)$ and $t\in\mathbb{N}$. Let $K$ be a Markov kernel with stationary distribution $\pi$. If $K$ admits an $L_2$-spectral gap of $(1-\gamma)$, then the following holds:
    \begin{equation}\label{eq:contractionMarkovChainLalpha}
            \left\lVert K^t-1_\pi \right\rVert_{L_p\to L_p} \leq \begin{cases}
                2^{2/p} \gamma^{2t \frac{p-1}{p}}, & p \in (1,2), \\
                2^{2 \frac{p-1}{p}} \gamma ^{2t/p}, &p \in (2,+\infty).
            \end{cases}
        \end{equation}
        Moreover, one has that:
    \begin{equation}
             \left\lVert K^t-1_\pi \right\rVert_{L_2\to L_2} \leq \gamma^t \text{ , }  \quad \left\lVert K^t-1_\pi \right\rVert_{L_1\to L_1} \leq 2 \text{ , }\quad  \left\lVert K^t-1_\pi \right\rVert_{L_\infty\to L_\infty} \leq 2. \label{eq:rieszThorin2-1-infty}
        \end{equation}
    \end{proposition}
As an example,  consider the $\text{BSC}(\lambda)$. Then, %
  $\gamma = |1-2\lambda|$ and \Cref{prop:L2toLalphaConvergence} with $p> 2$ 
yields:
   \begin{equation}
        \left\lVert K^t-1_\pi \right\rVert_{L_p\to L_p} \leq 2^{2 \frac{p-1}{p}} |1-2\lambda| ^{2t/p}.\label{eq:rieszThorinDSBS}
    \end{equation}
    In contrast, \Cref{thm:psiNormsContraction} gives:
 \begin{equation}
         \left\lVert K^t-1_\pi \right\rVert_{L_p\to L_p} \leq |1-2\lambda|^{t}, \text{ for every } p\geq 1, \label{eq:contractionDSBS}
    \end{equation}
    which improves over~\Cref{eq:rieszThorinDSBS} for every $p$. Notably,~\Cref{eq:contractionDSBS} holds even if $p\to 1$ and $p\to\infty$, while the result of~\Cref{prop:L2toLalphaConvergence} is vacuous in both cases (cf.~\Cref{eq:rieszThorin2-1-infty}).
\Cref{fig:comparisonOursVsInterpolation} considers Markov kernels induced by $5\times 5$ randomly generated stochastic matrices, and it shows that while the bound in~\Cref{prop:L2toLalphaConvergence} is always vacuous, the bound given by~\Cref{thm:alphaNormsContraction} is always below $1$ and often very close to $0$.

Moreover, given that we are in a discrete setting, one can explicitly compute the $L_\infty$/$L_1$-operator norm for a specific kernel $K$ and interpolate between the $L_\infty$/$L_1$-operator and the spectral gap. On average over the 100 random trials, our upper bound is roughly 50\% smaller than Riesz-Thorin interpolation for $n=5, t=20, p=10$; it is 70\% smaller for $n=5$, $t=200$, $p=100$; it is 23\% smaller for $n=3, t=4, p=1.100$; and it is 20\% smaller for $n=3, t=4, p=1.500$.

 \Cref{fig:comparisonPowerLaw} illustrates how a suitable choice of the Young function allows one to treat stationary distributions with heavy tails.
\begin{example}[An Orlicz bound when no $L_p$ bound with $p>1$ is available]\label{ex:heavyTailOrlicz}
Let
\[
    \pi(\mathrm dx)
    =
    2.100x^{-3.100}\mathbbm 1_{\{x\geq1\}}\,\mathrm dx
\]
and let the initial distribution be
\[
    \mu(\mathrm dx)
    =
    \frac{2}{x(1+\log x)^3}
    \mathbbm 1_{\{x\geq1\}}\,\mathrm dx .
\]
Both are probability measures on $[1,\infty)$. Moreover, for $E_n=[n,\infty), n\geq1,$
their tail probabilities are
\[
    \pi(E_n)=n^{-2.100},
    \qquad
    \mu(E_n)=\frac{1}{(1+\log n)^2}.
\]
The density of $\mu$ with respect to $\pi$ is
\[
    \frac{\mathrm d\mu}{\mathrm d\pi}(x)
    =
    \frac{2}{2.100}
    \frac{x^{2.100}}{(1+\log x)^3}.
\]
For every $p>1$,
\begin{align*}
    \int
    \left(
        \frac{\mathrm d\mu}{\mathrm d\pi}
    \right)^p
    \mathrm d\pi
    &=
    2.100\left(\frac{2}{2.100}\right)^p
    \int_1^\infty
    \frac{x^{2.100(p-1)-1}}
         {(1+\log x)^{3p}}
    \,\mathrm dx\\
    &=
    2.100\left(\frac{2}{2.100}\right)^p
    \int_0^\infty
    \frac{\exp(2.100(p-1)y)}
         {(1+y)^{3p}}
    \,\mathrm dy =\infty.
\end{align*}
Consequently,
\[
    \frac{\mathrm d\mu}{\mathrm d\pi}
    \notin L_p(\pi)
    \qquad\text{for every }p>1.
\]

Consider now the kernel \(K_\rho\), a lazy refreshment kernel that retains the current state with probability \(\rho\) and otherwise replaces it with an independent draw from \(\pi\)~\cite[Section 1.3]{levin2017markov}, \textit{i.e.}, $
    K_\rho(x,D)
    =
    \rho\delta_x(D)+(1-\rho)\pi(D),
    \text{ } 0<\rho<1.$ For every integer $t\in\mathbb N_0$,
$
    \mu K_\rho^t
    =
    \rho^t\mu+(1-\rho^t)\pi$ and hence
\[
    \frac{\mathrm d\mu K_\rho^t}{\mathrm d\pi}(x)-1
    =
    \rho^t
    \left(
        \frac{2}{2.100}
        \frac{x^{2.100}}{(1+\log x)^3}
        -1
    \right).
\]
It follows that, for every finite $t$,
\[
    \frac{\mathrm d\mu K_\rho^t}{\mathrm d\pi}-1
    \notin L_p(\pi)
    \qquad\text{for every }p>1.
\]

The only available $L_p$ norm is with $p=1$, which gives
\begin{align}
    \mu K_\rho^t(E_n)
    \leq
    n^{-2.100}
    +
    \left\lVert
        \frac{\mathrm d\mu K_\rho^t}{\mathrm d\pi}-1
    \right\rVert_{L_1(\pi)} &=
    n^{-2.100}
    +
    \rho^t
    \left\lVert
        \frac{2}{2.100}
        \frac{x^{2.100}}{(1+\log x)^3}
        -1
    \right\rVert_{L_1(\pi)}\\
    &\simeq
    n^{-2.100}+0.255 \,\rho^t. \label{eq:heavyTailL1NormDecay}
\end{align}
In particular, the second term in~\Cref{eq:heavyTailL1NormDecay} does
not decay with $n$.

Consider instead the Young function $
    \Psi(u)
    =
    u\left[\log(\mathrm e+u)\right]^{3/2}, 
    \text{ }
    u\geq0.$
This function is convex and nondecreasing, satisfies $\Psi(0)=0$,
and diverges as $u\to\infty$. Moreover,
$ \frac{\mathrm d\mu}{\mathrm d\pi}-1
    \in L_\Psi(\pi).$ Indeed, after the change of variables $y=\log x$, the tail of the
corresponding Orlicz modular behaves as
 $\int^\infty y^{-3/2}\,\mathrm dy<\infty.$

Numerical evaluation of the Luxemburg norm gives $\left\lVert
        \frac{\mathrm d\mu}{\mathrm d\pi}-1
    \right\rVert_{L_\Psi^L(\pi)}
    \simeq3.450.$
Using~\Cref{thm:probBoundPsiNorm} gives
\[
    \mu K_\rho^t(E_n)
    \leq
    n^{-2.100}
    +
    3.450 \,\rho^t
    n^{-2.100}\Psi^{-1}\left(n^{2.100}\right).
\]
Since $
    \Psi^{-1}(v)
    \sim
    \frac{v}{(\log v)^{3/2}},$
we obtain $n^{-2.100}\Psi^{-1}\left(n^{2.100}\right)
    \sim
    \frac{1}{(2.100\log n)^{3/2}}.$
Consequently, the Orlicz bound behaves as
\[
    \mu K_\rho^t(E_n)
    \leq
    n^{-2.100}
    +
    O\left(
        \frac{\rho^t}{(\log n)^{3/2}}
    \right).
\]

Thus, for every finite $t$, all $L_p$ bounds with $p>1$ are
unavailable and the bound induced by the $L_1$-norm has an additive term that is constant in $n$, whereas the
Orlicz correction converges to zero as $(\log n)^{-3/2}$ for every $t>0$.

\end{example}

    \subsection{Markov Chain Monte Carlo}\label{sec:MCMC}\Cref{prop:L2toLalphaConvergence} was used by~\cite{dependentViaStatAndChange} to prove %
    concentration %
    for MCMC methods.
    Here, %
    we show how an application of \Cref{thm:psiNormsContraction} improves over such results.
   Let us briefly recall the setup. Given a function $f$, we want to estimate the mean $\pi(f)$ with respect to a measure $\pi$ which cannot be directly sampled. 
A common approach is to consider a Markov chain $\{X_i\}_{i\geq 1}$ with stationary distribution $\pi$, and estimate $\pi(f)$ via empirical averages of samples $\{X_i\}_{t_0+1}^{t_0+t}$, where the burn-in period $t_0$ ensures that the Markov chain is close to $\pi$.
    In particular, consider a Markov kernel $K$ with stationary distribution $\pi$, let $X_1 \sim \nu$, and let \(f:\mathsf X\to[0,1]\). After a burn-in
period \(t_0\), consider the estimator $\frac1t \sum_{i=t_0+1}^{t_0+t} f(X_i)$ of the integral $\int f\pi(\mathrm{d}x)$. Let \(\mathbb P_{\nu,t_0}^{(t)}\) denote the law of
\((X_{t_0+1},\ldots,X_{t_0+t})\) if $X_1\sim \nu$ and after a burn-in period of $t_0$, and let
\(\mathbb P_{\pi}^{(t)}\) denote the corresponding stationary path such that $X_1 \sim \pi$. One has that \[
    \left\|
    \frac{d\mathbb P_{\nu,t_0}^{(t)}}
         {d\mathbb P_{\pi}^{(t)}}-1
    \right\|_{L_p(\mathbb P_{\pi}^{(t)})}
    =  \left\|
    \frac{d(\nu K^{t_0})}
         {d\pi}-1
    \right\|_{L_p(\pi)} = 
    \left\|
    (K^\star)^{t_0}
    \left(\frac{d\nu}{d\pi}-1\right)
    \right\|_{L_p(\pi)}.
\]

\begin{figure}[t]
  \centering
  \includegraphics[width=0.55\textwidth]{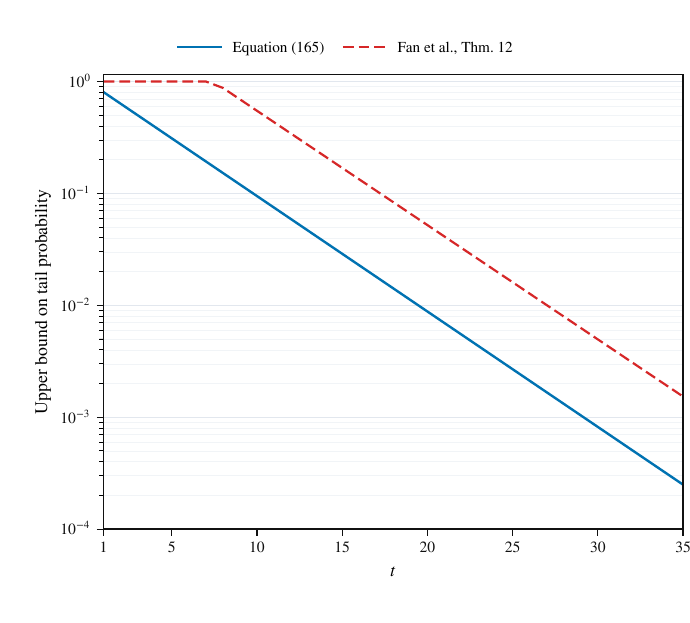}
    \caption{Comparison of the bound obtained from~\Cref{eq:mcmcPostBurn} using~\cite[Theorem 3]{dependentViaStatAndChange} and~\cite[Theorem 12]{dependentViaStatAndChange} as a function of $t$. We set $\eta=\frac12$, $p=100$ and $t_0=10$. The starting distribution and the rows of the $10\times 10$ stochastic matrix are generated by normalising i.i.d. Uniform$(0,1)$ entries with seed $2024$. Both bounds are truncated at $1$ and the relevant spectral quantities are computed numerically. The $y$-axis is logarithmic.}
  \label{fig:comparisonMCMC}
\end{figure}

Then, for \(p>1\) and \(q=p/(p-1)\),
\Cref{thm:probBoundAlphaNorms} gives
\begin{align}
    \frac{\mathbb P_{\nu,t_0}^{(t)}\left(\frac1t\sum_{i=t_0+1}^{t_0+t}f(X_i)-\pi(f)\geq \eta \right)}{\mathbb P_{\pi}^{(t)}\left(\frac1t\sum_{i=t_0+1}^{t_0+t}f(X_i)-\pi(f)\geq \eta \right)}
    &\leq
    1+
    \mathbb P_{\pi}^{(t)}\left(\frac1t\sum_{i=t_0+1}^{t_0+t}f(X_i)-\pi(f)\geq \eta \right)^{-1/p}
    \left\|
    (K^\star)^{t_0}
    \left(\frac{d\nu}{d\pi}-1\right)
    \right\|_{L_p(\pi)}
    \notag\\
    &\leq
      1+
    \mathbb P_{\pi}^{(t)}\left(\frac1t\sum_{i=t_0+1}^{t_0+t}f(X_i)-\pi(f)\geq \eta \right)^{-1/p}
    \left\|K^\star\right\|_{L_p^0\to L_p^0}^{t_0}
    \left\|\frac{d\nu}{d\pi}-1\right\|_{L_p(\pi)}.
    \label{eq:mcmcPostBurn}
\end{align}
Thus, a non-trivial \(L_p\)-contraction estimate produces an explicit and quantifiable
geometric improvement with the burn-in period. 
Let us now compare~\Cref{eq:mcmcPostBurn} with~\cite{dependentViaStatAndChange} in a Gibbs sampling for an Ising model example.
Moreover, \Cref{fig:comparisonMCMC} compares the bounds when the kernel is a $10\times10$ randomly generated stochastic matrix, and it clearly shows the improvement of our approach (in blue) over the interpolation method used by \cite{dependentViaStatAndChange} (in red). 
\begin{example}[A two-spin heat-bath Gibbs sampler]
\label{ex:twoSpinGibbs}
Let \(\mathsf X=\{-1,+1\}^2\). We represent a configuration
\(x=(x_1,x_2)\) by the two-symbol string \(x_1x_2\), so that the first
symbol denotes \(x_1\) and the second denotes \(x_2\):
\[
++=(+1,+1),\qquad
+-=(+1,-1),\qquad
-+=(-1,+1),\qquad
--=(-1,-1).
\]
Throughout this example, we use the state ordering $
(++,+-,-+,--).$ Consider the two-spin Ising distribution
\[
\pi_\beta(x_1,x_2)
=
\frac{\exp(\beta x_1x_2)}{Z_\beta},
\qquad
Z_\beta
=
\sum_{x_1,x_2\in\{-1,+1\}}
\exp(\beta x_1x_2).
\]
We choose \(\beta=\frac12\log 3\). Since \(x_1x_2=1\) for \(++\)
and \(--\), whereas \(x_1x_2=-1\) for \(+-\) and \(-+\), we have
\[
Z_\beta
=
2e^\beta+2e^{-\beta}
=
2\sqrt 3+\frac{2}{\sqrt 3}
=
\frac{8}{\sqrt 3}.
\]
Consequently, in the ordering \((++,+-,-+,--)\),
\[
\begin{split}
\pi
&=
\bigl(\pi(++),\pi(+-),\pi(-+),\pi(--)\bigr)=
\left(\frac38,\frac18,\frac18,\frac38\right).
\end{split}
\]

We consider the random-scan single-site heat-bath kernel, also known
as the Glauber dynamics or Gibbs sampler, targeting \(\pi\)
\cite[Sec.~3.3]{levinPeres2017}. At each step, one of the two
coordinates is selected uniformly and resampled from its conditional
distribution under \(\pi\), while the other coordinate is left
unchanged. For \(s\in\{-1,+1\}\) and \(i\neq j\),
\[
\pi_\beta(x_i=s\mid x_j=s)
=
\frac{e^\beta}{e^\beta+e^{-\beta}}
=
\frac34,
\qquad
\pi_\beta(x_i=-s\mid x_j=s)
=
\frac14.
\]
Therefore, in the same state ordering,
\[
K=
\begin{pmatrix}
 \frac34 & \frac18 & \frac18 & 0\\
 \frac38 & \frac14 & 0        & \frac38\\
 \frac38 & 0        & \frac14 & \frac38\\
 0        & \frac18 & \frac18 & \frac34
\end{pmatrix}.
\]
One verifies directly that \(\pi K=\pi\).

Suppose that \(X_1\sim\delta_{++}\), this is known in the literature as   ``cold start'', write
\(X_j=(X_{j,1},X_{j,2})\), and retain $X_{t_0+1},\ldots,X_{t_0+t}$ 
after a burn-in period of \(t_0\). Let \(g:\mathsf X\to[0,1]\) be arbitrary. We make this
normalisation only for simplicity; the same argument applies to
functions taking values in any bounded interval \([a,b]\), as
explained below.

Let \(\mathbb P_{\pi}^{(t)}\) denote the stationary path law and \(\mathbb P_{\delta_{++}}^{(t)}\) the path law when $X_1\sim \delta_{++}$.

For every \(\eta>0\), the stationary Hoeffding inequality~\cite[Theorem 3]{dependentViaStatAndChange} for this
kernel gives
\[
\mathbb P_{\pi}^{(t)}\left(
\frac1t\sum_{j=1}^{t}g(X_j)
-
\pi(g)
\geq\eta
\right)
\leq
\exp\left(-\frac{2t\eta^2}{7}\right).
\]

Moreover, in the ordering \((++,+-,-+,--)\),
\[
\frac{d\delta_{++}}{d\pi}-1
=
\left(\frac53,-1,-1,-1\right),
\]
and hence, for \(1<p<\infty\),
\[
\left\lVert
\frac{d\delta_{++}}{d\pi}-1
\right\rVert_{L_p(\pi)}
=
\left[
\frac38\left(\frac53\right)^p+\frac58
\right]^{1/p}.
\]
The contraction estimate obtained from
\Cref{thm:psiNormsContraction} therefore yields
\[
\left\lVert
\frac{d(\delta_{++}K^{t_0})}{d\pi}-1
\right\rVert_{L_p(\pi)}
\leq
\left[
\left(\frac34\right)^p+\left(\frac14\right)^p
\right]^{t_0/p}
\left[
\frac38\left(\frac53\right)^p+\frac58
\right]^{1/p}.
\]

Applying \Cref{thm:probBoundAlphaNorms} to the law of the retained
path, for \(1<p<\infty\) and \(q=p/(p-1)\), gives

\begin{align}
&\mathbb P_{\delta_{++}}\left(
\frac1t\sum_{j=t_0+1}^{t_0+t}g(X_j)
-
\pi(g)
\geq\eta
\right)\leq
\exp\left(-\frac{2t\eta^2}{7}\right)
+
\exp\left(-\frac{2t\eta^2}{7q}\right)
\left[
\left(\frac34\right)^p+\left(\frac14\right)^p
\right]^{t_0/p}
\left[
\frac38\left(\frac53\right)^p+\frac58
\right]^{1/p}. \label{eq:ourBoundGibbsSampler}
\end{align}

Taking the limit $p\to\infty$
\[
\begin{aligned}
&\mathbb P_{\delta_{++}}\left(
\frac1t\sum_{j=t_0+1}^{t_0+t}g(X_j)
- \pi(g)
\geq\eta
\right)\leq
\exp\left(-\frac{2t\eta^2}{7}\right)
\left[
1+\frac53\left(\frac34\right)^{t_0}
\right].
\end{aligned}
\]

For comparison, \cite[Thm.~12]{dependentViaStatAndChange},
specialised to the same kernel and initial distribution, gives
\begin{align}
&\mathbb P_{\delta_{++}}\left(
\frac1t\sum_{j=t_0+1}^{t_0+t}g(X_j)-\pi(g)\geq\eta
\right)
\leq
\begin{cases}
\displaystyle
\exp\left(-\frac{2t\eta^2}{7q}\right)
\left[
1+
2^{2/\min\{p,q\}}
\left(\frac34\right)^{2t_0/\max\{p,q\}}
\left(
\frac38\left(\frac53\right)^p+\frac58
\right)^{1/p}
\right],
& p\in(1,\infty)\setminus\{2\},
\\[1em]
\displaystyle
\exp\left(-\frac{t\eta^2}{7}\right)
\left[
1+\left(\frac34\right)^{t_0}\sqrt{\frac53}
\right],
& p=2,
\\[1em]
\displaystyle
\frac83\exp\left(-\frac{2t\eta^2}{7}\right),
& p=\infty.
\end{cases}
\label{eq:FanEtAlGibbsSampler}
\end{align}

All the displayed probability bounds are understood to be truncated
at \(1\).
\Cref{fig:gibbsFanComparison} compares the two upper bounds as a
function of the retained sample size \(t\), namely, the number of
samples collected after the \(t_0\)-step burn-in period. The vertical
axis is displayed on a logarithmic scale. For each value of \(t\), we numerically minimise~\Cref{eq:ourBoundGibbsSampler} and the right-hand side of~\Cref{eq:FanEtAlGibbsSampler} over
\(p\in(1,\infty]\). Thus, neither curve is obtained by imposing a
common or preselected value of \(p\). 
\begin{figure}[t]
    \centering
    \includegraphics[width=0.55\textwidth]
    {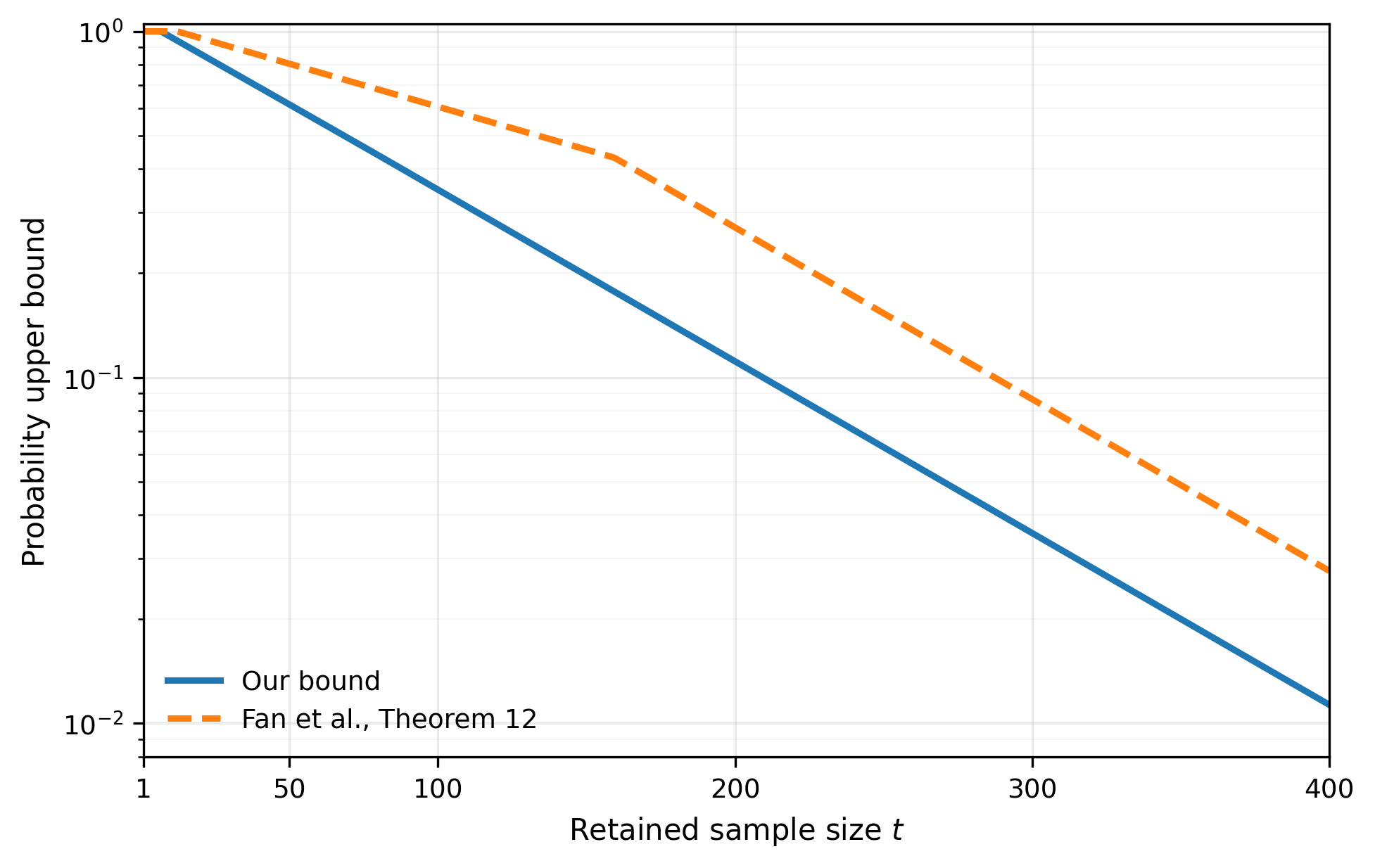}
    \caption{Comparison of our bound with the bound obtained from
    \cite[Thm.~12]{dependentViaStatAndChange} for the two-spin
    random-scan heat-bath Gibbs sampler. The chain starts from
    \(\delta_{++}\), with \(\eta=0.200\) and \(t_0=10\).
    Each curve is independently minimised over
    \(p\in(1,\infty]\) and truncated at \(1\).
    The comparison is uniform over \(g:\mathsf X\to[0,1]\).
    The vertical axis is logarithmic.}
    \label{fig:gibbsFanComparison}
\end{figure}
\end{example}
\subsection{Concentration of measure without independence}\label{sec:concentration}
Another application comes from leveraging~\Cref{thm:probBoundAlphaNorms} along with~\Cref{thm:alphaNormsContraction} in proving concentration of measure bounds for non-independent random variables. %
In particular, the following result holds.
\begin{corollary}\label{thm:sdpiConcentrationBound}
     For $i\geq 2$, let $K_i$ be a discrete-valued Markov kernel. Let $X_1,\ldots,X_t$  be a sequence of random variables with joint measure $\Pm_{X^t}$ and marginals $\Pm_{X_i}$ with $1\leq i \leq t$. Assume that $\Pm_{X_i} = \Pm_{X_{i-1}}K_{i}$ \textit{i.e.}, the random variables are Markovian. Then, for every $p>1$, $q=p/(p-1)$, and every $f$ s.t.\ $|f(x)-f(\hat{x})|\leq \frac1t$ with $x$ and $\hat{x}$ two $t$-dimensional vectors differing only in the $i$-th position,
        \begin{equation}
               \mathbb{P}\left(\left\lvert f(X_1,\ldots,X_t)-\Pm_{\bigotimes_{i=1}^t X_i}(f)\right\rvert \geq \eta \right) \leq 2^\frac1q \exp\left(-\frac{2t\eta^2}{q}\right)\left(\prod_{i=2}^t \left(\left\lVert K_i^\star \right\rVert_{L_p^0\to L_p^0}\omega_p^i+1\right)\right),\label{eq:concentrationSDPI}
        \end{equation}
   where $\omega_p^i = \max_{x} 
((1-\Pm_{X_{i-1}}(\{x\}))^p \Pm_{X_{i-1}}(\{x\})^{1-p}$ $+ (1-\Pm_{X_{i-1}}(\{x\})) )^\frac1p$ and $\Pm_{\bigotimes_{i=1}^t X_i}(f)$ is mean of $f$ under the product of the marginals.  
\end{corollary}
The proof of~\Cref{thm:sdpiConcentrationBound} can be found in~\Cref{app:proofConcentrationWithoutIndependence}. 
Let us compare with existing results in the following example. 
\begin{example}
\label{ex:nonStationaryTimeDependentConcentration}
Let
\[
\mathcal X=\{1,2,3,4\},
\qquad
\pi=
\left(\frac14,\frac14,\frac14,\frac14\right),
\]
and consider the transition matrix
\begin{equation}
K=
\frac1{20}
\begin{pmatrix}
    7&3&7&3\\
    3&7&3&7\\
    5&5&5&5\\
    5&5&5&5
\end{pmatrix}.
\label{eq:nonStationaryTimeDependentKernel}
\end{equation}
The matrix \(K\) is doubly stochastic, and hence \(\pi\) is
stationary. We take the non-stationary initial distribution $
X_1\sim\nu=
\left(0,0,\frac14,\frac34\right).
\label{eq:nonStationaryInitialDistribution}$
In particular, \(\nu\neq\pi\). Since the third and fourth rows of
\(K\) are both equal to \(\pi\), one nevertheless has $X_i\sim\pi
\text{ for every }i\geq2.$ For every \(t\geq2\), let $f_i:\mathsf X\longrightarrow[0,1], i=1,\ldots,t,$ 
be arbitrary and not necessarily identical. Define $X^t:=(X_1,\ldots,X_t)$
and $
f(X^t):=\frac1t\sum_{i=1}^t f_i(X_i).
\label{eq:nonStationaryTimeDependentFunction}$
No particular choice of the functions \(f_i\) is made. All the
bounds below hold uniformly over every collection
\(f_1,\ldots,f_t:\mathsf X\to[0,1]\).

If \(x^t,y^t\in\mathsf X^t\) have Hamming distance one, then $
\left|f(x^t)-f(y^t)\right|\leq\frac1t.$
Thus \(f(X^t)\) satisfies the bounded-difference assumptions of the
concentration inequalities considered below. Since \(f(X^t)\) is additive, linearity gives
\begin{align}
\mathbb P_{X^t}(f)
&=
\int_{\mathsf X^t}
f(x^t)\,\mathbb P_{X^t}(dx^t) =
\frac1t\sum_{i=1}^t
\int_{\mathsf X}f_i(x_i)\,\mathbb P_{X_i}(dx_i) =
\int_{\mathsf X^t}
f(x^t)\,
\left(\bigotimes_{i=1}^t\mathbb P_{X_i}\right)(dx^t)=
\mathbb P_{\bigotimes_{i=1}^tX_i}(f).
\label{eq:nonStationaryPathProductMean}
\end{align}

Thus, the two means match and we can easily compare with the rest of the results. We consider the one-sided deviation $
\mathbb P\left(
    f(X^t)-\mathbb P_{X^t}(f)\geq\eta
\right).$ For the first transition, \(K(\cdot\mid x)=\pi\) for every
\(x\) in the support of \(\nu\). Hence, the first transition
contributes no density factor. For every subsequent transition,
\Cref{thm:psiNormsContraction} gives, sharply,
\begin{equation}
\left\|K^\star\right\|_
    {L_\infty^0(\pi)\to L_\infty^0(\pi)}
=
\left\|
    \left\|
        \frac{K(Y\mid X)}{\pi(Y)}-1
    \right\|_{L_1(\pi_X)}
\right\|_{L_\infty(\pi_Y)}
=
\frac15.
\label{eq:nonStationaryTimeDependentContraction}
\end{equation}
Moreover,
\[
\max_{x\in\mathsf X}
\left\|
    \frac{d\delta_x}{d\pi}-1
\right\|_{L_\infty(\pi)}
=3.
\]
Therefore,~\Cref{thm:sdpiConcentrationBound} with $p\to\infty$ (which maximises the exponential rate of convergence, at the cost of a higher Hellinger integral) gives
\begin{align}
\mathbb P\left(
    f(X^t)-\mathbb P_{X^t}(f)\geq\eta
\right) \leq
\exp\left(-2t\eta^2\right)\left(\frac85\right)^{t-2}.
\label{eq:nonStationaryTimeDependentOurBound}
\end{align}

We now compare this expression with~\cite{dependentViaMartingale, dependentViaStatAndChange, Marton1996BoundingB}. Since~\cite{Marton1996BoundingB} provides concentration around the median, we translate said result in a concentration around the mean bound first. 
Let
\(\operatorname{med}_{\mathbb P_{X^t}}(f)\) denote a median of
\(f(X^t)\). Marton's inequality gives
\begin{align}
\mathbb P\left(
    \left|
        f(X^t)
        -
        \operatorname{med}_{\mathbb P_{X^t}}(f)
    \right|>u
\right) \leq
\min\left\{
    1,\,
    2\exp\left[
        -2t\left(
            \frac35u-\sqrt{\frac{\log 2}{2t}}
        \right)_+^2
    \right]
\right\}.
\label{eq:nonStationaryTimeDependentMartonMedian}
\end{align}
Integrating this inequality and using that probabilities are at most
one gives
\begin{align}
\left|
    \mathbb P_{X^t}(f)
    -
    \operatorname{med}_{\mathbb P_{X^t}}(f)
\right|
\leq
\frac{
    2\sqrt{\log 2}
    +
    \sqrt{\pi}\,
    \operatorname{erfc}\!\left(\sqrt{\log 2}\right)
}{
    (3/5)\sqrt{2t}
}.
\label{eq:nonStationaryTimeDependentMartonMeanMedian}
\end{align}
Consequently, Marton's approach yields
\begin{align}
\mathbb P\left(
    f(X^t)-\mathbb P_{X^t}(f)\geq\eta
\right) \leq
\min\left\{
    1,\,
    2\exp\left[
        -2t\left(
            \frac35\eta
            -
            \frac{
                3\sqrt{\log 2}
                +
                \sqrt{\pi}\,
                \operatorname{erfc}\!\left(\sqrt{\log 2}\right)
            }{\sqrt{2t}}
        \right)_+^2
    \right]
\right\}.
\label{eq:nonStationaryTimeDependentMartonMean}
\end{align}
The bound in~\cite{dependentViaStatAndChange} gives, with
\(q=p/(p-1)\),
\begin{align}
\mathbb P_\nu\left(
    f(X^t)-\mathbb P_{X^t}(f)\geq\eta
\right) \leq
\min\left\{
    1,\,
    \inf_{1<p\leq\infty}
    C(\nu,0,p)
    \exp\left[
        -\frac{2t}{q}
        \frac{5-\sqrt2}{5+\sqrt2}
        \left(\eta-\frac1{2t}\right)^2
    \right]
\right\},
\label{eq:nonStationaryTimeDependentFanBound}
\end{align}
where
\[
C(\nu,0,p)=
\begin{cases}
\displaystyle
1+2^{\,2(1-2/p)_+}(2+2^p)^{1/p},
& p\in(1,\infty)\setminus\{2\},\\[0.6em]
\displaystyle 1+\sqrt{\frac32},
& p=2,\\[0.6em]
3,
& p=\infty.
\end{cases}
\]
We adopted the same notation as in~\cite{dependentViaStatAndChange} and the second argument of $C$ denotes the fact that there is no burn-in period.
The Dobrushin contraction coefficient of
\Cref{eq:nonStationaryTimeDependentKernel} is \(2/5\).
Consequently, the one-sided Kontorovich--Ramanan inequality~\cite{dependentViaMartingale} gives
\begin{align}
\mathbb P\left(
    f(X^t)-\mathbb P_{X^t}(f)\geq\eta
\right) \leq
\min\left\{
    1,\,
    \exp\left[
        -\frac{9t\eta^2}{
            50\left(1-(2/5)^t\right)^2
        }
    \right]
\right\}.
\label{eq:nonStationaryTimeDependentKRBound}
\end{align}
\begin{figure}[t]
    \centering
    \subfloat[Moderate deviation, $\eta=0.50$.
    \label{fig:nonStationaryModerateEta}]
    {
        \includegraphics[width=0.48\textwidth]
        {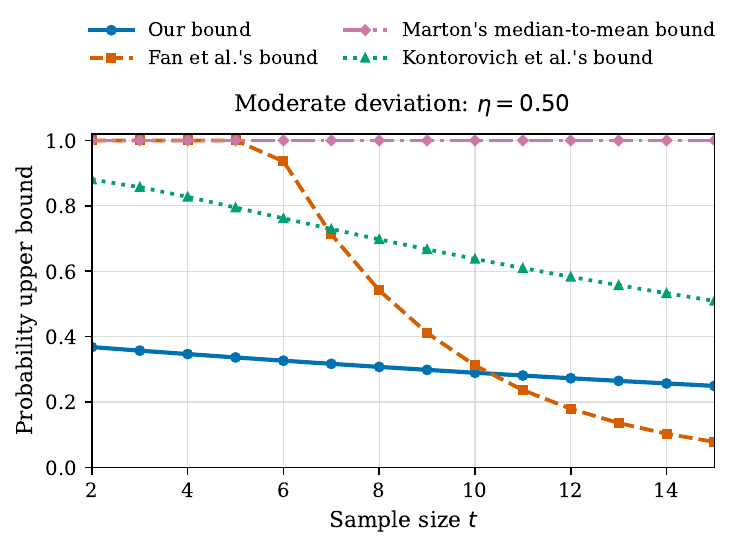}
    }
    \hfill
    \subfloat[Larger deviation, $\eta=0.74$.
    \label{fig:nonStationaryLargeEta}]
    {
        \includegraphics[width=0.48\textwidth]
        {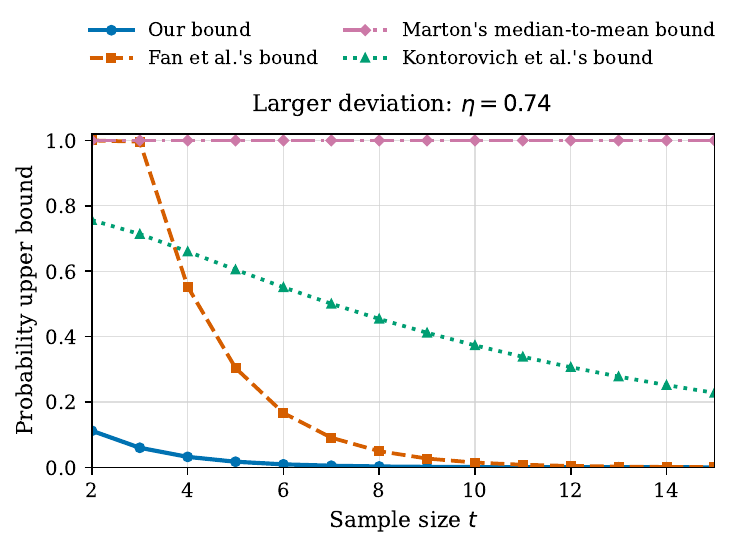}
    }
    \caption{
        Comparison of the one-sided probability bounds for
        $t=2,\ldots,15$. The solid blue curve is our bound, the
        dashed orange curve is Fan et al.'s bound
        from~\cite{dependentViaStatAndChange}, the dash-dotted pink
        curve is Marton's median-to-mean bound obtained
        from~\cite{Marton1996BoundingB}, and the dotted green curve
        is Kontorovich et al.'s bound
        from~\cite{dependentViaMartingale}. The value $\eta=0.74$
        in panel~\textup{(b)} is chosen so that our bound is
        analytically guaranteed to be strictly smaller than each
        of the other three displayed bounds for every integer
        $t\geq 2$. Both vertical axes use a linear scale.
    }
    \label{fig:nonStationaryTwoEtaBounds}
\end{figure}
\Cref{fig:nonStationaryTwoEtaBounds} compares the various bounds for moderate deviations and $\eta=0.5$ as well as in the large deviation regime with $\eta=0.74$, showing how our bounds improve over the rest, respectively, for a restricted set of values of $t$ and for all $t\geq 2$. 
The improvement is uniform over all collections
\(f_1,\ldots,f_t:\mathsf X\to[0,1]\); no particular additive
function is selected. Notice, moreover, that~\Cref{thm:sdpiConcentrationBound} is more general as it does not require $f$ to be additive while~\cite{dependentViaStatAndChange} imposes such a restriction. 
\end{example}
\section{Conclusions}

We developed a duality-based framework for bounding linear and nonlinear contraction of information divergences under Markov kernels. By expressing the evolution of densities through the reverse kernel, contraction reduces to studying the reverse operator on centred functions. Combined with Orlicz-space duality, this yields explicit mixed-norm bounds that apply directly to \(L_p\) and Orlicz spaces, without relying on spectral gaps or \(L_2\) interpolation.

For norm-generated divergences, \Cref{thm:psiNormsContraction,cor:chiSquareBound} provide computable distribution-dependent bounds on contraction and SDPI coefficients. On finite alphabets, they can be evaluated directly from the reference measure and kernel, without optimisation over input distributions. The bounds are tight for several nontrivial settings, including the \(\chi^2\)-divergence for important binary and low-rank channels, and can improve existing distribution-independent estimates for full-rank kernels. For general \(\varphi\)-divergences, our polar-gauge formulation replaces the infinite-dimensional optimisation defining \(F_\varphi\) with closed-form quantities or one-dimensional convex problems. Applied to the squared Hellinger distance and relative entropy, it yields nonlinear contraction and linear SDPI bounds, with Bennett, Bernstein, and sub-Gaussian relaxations offering different trade-offs between sharpness and simplicity.

The resulting bounds yield mixing guarantees, remain effective for heavy-tailed stationary distributions, improve burn-in and high-probability MCMC estimates, and strengthen concentration inequalities for Markov-dependent variables. Thus, reverse-kernel duality provides a unified and computationally tractable link between operator contraction, information inequalities, mixing, and concentration. Future work includes optimising the underlying Young functions, extending the framework to broader continuous-state settings, and applying the resulting concentration bounds to problems such as multi-armed bandits with Markovian rewards.



\newpage

\appendix
\crefalias{section}{appendix}
\crefalias{subsection}{appendix}
\section{Additional preliminaries and results}
\subsection{Orlicz spaces} \label{app:orliczSpaces}
Orlicz spaces are quite general, and they include the well-known $L_p$-spaces. Indeed, if $\psi(x)= |x|^p$, then $L_\psi = L_p$ and $\left\lVert\cdot\right\rVert_{L^L_\psi(\mu)}$ is the $L^p$-norm, see~\Cref{eq:amemiyaNorm}.
We note that different choices of $\psi$ categorise different tail behaviours of random variables. 
Given a function and its %
Young's dual, the following  generalised H\"older's inequality holds, see \cite[Appendix A]{fullVersionGeneralization} for a proof.
\begin{lemma}\label{thm:generalisedHolder}
Let $\psi$ be an Orlicz function and $\psi^\star$ %
its complementary function. For every pair of non-negative random variables $U,V$ respectively defined over $(\Omega_U,\F_U,\Pm_U),(\Omega_V,\F_V,\Pm_V)$, given a coupling  $\Pm_{UV}(UV)$, one has that
    \begin{equation}\label{eq:genHoldIneq}
        \Pm_{UV}(UV) \leq \lVert U\rVert_{L^L_\psi(\Pm_U)} \lVert V\rVert_{L^A_{\psi^\star}(\Pm_V)}.
    \end{equation}
\end{lemma}
Choosing $\psi(x)=|x|^p$, which implies ${\psi}^\star(x^\star)=|x^\star|^q\frac{(p-1)}{p^q}$ with $\frac1p+\frac1q=1$, Lemma~\ref{thm:generalisedHolder} yields the classical H\"older's inequality:
    \begin{equation}
    \Pm_{UV}(UV) \leq \left\lVert U\right\rVert_{L^p(\Pm_U)} \cdot \left\lVert V\right\rVert_{L^q(\Pm_V)}.
    \end{equation}
\subsection{Proof of Lemma~\ref{lem:kStarRND}}\label{app:kStarRND}
 \begin{proof}
    Given that $\nu\ll\mu$, one has that $\nu K\ll \mu K$. %
    Indeed, let $E$ be an event such that $\mu K(E)=0$. Then, %
    \begin{equation}
        \mu K(E) = \int \mathbbm{1}_E d\mu K = \int K(\mathbbm{1}_E) d\mu = 0 .
    \end{equation} Hence, $K(\mathbbm{1}_E)$ is $0$ $\mu$-a.e., and by absolute continuity between $\nu$ and $\mu$, $K(\mathbbm{1}_E)$ is also $0$ $\nu$-a.e. and $\nu K(E)=0$. Thus, the function $g$ is well-defined. Moreover, given any function $h$,
    \begin{align}
        \langle h, g \rangle_{\mu K} = \mu K( h\cdot g) &=  \int h \frac{d\nu K}{d\mu K} d\mu K\\ &=\int h d\nu K \\ &= \int Kh d\nu \\
        &= \int (Kh)f d\mu =  \langle Kh,f \rangle_{\mu} = \langle h,K^\star_\mu f \rangle_{\mu K}, 
    \end{align}
    which implies that $g=K^\star_\mu f$, $\mu K$-almost everywhere.
    \end{proof}
    An important result allows us to characterise the contraction coefficient of the  $t$-fold application $K^t$ of a Markovian operator, leveraging the contraction coefficient of $K$ itself.
\begin{theorem}\label{thm:recursionOperatorNorm}
    Let $K$ be a Markov kernel with stationary distribution $\pi$, let $\psi$ be a Young functional, and let $N\in\{A,L\}$. Then, for every $f \in L_\psi^0(\pi)$ and for every $t\in\mathbb N$, we have %
    \begin{equation}
        \left\lVert K^t f\right\rVert_{L_\psi^N(\pi)} \leq \left\lVert K\right\rVert_{L_\psi^{N,0}\to L_\psi^{N,0}} \cdot  \left\lVert K^{t-1} f\right\rVert_{L_\psi^N(\pi)}.
    \end{equation}
    Consequently, one has that:
    \begin{equation}
        \left\lVert K^t \right\rVert_{L_\psi^{N,0}\to L_\psi^{N,0}} \leq \left\lVert K\right\rVert_{L_\psi^{N,0}\to L_\psi^{N,0}}^t.
    \end{equation}
\end{theorem}
\begin{proof}
    For a given $x$ and measurable event $E$, $K^t(E|x) = \int K^{t-1}(E|z) dK(z|x)$ and consequently $K^tf = Kg$ $\pi$-a.e., where $g(z) = K^{t-1}f(z)$. Indeed, it follows by Fubini's theorem that
    \begin{align}
        K^tf(x) = \int_y f(y) dK^t(y|x) &= \int_y f(y) \int_z K^{t-1}(y|z) dK(z|x)  \\ &=  \int_z dK(z|x) \int_y f(y) K^{t-1}(y|z) \\ &= \int_z dK(z|x) K^{t-1}f(z) = Kg(x).
    \end{align}
    Hence, one can conclude that
    \begin{align}
         \left\lVert K^t f\right\rVert_{L_\psi^N(\pi)} = \left\lVert Kg \right\rVert_{L_\psi^N(\pi)} \leq \left\lVert K\right\rVert_{L_\psi^{N,0}\to L_\psi^{N,0}} \cdot  \left\lVert  g \right\rVert_{L_\psi^N(\pi)} = \left\lVert K\right\rVert_{L_\psi^{N,0}\to L_\psi^{N,0}} \cdot  \left\lVert  K^{t-1}f\right\rVert_{L_\psi^N(\pi)}, \label{eq:contractionNthPower}
    \end{align}
 where the inequality in~\Cref{eq:contractionNthPower} follows from the fact that $g \in L_\psi^0(\pi)$. Indeed, $\pi(g) = \int g d\pi = \int K^{t-1}f d\pi = \int f d(\pi K^{t-1}) = \int f d\pi = 0$, as $f\in L_\psi^0(\pi).$ 
\end{proof}
\subsection{Contraction in Orlicz Spaces}
 The %
    contractivity of Markov operators has been %
    studied in particular for $L^p$ spaces, with \cite{contractionLedoux, hypercontractivityMarkovSemigroupsOrlicz} %
    considering %
    Orlicz spaces as well. As we %
     prove results for both Amemiya and Luxemburg norms, %
    we use $\left\lVert \cdot \right\rVert_{L_\psi^N(\mu)}$ with $N\in \{A,L\}$ to denote both norms and we use $\left\lVert \cdot \right\rVert_{L_\psi^{N^\star}(\mu)}$ for the corresponding dual norm (\textit{e.g.}, if $N=A$ and one has an Amemiya norm, then $N^\star = L$ and the dual is the Luxemburg norm).

    \begin{definition}\label{def:contractionCoefficients}
        Let $\psi$ be a Young functional, $N\in\{A,L\}$ and $K$ a Markov kernel with stationary distribution $\mu$. The contraction coefficients of $K$ in the spaces $L_\psi(\mu), L_\psi^0(\mu)$ are denoted as follows:
    \begin{equation}
            \sup_{f\in L_\psi(\mu)} \frac{\left\lVert Kf \right\rVert_{L_\psi^N(\mu)}}{\left\lVert f \right\rVert_{L_\psi^N(\mu)}} = \left\lVert K \right\rVert_{L_\psi^N(\mu)\to L_\psi^N(\mu)},\qquad \sup_{f\in L_\psi^0(\mu)} \frac{\left\lVert Kf \right\rVert_{L_\psi^N(\mu)}}{\left\lVert f \right\rVert_{L_\psi^N(\mu)}} = \left\lVert K \right\rVert_{L_\psi^{N,0}(\mu)\to L_\psi^{N,0}(\mu)},
        \end{equation}
        where $L_\psi^0(\mu)$ denotes the closed subspace of functions with mean $0$, \textit{i.e.}, $L_\psi^0(\mu)= \{f \in L_\psi(\mu): \mu(f)=0\}$. Whenever the %
        measure is clear from the context, for ease of notation, we denote the contraction coefficient as $\left\lVert K \right\rVert_{L_\psi^N\to L_\psi^N}$. %
    \end{definition}
A %
characterisation of how 
Markovian operators behave 
in Orlicz spaces is stated below and proved in Appendix \ref{app:contractionInOrliczSpaces}. %
    \begin{theorem}\label{thm:contractionInOrliczSpaces}
        Let $\psi$ be a Young functional, $N\in \{A,L\}$, and $K$ a Markov kernel with stationary distribution $\pi$. Then, for every $f\in L_\psi(\pi)$, %
    \begin{align}
            \left\lVert K f\right\rVert_{L^N_\psi(\pi)} \leq 
            \left\lVert f\right\rVert_{L^N_\psi(\pi)}\hspace{2em}&\text{and}\hspace{2em} \left\lVert K \right\rVert_{L^N_\psi \to L^N_\psi} =1.
        \end{align}
    \end{theorem}

   The key takeaway of \Cref{thm:contractionInOrliczSpaces} is that, %
    as in $L_p$ spaces, Markov kernels are contractive, but on the general space of $L_\psi$-integrable functions said contraction coefficient is trivially $1$. Restricting %
    to the closed subspace of functions with mean $0$ (\textit{i.e.}, $L^0_\psi(\mu)$) does yield an improvement. In fact, %
    the contraction coefficient of a Markov kernel $K$ in %
    $L_\psi^0$ is typically less than $1$. This observation is fundamental to characterising convergence to the stationary distribution. 
    
    The purpose of this work is to relate the contractivity of Markovian operators %
    on Orlicz spaces to the convergence of the corresponding Markov chain to its stationary distribution. %
    To do so, we leverage~Lemma~\ref{lem:kStarRND} and show %
    convergence in $L_\psi^N$-norm, as soon as the contraction coefficient of the dual kernel ${K}^\star$ is $<1$. This is formalised in the result below, which is proved in Appendix \ref{app:convergenceInNorm}. %
    \begin{theorem}\label{thm:convergenceInNorm}
        Let $K$ be a Markov kernel with stationary distribution $\pi$, $N\in\{A,L\}$, and $\psi$ a Young functional. %
        For every measure $\nu\ll\pi$ and $t\in\mathbb{N}$, we have %
    \begin{equation}
            \left\lVert \frac{d\nu K^t}{d\pi} - 1 \right\rVert_{L^N_\psi(\pi)} \leq \left\lVert {K}^\star\right\rVert_{L_\psi^{N,0}\to L_\psi^{N,0} }^t\left\lVert \frac{d\nu}{d\pi}-1\right\rVert_{L_\psi^{N}(\pi)} .\label{eq:convergenceInNorm}
        \end{equation}
    \end{theorem}
   \Cref{thm:convergenceInNorm} links the convergence of a Markovian process %
   to the contraction coefficient of the dual kernel $K^\star$. Indeed, if %
   $\left\lVert {K}^\star\right\rVert_{L_\psi^{N,0}}<1$, then %
   the right-hand side of~\Cref{eq:convergenceInNorm} goes to $0$ exponentially fast in the number of steps $t$ and, hence, %
   $\nu K^{\infty}=\pi$. %
   We also note that %
   the proof can be adapted to replace $\left\lVert {K}^\star\right\rVert_{L_\psi^{N,0}\to L_\psi^{N,0} }^t$ with $\left\lVert {K}^\star-1_\pi \right\rVert^t_{L_\psi^{N}\to L_\psi^{N}}$, where $1_\pi$ is the operator mapping a function $f$ to $\int f d\pi$. %
   These two bounds are closely related:
   $\left\lVert {K}^\star\right\rVert_{L_\psi^{N,0}\to L_\psi^{N,0} }^t$ focuses on $0$-mean functions, while in $\left\lVert {K}^\star-1_\pi \right\rVert^t_{L_\psi^{N}\to L_\psi^{N}}$ we directly subtract the mean $1_\pi$. 
 More generally, we can relate the contraction coefficient of $K^\star$ on $L_\psi^0$ to that of $K^\star-1_\pi$ on $L_\psi$, see~\Cref{app:contractionCoefficientsRelationship}. %

\subsection{Relationship between contraction coefficients}\label{app:contractionCoefficientsRelationship}

\begin{theorem} \label{thm:contractionCoefficientsRelationship}
        Let $K$ be a Markov kernel with stationary distribution $\pi$, let $\psi$ be a Young functional, $N\in\{A,L\}$, and let $t\in\mathbb{N}$. 
        Then, the following holds:
        \begin{align}
            \left\lVert K^t \right\rVert_{L_\psi^{N,0} \to L_\psi^{N,0}} &\leq \left\lVert (K - 1_\pi)^t \right\rVert_{L_\psi^N \to L_\psi^N}  \leq 2 \left\lVert K^t \right\rVert_{L_\psi^{N,0} \to L_\psi^{N,0}}. \label{eq:contractionOrliczMean0AndNot}
            \end{align}
    \end{theorem}
    \begin{remark}
    Taking %
    $\psi(x)=|x|^p$, which gives $L_p$-norms, we recover Lemma 3.16 by %
    \cite{mcmcThesis}. 
    \end{remark}

\begin{proof}
The first inequality is obtained via the passages below:
    \begin{align}
       \left\lVert K^t \right\rVert_{L_\psi^{N,0} \to L_\psi^{N,0}} &= \sup_{g : \left\lVert g \right\rVert_{L_\psi^N(\pi)} \leq 1, \pi(g)=0} \left\lVert K^t g \right\rVert_{L_\psi^N(\pi)} \\
       &= \sup_{g : \left\lVert g \right\rVert_{L_\psi^N(\pi)} \leq 1, \pi(g)=0} \left\lVert (K^t-1_\pi) g \right\rVert_{L_\psi^N(\pi)} \\
       &\leq  \sup_{f : \left\lVert f \right\rVert_{L_\psi^N(\pi)} \leq 1, } \left\lVert (K^t-1_\pi) f \right\rVert_{L_\psi^N(\pi)} \\
       &=  \left\lVert K^t-1_\pi \right\rVert_{L_\psi^{N} \to L_\psi^{N}}.
    \end{align}
    
    Moreover, one has that there exists a constant $c_N$ such that the following sequence of steps holds:
    \begin{align}
        \left\lVert K^t - 1_\pi \right\rVert_{L_\psi^N \to L_\psi^N} &= \sup_{f : \left\lVert f \right\rVert_{L_\psi^N(\pi)} \leq 1} \left\lVert (K^t - 1_\pi) f \right\rVert_{L_\psi^N(\pi)} \\ &= \sup_{f : \left\lVert f \right\rVert_{L_\psi^N(\pi)} \leq 1} \left\lVert K^t (f-\pi(f)) \right\rVert_{L_\psi^N(\pi)} \\
        &= c_N \sup_{f : \left\lVert f \right\rVert_{L_\psi^N(\pi)} \leq 1} \left\lVert K^t\left( \frac{f-\pi(f)}{c_N}\right) \right\rVert_{L_\psi^N(\pi)} \\
        &\leq c_N \sup_{g : \left\lVert g \right\rVert_{L_\psi^N(\pi)} \leq 1, \pi(g)=0} \left\lVert K^t g \right\rVert_{L_\psi^N(\pi)} \label{eq:constantStepContraction}\\ &= c_N \left\lVert K^t \right\rVert_{L_\psi^{N,0} \to L_\psi^{N,0}}.
    \end{align}
    It remains to upper bound the constant $c_N$. In order for~\Cref{eq:constantStepContraction} to hold, $c_N$ needs to satisfy %
    $$\left\lVert \frac{f-\pi(f)}{c_N} \right\rVert_{L^N_\psi(\pi)}\leq 1.$$ %
    One clearly has that
    \begin{equation}
        \left\lVert \frac{f-\pi(f)}{c_N} \right\rVert_{L^N_\psi(\pi)}\leq \frac{1}{|c_N|} \left( \left\lVert f\right\rVert_{L^N_\psi(\pi)} +  \left\lVert \pi(f) \right\rVert_{L^N_\psi(\pi)}\right).
    \end{equation}
    In particular, an application of Jensen's inequality gives
    \begin{align}
        \left\lVert \pi(f) \right\rVert_{L_\psi^N(\pi)} \leq   \pi(\left\lVert f\right\rVert_{L_\psi^N(\pi)}) \leq 1.    \end{align}
        Thus, the optimal admissible constant satisfies \(c_N\leq2\), which concludes the proof. %
    \end{proof}
    To conclude our characterisation of the contraction coefficients, we prove the equivalence between $\left\lVert (K - 1_\pi)^\star \right\rVert_{L_\psi^{N}\to L_\psi^{N}}$ and $\left\lVert K^\star - 1_\pi \right\rVert_{L_\psi^{N}\to L_\psi^{N}}$.
    \begin{lemma} Let $K$ be a Markov kernel with stationary distribution $\pi$, and $K^\star=K^\star_\pi$ the dual kernel. Then, one has that 
    \begin{equation}
        (K-1_\pi)^\star = K^\star -1_\pi, \pi-a.e.
    \end{equation}
    \end{lemma}
    \begin{proof} 
    We first show that 
        $1_\pi^\star = 1_\pi$, \textit{i.e.}, that for every function $h$ and $f$, $\langle 1_\pi h,f\rangle_\pi = \langle h, 1_\pi f\rangle_\pi$. 
        Indeed, we have
        \begin{align}
           \langle 1_\pi h, f \rangle_\pi = \int d\pi f \pi(h) = \pi(f)\pi(h) = \int d\pi h \pi(f) = \langle h, 1_\pi f\rangle_\pi .
        \end{align}
        From this, one can also conclude that, if $K$ is stationary with respect to $\pi$, then $(K-1_\pi)^\star = K^\star-1_\pi$.
       Indeed, we have
       \begin{align}
             \langle (K-1_\pi) h, f \rangle_\pi = \langle Kh, f \rangle_\pi - \langle 1_\pi h, f \rangle_\pi = \langle h, K^\star f \rangle_\pi - \langle  h, 1_\pi f \rangle_\pi = \langle h, (K^\star-1_\pi) f \rangle_\pi.
       \end{align}
    \end{proof}

\subsection{Proof of Theorem~\ref{thm:contractionInOrliczSpaces}}\label{app:contractionInOrliczSpaces}
\begin{proof}
    Assume that, for every $\epsilon>0$, there exists a constant $\lambda$ such that 
     \begin{equation}
        \lambda = \left\lVert f \right\rVert_{L^L_\psi(\pi)} + \epsilon
        \hspace{2em} \text{and} \hspace{2em}
        \pi\left(\psi\left(\frac{|f|}{\lambda}\right)\right)\leq 1.
    \end{equation}
    Moreover, the following steps hold:
    \begin{align}
         \pi\left(\psi\left(\frac{|Kf|}{\lambda}\right)\right) 
        &\leq \pi\left(\psi\left(\frac{K|f|}{\lambda}\right)\right)
        \\
        &\leq \pi K\left(\psi\left(\frac{|f|}{\lambda}\right)\right) \\
        &= \pi\left(\psi\left(\frac{|f|}{\lambda}\right)\right) \leq 1.
    \end{align}
    Hence, since $\left\lVert Kf \right\rVert_{L_\psi^L(\pi)} = \inf\left\{\sigma>0: \pi\left(\psi\left(\frac{|Kf|}{\sigma}\right)\right) \leq 1 \right\} $ one has that $\left\lVert Kf \right\rVert_{L_\psi^L(\pi)} \leq \lambda = \left\lVert f \right\rVert_{L_\psi^L(\pi)} + \epsilon$. The statement follows from the arbitrariness of $\epsilon$. 
    To conclude the argument for the Luxemburg norm, select the function $g$ identically equal to $\psi^{-1}(1)>0$. One has that in order for $\pi\left(\psi\left(\frac{g}{\lambda}\right)\right) = \pi\left(\psi\left(\frac{\psi^{-1}(1)}{\lambda}\right)\right) \leq 1$ the following needs to hold  
    \begin{equation}
        \lambda \geq \frac{\psi^{-1}(1)}{\psi^{-1}(1)} = 1,
    \end{equation}  
    and thus $\left\lVert g \right\rVert_{L^L_\psi(\pi)} = 1$. Hence, since $Kg=g$,  one has that $\left\lVert K \right\rVert_{L_\psi^L \to L_\psi^L} = 1$.
    
    As for the Amemiya norm, assume that, for every $\epsilon>0$, there exists $\lambda$ such that
    \begin{equation}
        \frac{1+\pi(\psi(\lambda|f|))}{\lambda} \leq \left\lVert f \right\rVert_{L^A_\psi(\pi)} + \epsilon,
    \end{equation}
    then one has that for every Markov kernel $K$ 
    \begin{align}
       \left\lVert Kf \right\rVert_{L^A_\psi(\pi)} \leq  \frac{1+\pi(\psi(\lambda |Kf|))}{\lambda} &\leq \frac{1+\pi(\psi(\lambda K |f|))}{\lambda} \\
       &\leq \frac{1+\pi K(\psi(\lambda  |f|))}{\lambda} \\
       &= \frac{1+\pi(\psi(\lambda  |f|))}{\lambda} \\
       &\leq \left\lVert f \right\rVert_{L^A_\psi(\pi)} + \epsilon.
    \end{align}
    The argument then follows from the arbitrariness of $\epsilon$. This implies that $\left\lVert K \right\rVert_{L_\psi^A \to L_\psi^A} \leq 1$.    Moreover,  if one selects $g$ to be the function identically equal to $\frac{1}{{\psi^{\star}}^{-1}(1)}$, then 
    \begin{align}
        \left\lVert g \right\rVert_{L^A_\psi(\pi)} =  \inf_{t>0} \frac{1+\pi\left(\psi\left(\frac{t}{{\psi^{\star}}^{-1}(1)}\right)\right)}{t} 
        &= \inf_{t>0} \frac{1+\psi\left(\frac{t}{{\psi^{\star}}^{-1}(1)}\right)}{t} = 1.
    \end{align}
    Indeed, denoting $\varphi(x) = \psi(x/{\psi^\star}^{-1}(1))$, one has that $\varphi^\star(x) = \psi^\star(x{\psi^\star}^{-1}(1) )$ and ${\varphi^\star}^{-1}(y) = {\psi^\star}^{-1}(y)/{\psi^\star}^{-1}(1)$. Given that
    \begin{equation}
        \inf_{t>0} \frac{1+\psi\left(\frac{t}{{\psi^{\star}}^{-1}(1)}\right)}{t} = \inf_{t>0} \frac{1+\varphi(t)}{t} = {\varphi^\star}^{-1}(1) = \frac{{\psi^\star}^{-1}(1)}{{\psi^\star}^{-1}(1)},
    \end{equation}
    the argument follows.
    Hence, similarly to before, since $Kg=g$, one has that $\left\lVert K \right\rVert_{L_\psi^A \to L_\psi^A} = 1$.
\end{proof}
    
    \subsection{Proof of Theorem~\ref{thm:convergenceInNorm}}\label{app:convergenceInNorm}
\begin{proof}
        Let $t\in\mathbb{N}$ and let $\nu\ll\pi$. Then,
         \begin{align}
            \left\lVert \frac{d\nu K^t}{d\pi} -1 \right\rVert_{L_\psi^{N}(\pi)}
            &= \left\lVert {K^t}^\star \left(\frac{d\nu}{d\pi} -1 \right) \right\rVert_{L_\psi^{N}(\pi)} \\
            &\leq \left\lVert {K^t}^\star\right\rVert_{L_\psi^{N,0}\to L_\psi^{N,0} } \left\lVert \frac{d\nu}{d\pi}-1\right\rVert_{L_\psi^{N}(\pi)} \label{eq:boundOnNormPsiMixing} \\
            &\leq \left\lVert {K}^\star\right\rVert_{L_\psi^{N,0}\to L_\psi^{N,0} }^t \left\lVert \frac{d\nu}{d\pi}-1\right\rVert_{L_\psi^{N}(\pi)}\label{eq:boundOnNormTensorPsi}, 
        \end{align}
    where~\Cref{eq:boundOnNormPsiMixing} follows by the definition of $\left\lVert {K}^\star\right\rVert_{L_\psi^{N,0}\to L_\psi^{N,0} }$ (see~Definition~\ref{def:contractionCoefficients}) and the fact that if $f=\frac{d\nu}{d\pi}-1$ then $\pi\left(f\right)=0$; \Cref{eq:boundOnNormTensorPsi} follows from~\Cref{thm:recursionOperatorNorm}.
\end{proof}
\section{Proofs of results in the main text}

\subsection{Proof of Theorem~\ref{thm:psiNormsContraction}}\label{app:proofPsiNormsContraction}
      \begin{proof}
          As $f \in L_\varphi^0(\mu K)$, we have that $\mu K(f)=0$. Thus, 
    \begin{align}
        \left\lVert K f\right\rVert_{L_\psi^N(\mu)} &= \left\lVert |K f|\right\rVert_{L_\psi^N(\mu )} \label{eq:idealNorms}  \\
        &= \left\lVert\left| \int g_X(y) f(y) d\mu K \right| \right\rVert_{L_\psi^N(\mu )} \\
         &= \left\lVert\left| \int (g_X(y)-1) f(y) d\mu K \right| \right\rVert_{L_\psi^N(\mu )} \\ &\leq \left\lVert \int \left|(g_X(y)-1) f(y)\right| d\mu K \right\rVert_{L_\psi^N(\mu )} \label{eq:jensenMainThm}\\
         &\leq \left\lVert f\right\rVert_{L_\varphi^N(\mu K)}\left\lVert \left\lVert g_X-1\right\rVert_{L_{\varphi^\star}^{N^\star}(\mu K)}\right\rVert_{L_\psi^N(\mu )}\label{eq:genHolderMainThm},
    \end{align}
    where~\Cref{eq:idealNorms} uses the ideal property of the
    Amemiya and Luxemburg norms,
    \Cref{eq:jensenMainThm} is the triangle inequality for the
    integral, and~\Cref{eq:genHolderMainThm} follows from the
    generalised H\"older inequality (see~Lemma~\ref{thm:generalisedHolder}).
    The reverse-channel bound follows by the same argument adapted.
      \end{proof}

\subsection{Proof of Corollary~\ref{cor:chiSquareBound}}\label{app:proofChiSquareBound}
\begin{proof}
Let $h=d\nu/d\mu-1$.  Then $\mu(h)=0$ and
Lemma~\ref{lem:kStarRND} gives
$d\nu K/d\mu K-1=K_\mu^\star h$.  The reverse inequality
in~\Cref{thm:psiNormsContraction}, specialised to $L_\alpha$ and its
dual $L_\beta$, gives
\begin{equation}
H_\alpha(\nu K\|\mu K)
\leq
H_\alpha(\nu\|\mu)
\left\lVert
\left\lVert g_Y-1\right\rVert_{L_\beta(\mu)}
\right\rVert_{L_\alpha(\mu K)}^\alpha.
\end{equation}
Taking the supremum over $\nu$ and using the ordinary
data-processing inequality proves the claim.
\end{proof}

\subsection{Proof of Theorem~\ref{thm:alphaInfinityCase}}\label{app:proofAlphaInfinityCase}
\begin{proof}
    The inequality follows from~\Cref{thm:alphaNormsContraction} with $p\to\infty$. As for the equality, notice that, for every $x$, and $t\in\mathbb{N}$
\begin{align}
    \left\lVert g^t_x -1  \right\rVert_{L_1(\pi)}  = \left\lVert \frac{dK^t(Y|x)}{d\pi(Y)}  -1  \right\rVert_{L_1(\pi)} &= \int d\pi(y) \left| \frac{dK^t(y|x)}{d\pi(y)} -1 \right| =2 \left\lVert K^t(\cdot |x)- \pi \right\rVert_{TV}.
\end{align}
The last step follows from the definition of Total Variation distance. Hence, 
\begin{equation}
    \left\lVert \left\lVert g^t_X -1  \right\rVert_{L_1(\pi)} \right\rVert_{L_q(\pi)}\xrightarrow[q\to\infty]{} \esssup_\pi \left\lVert g^t_X -1  \right\rVert_{L_1(\pi)} =  \esssup_\pi 2 \left\lVert K^t(\cdot |X)- \pi \right\rVert_{TV}.
\end{equation}

\end{proof}

\subsection{Proof of Corollary~\ref{thm:ultraMixing}}\label{app:proofUltraMixing}
\begin{proof}
  In~\cite[Lemma 4.1]{contractionLedoux},
    the authors prove that if~\Cref{eq:ultraMixing} holds, then for every measure $\xi$ one has that:
    \begin{equation}
        \frac{dK_\xi^\star(\cdot|Y)}{d\xi}\geq \epsilon \hspace{2em} \xi K-a.e.,
    \end{equation}
     where $Y$ is distributed according to $\xi K$.
    Selecting $\xi=\mu$, define
    \begin{equation}
        R^\star(dx\mid y)=\frac{g_y(x)-\epsilon}{1-\epsilon}\,\mu(dx).
    \end{equation}
    Then $R^\star$ is a Markov kernel, $\mu K R^\star=\mu$, and hence it is contractive from $L_\psi^N(\mu)$ to $L_\psi^N(\mu K)$. For $h=d\nu/d\mu-1$, one has $\mu(h)=0$ and
    \begin{equation}
        \frac{d\nu K}{d\mu K}-1=K_\mu^\star h=(1-\epsilon)R^\star h.
    \end{equation}
    The claim follows.
\end{proof}

\subsection{Proof of Proposition~\ref{prop:hellingerPolarGauge}}\label{app:proofHellingerPolarGauge}
\begin{proof}
    For every $s<\frac{1}{2}$, direct maximisation over $u\geq0$ gives
    \[
        \varphi_{\mathrm H}^{\star}(s)
        =
        \sup_{u\geq0}
        \left\{
        su-\frac{1}{2}(\sqrt{u}-1)^2
        \right\}
        =
        \frac{s}{1-2s}.
    \]
    The supremum is infinite when $s\geq\frac{1}{2}$.

    Let $\nu\ll\mu$ satisfy
    $H^2(\nu,\mu)\leq c$. For every $t>0$ such that
    $t\operatorname*{ess\,sup}_{\mu}f<1$, the
    Fenchel--Young inequality gives
    \begin{align}
        \frac{t}{2}\nu(f)
        &=
        \mu\left[
        \frac{\mathrm d\nu}{\mathrm d\mu}\,\frac{t}{2}f
        \right]\leq
        H^2(\nu,\mu)
        +
        \mu\left[
        \varphi_{\mathrm H}^{\star}\left(\frac{t}{2}f\right)
        \right]\leq
        c+
        \mu\left(
        \frac{tf}{2(1-tf)}
        \right).
        \label{eq:hellingerFenchelYoung}
    \end{align}
    Since $\mu(f)=0$,
    \[
        \mu\left(
        \frac{tf}{2(1-tf)}
        \right)
        =
        \frac{t^2}{2}\mu\left(
        \frac{f^2}{1-tf}
        \right).
    \]
    Dividing~\eqref{eq:hellingerFenchelYoung} by $t/2$,
    taking the supremum over
    $\nu\in\mathcal B_{\varphi_{\mathrm H}}(\mu,c)$,
    and then taking the infimum over $t$ proves
    \eqref{eq:hellingerPolarVariational}.

    Moreover,
    \[
        \mu\left(
        \frac{f^2}{1-tf}
        \right)
        \leq
        \frac{
        \lVert f\rVert_{L^2(\mu)}^2
        }{
        1-t\operatorname*{ess\,sup}_{\mu}f
        }.
    \]
    Therefore,
    \[
        \rho_{\mu,\varphi_{\mathrm H}}^c(f)
        \leq
        \inf_{\substack{t>0\\
        t\operatorname*{ess\,sup}_{\mu}f<1}}
        \left\{
        \frac{2c}{t}
        +
        \frac{
        t\lVert f\rVert_{L^2(\mu)}^2
        }{
        1-t\operatorname*{ess\,sup}_{\mu}f
        }
        \right\}.
    \]
    The infimum is attained at
    \[
        t
        =
        \frac{
        \sqrt{2c}
        }{
        \lVert f\rVert_{L^2(\mu)}
        +
        \sqrt{2c}\operatorname*{ess\,sup}_{\mu}f
        },
    \]
    and its value is
    \[
        2\sqrt{2c}\,
        \lVert f\rVert_{L^2(\mu)}
        +
        2c\operatorname*{ess\,sup}_{\mu}f.
    \]
    This proves~\eqref{eq:hellingerPolarExplicit}.
    Applying the same argument to $-f$ proves
    \eqref{eq:hellingerPolarExplicitNegative}.
\end{proof}

\subsection{Proof of Theorem~\ref{thm:hellingerNonlinearContraction}}\label{app:proofHellingerNonlinearContraction}
\begin{proof}
Substituting
\Cref{eq:hellingerNegativeGaugeChannelClipped,%
eq:hellingerPositiveGaugeChannelClipped}
into the general contraction bound in
\Cref{eq:gaugeInFphi}
gives~\eqref{eq:hellingerNonlinearContraction}.
\end{proof}

\subsection{Proof of Corollary~\ref{cor:hellingerSDPI}}\label{app:proofHellingerSDPI}
\begin{proof}
Since $
    0
    \leq
    D_{\varphi_{\mathrm H}}(\nu\|\mu)
    \leq
    1,$ 
the linearisation identity for contraction curves gives
\begin{equation}
    \eta_{\varphi_{\mathrm H}}(\mu,K)
    =
    \sup_{0<c\leq 1}
    \frac{
        F_{\varphi_{\mathrm H}}(\mu,K;c)
    }{c}.
\end{equation}
Combining this identity with
\Cref{eq:hellingerNonlinearContraction} reduces the proof to the following pointwise linearisation.

For $A,s\geq0$, setting $r=\sqrt{2c}$ gives
\begin{equation}
    \sup_{c>0}
    \frac{
        \min\left\{
            A,\,
            2s\sqrt{2c}+2Ac
        \right\}
    }{
        \sqrt{2c}
    }
    =
    s+\sqrt{s^2+A^2}.
    \label{eq:normalizedClippedSquareRootLinearization}
\end{equation}
Moreover,
\begin{equation}
    \left(\sqrt{1+a}-1\right)^2
    =
    \frac{a^2}{
        \left(1+\sqrt{1+a}\right)^2
    }
    \leq
    \frac{a^2}{4}.
\end{equation}
For the negative part, since $0\leq a\leq1-m_y$,
\begin{equation}
    \left(1-\sqrt{1-a}\right)^2
    =
    \frac{a^2}{
        \left(1+\sqrt{1-a}\right)^2
    }
    \leq
    \frac{a^2}{
        \left(1+\sqrt{m_y}\right)^2
    }.
\end{equation}
Applying
\Cref{eq:normalizedClippedSquareRootLinearization}
with $A=M_y-1$ and $A=1-m_y$, respectively, and then moving the
supremum inside the integral proves
\Cref{eq:hellingerSDPIClosedFormBound}.
\end{proof}

\subsection{Proof of Theorem~\ref{thm:klNonlinearContraction}}\label{app:proofKLNonlinearContraction}
\begin{proof}
We first prove the equality for the polar gauge and the Amemiya-like functional. For every measurable $f$ whose centred exponential moment is
finite in a neighbourhood of the origin, define
\begin{equation}
    \rho_{\mu,\mathrm{KL}}^c(f)
    =
    \sup_{\nu\in\mathcal B_{\mathrm{KL}}(\mu,c)}
    \left\{
        \nu(f)-\mu(f)
    \right\}.
\end{equation}
Since $c>0$, the distribution $\mu$ is a strictly feasible point of
the relative-entropy ball. 
By Donsker-Varadhan and Lagrange duality one has the following:
\begin{align}
    \rho_{\mu,\mathrm{KL}}^c(f)
    &=
    \inf_{\lambda>0}
    \Bigg\{
        \lambda c
        +
        \sup_{\nu\ll\mu}
        \left[
            \nu\left(f-\mu(f)\right)
            -
            \lambda D_{\mathrm{KL}}(\nu\|\mu)
        \right]
    \Bigg\}=
    \inf_{\lambda>0}
    \left\{
        \lambda c
        +
        \lambda
        \log\mu\left(
            e^{(f-\mu(f))/\lambda}
        \right)
    \right\}.
\end{align}

Consequently, one has
\begin{align}
    \rho_{\mu,\mathrm{KL}}^c(f)
    &=
    \inf_{\lambda>0}
    \left\{
        \lambda c
        +
        \lambda
        \log\mu\left(
            e^{(f-\mu(f))/\lambda}
        \right)
    \right\}
    =
    \inf_{t>0}
    \frac{
        c+
        \log\mu\left(
            e^{t(f-\mu(f))}
        \right)
    }{t},
    \label{eq:klPolarGaugeMGF}
\end{align}
where the second equality follows from the change of variables
$t=1/\lambda$.

The last identity is the variational representation of the
log-moment-generating function.

If the relative-entropy ball is defined using the strict inequality
$D_{\mathrm{KL}}(\nu\|\mu)<c$, its polar gauge has the same value.
Indeed, for any $\nu$ satisfying
$D_{\mathrm{KL}}(\nu\|\mu)\leq c$, the mixture $
    \nu_\varepsilon
    =
    (1-\varepsilon)\nu+\varepsilon\mu$
 satisfies $
    D_{\mathrm{KL}}(\nu_\varepsilon\|\mu)
    \leq
    (1-\varepsilon)
    D_{\mathrm{KL}}(\nu\|\mu)
    <
    c,$
and $\nu_\varepsilon(f)\to\nu(f)$ as
$\varepsilon\downarrow0$.

We now apply this representation to the channel. Fix
$\nu\in\mathcal B_{\mathrm{KL}}(\mu,c)$ and write $
    h
    =
    \frac{\mathrm d\nu}{\mathrm d\mu}-1.
$
The output ratio satisfies
\begin{align}
    \frac{\mathrm d\nu K}{\mathrm d\mu K}(y)-1
    =
    K_\mu^\star h(y)
    =
    \mu\left(hg_y\right)
    =
    \mu\left(h(g_y-1)\right)
    =
    \nu(g_y-1),
    \label{eq:klOutputLikelihoodRatioPerturbation}
\end{align}
where we used $\mu(h)=0$ and $\mu(g_y)=1$.

By the definition of the polar gauge, $
    \nu(g_y-1)
    \leq
    \rho_{\mu,\mathrm{KL}}^c(g_y-1).$
Applying the same argument to $1-g_y$ gives$
    -\nu(g_y-1)
    =
    \nu(1-g_y)
    \leq
    \rho_{\mu,\mathrm{KL}}^c(1-g_y).$
Consequently,
\begin{equation}
    -\rho_{\mu,\mathrm{KL}}^c(1-g_y)
    \leq
    \frac{\mathrm d\nu K}{\mathrm d\mu K}(y)-1
    \leq
    \rho_{\mu,\mathrm{KL}}^c(g_y-1).
    \label{eq:klOutputLikelihoodRatioInterval}
\end{equation}

Since $\mu(g_y-1)=\mu(1-g_y)=0$,
\Cref{eq:klPolarGaugeMGF} gives the exact representations
\begin{align}
    \rho_{\mu,\mathrm{KL}}^c(g_y-1)
    &=
    \inf_{t>0}
    \frac{
        c+
        \log\mu\left(
            e^{t(g_y-1)}
        \right)
    }{t},
    \label{eq:klPositivePolarGaugeMGF}\\
    \rho_{\mu,\mathrm{KL}}^c(1-g_y)
    &=
    \inf_{t>0}
    \frac{
        c+
        \log\mu\left(
            e^{-t(g_y-1)}
        \right)
    }{t}.
    \label{eq:klNegativePolarGaugeMGF}
\end{align}

Because $\varphi_{\mathrm{KL}}$ is convex, its maximum on the interval$ 
    \left[
        1-\rho_{\mu,\mathrm{KL}}^c(1-g_y),
        \,
        1+\rho_{\mu,\mathrm{KL}}^c(g_y-1)
    \right]$ is attained at one of its boundary points. Therefore,
\begin{align}
    D_{\mathrm{KL}}(\nu K\|\mu K)
    \leq
    \int
    \max\Bigg\{&
        \varphi_{\mathrm{KL}}
        \left(
            1-\rho_{\mu,\mathrm{KL}}^c(1-g_y)
        \right),
        \varphi_{\mathrm{KL}}
        \left(
            1+\rho_{\mu,\mathrm{KL}}^c(g_y-1)
        \right)
    \Bigg\}
    \,\mu K(\mathrm dy).
    \label{eq:klPolarGaugeOutputBound}
\end{align}
Data processing also gives
\begin{equation}
    D_{\mathrm{KL}}(\nu K\|\mu K)
    \leq
    D_{\mathrm{KL}}(\nu\|\mu)
    \leq c.
\end{equation}
Taking the supremum over
$\nu\in\mathcal B_{\mathrm{KL}}(\mu,c)$ proves
\begin{align}
F_{\mathrm{KL}}(\mu,K;c)
\leq
\min\Bigg\{
c,\,
\int
\max\Bigg\{&
\varphi_{\mathrm{KL}}
\left(
    1-\rho_{\mu,\mathrm{KL}}^c(1-g_y)
\right),
\varphi_{\mathrm{KL}}
\left(
    1+\rho_{\mu,\mathrm{KL}}^c(g_y-1)
\right)
\Bigg\}
\,\mu K(\mathrm dy)
\Bigg\}.
\end{align}
\end{proof}

\subsection{Proof of Theorem~\ref{thm:klBennettContraction}}\label{app:proofKLBennettContraction}
\begin{proof}
For $\mu K$-almost every $y$, the random variable $g_y-1$ is centred
under $\mu$, is bounded above by $M_y-1$, and satisfies
\begin{equation}
    \mu\left((g_y-1)^2\right)
    =
    \lVert g_y-1\rVert_{L^2(\mu)}^2.
\end{equation}
Bennett's moment-generating-function inequality therefore gives, for
every $t>0$,
\begin{align}
\log\mu\left(e^{t(g_y-1)}\right)
\leq
\frac{
    \lVert g_y-1\rVert_{L^2(\mu)}^2
}{
    (M_y-1)^2
}
\left(
    e^{t(M_y-1)}
    -
    t(M_y-1)
    -
    1
\right).
\label{eq:klPositiveMGFBennett}
\end{align}

Let us compute first the scalar minimisation. For
$A>0$, $\sigma^2>0$, and $c>0$,
\begin{align}
\inf_{t>0}
\frac{1}{t}
\left[
c+
\frac{\sigma^2}{A^2}
\left(
    e^{tA}-tA-1
\right)
\right]
=
\frac{\sigma^2}{A}
\left[
    \varphi_{\mathrm{KL}}(1+\cdot)
\right]^{-1}
\left(
    \frac{cA^2}{\sigma^2}
\right).
\label{eq:bennettGaugeMinimization}
\end{align}
Indeed, setting $u = tA$, its first-order optimality condition is
\begin{equation}
\frac{cA^2}{\sigma^2}=e^u(u-1)+1.
\label{eq:bennettOptimalityCondition}
\end{equation}
The right-hand side is strictly increasing from zero to infinity
on $u>0$, so this equation has a unique solution.

Set $z=e^u-1$. Since $u=\log(1+z)$,
\begin{align}
\varphi_{\mathrm{KL}}(1+z)
=
(1+z)\log(1+z)-z =
e^u u-e^u+1 =
e^u(u-1)+1.
\end{align}
Consequently, from~\Cref{eq:bennettOptimalityCondition} one has $
z^\star
=
\left[
    \varphi_{\mathrm{KL}}(1+\cdot)
\right]^{-1}
\left(
    \frac{cA^2}{\sigma^2}
\right), u^\star = \log(1+z^\star).$
Moreover, the first-order condition implies
\[
\frac{cA^2}{\sigma^2}+e^{u^\star}-u^\star-1
=
u^\star(e^{u^\star}-1)
=
u^\star z^\star.
\]
Hence, at the optimiser,
\[
\frac{\sigma^2}{A}
\frac{
\frac{cA^2}{\sigma^2}+e^{u^\star}-u^\star-1
}{
u^\star
}
=
\frac{\sigma^2}{A}z^\star,
\]
which proves~\eqref{eq:bennettGaugeMinimization}.

Using the Amemiya-like representation of the KL polar gauge,
\begin{equation}
    \rho_{\mu,\mathrm{KL}}^c(g_y-1)
    =
    \inf_{t>0}
    \frac{
        c+\log\mu\left(e^{t(g_y-1)}\right)
    }{t},
\end{equation}
and applying
\Cref{eq:klPositiveMGFBennett,eq:bennettGaugeMinimization}
with $
    A=M_y-1,
   \text{ }
    \sigma^2
    =
    \lVert g_y-1\rVert_{L^2(\mu)}^2,$
gives
\begin{align}
\rho_{\mu,\mathrm{KL}}^c(g_y-1)
\leq
\frac{
    \lVert g_y-1\rVert_{L^2(\mu)}^2
}{
    M_y-1
}
\left[
    \varphi_{\mathrm{KL}}(1+\cdot)
\right]^{-1}
\left(
    \frac{
        c(M_y-1)^2
    }{
        \lVert g_y-1\rVert_{L^2(\mu)}^2
    }
\right).
\end{align}
Since the exact polar gauge also satisfies$
    \rho_{\mu,\mathrm{KL}}^c(g_y-1)
    \leq
    M_y-1,$
we obtain
\begin{align}
\rho_{\mu,\mathrm{KL}}^c(g_y-1)
\leq
\min\Bigg\{
M_y-1,\,
\frac{
    \lVert g_y-1\rVert_{L^2(\mu)}^2
}{
    M_y-1
}
\left[
    \varphi_{\mathrm{KL}}(1+\cdot)
\right]^{-1}
\left(
    \frac{
        c(M_y-1)^2
    }{
        \lVert g_y-1\rVert_{L^2(\mu)}^2
    }
\right)
\Bigg\}.
\label{eq:klPositivePolarGaugeBennett}
\end{align}

For the negative direction, $1-g_y$ is centred under $\mu$, is
bounded above by $1-m_y$, and has the same second moment:$
    \mu\left((1-g_y)^2\right)
    =
    \lVert g_y-1\rVert_{L^2(\mu)}^2.$
Bennett's inequality gives
\begin{align}
\log\mu\left(e^{-t(g_y-1)}\right)
\leq
\frac{
    \lVert g_y-1\rVert_{L^2(\mu)}^2
}{
    (1-m_y)^2
}
\left(
    e^{t(1-m_y)}
    -
    t(1-m_y)
    -
    1
\right).
\label{eq:klNegativeMGFBennett}
\end{align}
Using
\begin{equation}
    \rho_{\mu,\mathrm{KL}}^c(1-g_y)
    =
    \inf_{t>0}
    \frac{
        c+\log\mu\left(e^{-t(g_y-1)}\right)
    }{t}
\end{equation}
and applying
\Cref{eq:bennettGaugeMinimization}
with $A=1-m_y$ gives
\begin{align}
\rho_{\mu,\mathrm{KL}}^c(1-g_y)
\leq
\min\Bigg\{
1-m_y,\,
\frac{
    \lVert g_y-1\rVert_{L^2(\mu)}^2
}{
    1-m_y
}
\left[
    \varphi_{\mathrm{KL}}(1+\cdot)
\right]^{-1}
\left(
    \frac{
        c(1-m_y)^2
    }{
        \lVert g_y-1\rVert_{L^2(\mu)}^2
    }
\right)
\Bigg\}.
\label{eq:klNegativePolarGaugeBennett}
\end{align}

The functions $
     a\longmapsto
    \varphi_{\mathrm{KL}}(1+a),
   \text{ } a\geq0,$
and $ 
    a\longmapsto
    \varphi_{\mathrm{KL}}(1-a),
    \text{ } 0\leq a\leq1,$
are nondecreasing. Substituting
\Cref{eq:klPositivePolarGaugeBennett,%
eq:klNegativePolarGaugeBennett}
into the exact polar-gauge contraction bound of
\Cref{thm:klNonlinearContraction}
proves~\eqref{eq:klBennettContractionCurve}. Finally, $
    \eta_{\mathrm{KL}}(\mu,K)
    =
    \sup_{c>0}
    \frac{
        F_{\varphi_{\mathrm{KL}}}(\mu,K;c)
    }{c}.$
Combining this identity with
\Cref{eq:klBennettContractionCurve}
proves~\eqref{eq:klBennettSDPI}.
\end{proof}

\subsection{Proof of Corollary~\ref{cor:klBernsteinClosedFormSDPI}}\label{app:proofKLBernsteinClosedFormSDPI}
\begin{proof}
For $A,s\geq0$, one has
\begin{align}
&\sup_{c>0}
\frac{
    \min\left\{
        A,\,
        \sqrt{2c}\,s+\dfrac{Ac}{3}
    \right\}
}{
    \sqrt c
}
=
\frac{1}{\sqrt{2}}
\left(
    s+\sqrt{s^2+\frac{2A^2}{3}}
\right).
\label{eq:bernsteinClippedGaugeLinearization}
\end{align}
Indeed, setting $r=\sqrt c$, the expression inside the supremum is$
    \min\left\{
        \frac{A}{r},\,
        \sqrt{2}\,s+\frac{Ar}{3}
    \right\}.$
The first function is decreasing in $r$, whereas the second is increasing.  Hence, their minimum is maximised at the unique point
where the two terms coincide. Solving
$
\frac{A}{x}
=
\sqrt{2}\,s+\frac{A}{3}x$
and substituting the positive solution yields~\Cref{eq:bernsteinClippedGaugeLinearization}.

Moreover, one has that $
    \varphi_{\mathrm{KL}}(1+x)
    \leq
    \frac{x^2}{2}, \text{ }x\geq0.
    \label{eq:klPositiveQuadraticEnvelope}$
Applying
\Cref{eq:bernsteinClippedGaugeLinearization}
with $A=M_y-1$ and
$s=\lVert g_y-1\rVert_{L^2(\mu)}$ gives
\begin{align}
&\sup_{c>0}
\frac{1}{c}
\varphi_{\mathrm{KL}}
\left(
1+
\min\left\{
M_y-1,\,
\sqrt{2c}\,
\lVert g_y-1\rVert_{L^2(\mu)}
+
\frac{c(M_y-1)}{3}
\right\}
\right)
\nonumber\\
&\qquad\leq
\frac{1}{4}
\left(
    \lVert g_y-1\rVert_{L^2(\mu)}
    +
    \sqrt{
        \lVert g_y-1\rVert_{L^2(\mu)}^2
        +
        \frac{2}{3}(M_y-1)^2
    }
\right)^2.
\label{eq:klPositiveBernsteinLinearization}
\end{align}

For the negative part, one has that
$
    x
    \longmapsto
    \frac{
        \varphi_{\mathrm{KL}}(1-x)
    }{x^2}$
is nondecreasing on $[0,1)$. Consequently, for
$0\leq x\leq1-m_y$, $
    \varphi_{\mathrm{KL}}(1-x)
    \leq
    \frac{
        \varphi_{\mathrm{KL}}(m_y)
    }{
        (1-m_y)^2
    }
    x^2.
    \label{eq:klNegativeQuadraticEnvelope}$
Applying
\Cref{eq:bernsteinClippedGaugeLinearization}
with $A=1-m_y$ gives
\begin{align}
&\sup_{c>0}
\frac{1}{c}
\varphi_{\mathrm{KL}}
\left(
1-
\min\left\{
1-m_y,\,
\sqrt{2c}\,
\lVert g_y-1\rVert_{L^2(\mu)}
+
\frac{c(1-m_y)}{3}
\right\}
\right)
\nonumber\\
&\qquad\leq
\frac{
    \varphi_{\mathrm{KL}}(m_y)
}{
    2(1-m_y)^2
}
\left(
    \lVert g_y-1\rVert_{L^2(\mu)}
    +
    \sqrt{
        \lVert g_y-1\rVert_{L^2(\mu)}^2
        +
        \frac{2}{3}(1-m_y)^2
    }
\right)^2.
\label{eq:klNegativeBernsteinLinearization}
\end{align}

Substituting
\Cref{eq:klPositiveBernsteinLinearization,%
eq:klNegativeBernsteinLinearization}
into
\Cref{eq:klNonlinearContraction}, then taking the supremum over $c>0$
and using
\begin{equation}
    \eta_{\mathrm{KL}}(\mu,K)
    =
    \sup_{c>0}
    \frac{
        F_{\varphi_{\mathrm{KL}}}(\mu,K;c)
    }{c}
\end{equation}
gives~\eqref{eq:klBernsteinClosedFormSDPI}.
\end{proof}

\subsection{Proof of Theorem~\ref{thm:subGaussianKLContraction}}\label{app:proofSubGaussianKLContraction}
\begin{proof}
Since $\mu(g_Y-1)=0$, the definition of the sub-Gaussian norm gives
\[
\log\mu\left(e^{t(g_Y-1)}\right)
\leq
\frac{t^2}{2}
\lVert g_Y-1\rVert_{\mathrm{SG}(\mu)}^2,
\qquad t>0.
\]
Because the sub-Gaussian norm is symmetric, the bound holds also for $1-g_Y$.
It follows from the variational representation of the KL polar
gauge that
\begin{align}
\rho_{\mu,\mathrm{KL}}^c(g_Y-1)
&\leq
\inf_{t>0}
\left\{
\frac{c}{t}
+
\frac{t}{2}
\lVert g_Y-1\rVert_{\mathrm{SG}(\mu)}^2
\right\} =
\sqrt{2c}\,
\lVert g_Y-1\rVert_{\mathrm{SG}(\mu)},
\label{eq:subGaussianPositiveGauge}
\end{align}
and, similarly, for $\rho_{\mu,\mathrm{KL}}^c(1-g_Y).$
The pointwise bounds on $g_Y$ also give
\begin{align}
\rho_{\mu,\mathrm{KL}}^c(g_Y-1)
\leq M_Y-1, \qquad
\rho_{\mu,\mathrm{KL}}^c(1-g_Y)
\leq 1-m_Y.
\end{align}
Substitution in the general polar-gauge contraction bound proves
\eqref{eq:subGaussianKLNonlinearContraction}. For $a\geq0$, $\varphi_{\mathrm{KL}}(1+a)
\leq
\frac{a^2}{2},$
whereas, for $0\leq a\leq1-m_Y$, one has $
\varphi_{\mathrm{KL}}(1-a)
\leq
\kappa(m_Y)a^2.
\label{eq:klNegativeQuadraticBound}$
Moreover,
\(
\frac12\leq\kappa(m_Y)\leq1.
\)
Therefore, each term in the maximum in
\eqref{eq:subGaussianKLNonlinearContraction} is bounded by
\(
2c\,\kappa(m_Y)
\lVert g_Y-1\rVert_{\mathrm{SG}(\mu)}^2.
\)
Dividing by $c$ and taking the supremum over $c>0$ proves
\eqref{eq:subGaussianKLSDPI}.
\end{proof}

\subsection{Proof of Corollary~\ref{cor:subGaussianKLRelaxations}}\label{app:proofSubGaussianKLRelaxations}
\begin{proof}
Since $\kappa(m_Y)\leq1$,
\Cref{eq:subGaussianKLSDPI} immediately gives
\eqref{eq:raginskySubGaussianBound}.

Convexity of the exponential and $\mu(g_Y)=1$ give
\begin{align}
\mu\left(e^{t(g_Y-1)}\right)
\leq&
\frac{M_Y-1}{M_Y-m_Y}
e^{-t(1-m_Y)}+
\frac{1-m_Y}{M_Y-m_Y}
e^{t(M_Y-1)}.
\end{align}
Applying the Kearns--Saul inequality to the Bernoulli
moment-generating function on the right-hand side proves
\eqref{eq:kearnsSaulSubGaussianNorm}. Hoeffding's lemma gives
\eqref{eq:hoeffdingSubGaussianNorm}. Substitution of these two
bounds in~\eqref{eq:raginskySubGaussianBound} proves
\eqref{eq:kearnsSaulKLBound} and
\eqref{eq:hoeffdingKLBound}.
\end{proof}

\subsection{Proof of Theorem~\ref{thm:probBoundPsiNorm}}\label{app:proofProbBoundPsiNorm}
    \begin{proof}
        The proof leverages the generalised H\"older's inequality. Let $\mu K^t$ be the measure obtained in the Markovian process when $\mu$ is the starting distribution and after $t$ applications of the Markov kernel $K$. Assume that $\mu K^t \ll \pi$, with $\pi$ being the stationary distribution. 
The following sequence of steps hold for every Young function $\psi$:
\begin{align}
    \mu K^t(E) = \int_E d\mu K^t &= \int \mathbbm{1}_E  d\mu K^t\\
               &= \int \mathbbm{1}_E  \frac{d\mu K^t}{d\pi} d\pi \\
               &\leq \left\lVert \mathbbm{1}_E\right\rVert_{L_{{\psi}^\star}^{N^\star}(\pi)}\left(\left\lVert \frac{d\mu K^t}{d\pi}-1\right\rVert_{L_\psi^N(\pi)}+\left\lVert 1\right\rVert_{L_\psi^N(\pi)}\right) \\
               &=\begin{cases}
                   1/{\psi^\star}^{-1}(1/\pi(E))\left(\left\lVert \frac{d\mu K^t}{d\pi}-1\right\rVert_{L_\psi^A(\pi)}+{\psi^\star}^{-1}(1)\right) \\
                   \pi(E)\psi^{-1}(1/\pi(E))\left(\left\lVert \frac{d\mu K^t}{d\pi}-1\right\rVert_{L_\psi^L(\pi)}+1/\psi^{-1}(1)\right)
               \end{cases}.
\end{align}

    \end{proof}

\subsection{Proof of Corollary~\ref{thm:probBoundAlphaNorms}}\label{app:proofProbBoundAlphaNorms}
\begin{proof}
    Since we know the exact closed-form expression of the norm when $\psi(x)=|x|^p/p$, it is possible to prove a stronger result directly.
Consider the function $\hat{\psi}(x)=|x-1|^p/p$. This is a convex function such that $\hat{\psi}(1)=0$ and is no longer a Young function. One can thus define a corresponding $f$-divergence~\cite{fDiv} with $f(x) =\hat{\psi}(x)$. \textit{I.e.}, given two measures $\mu\ll\nu$,
\begin{equation}
    D_{\hat{\psi}}(\mu\|\nu) = \int \hat\psi\left(\frac{d\mu}{d\nu}\right) d\nu =  \frac1p \left(\left\lVert \frac{d\mu}{d\nu}-1\right\rVert_{L_p(\nu)}\right)^p.
\end{equation} As such, it will satisfy the data-processing inequality, which we can use to prove~\Cref{thm:probBoundAlphaNorms}. Following the same steps undertaken to prove~\cite[Theorem 3]{fullVersionGeneralization}, for every event $E$ one has that 
\begin{align}
     \frac{D_{\hat{\psi}}(\mu\|\nu)+\nu(E^c){\hat{\psi}}^\star(0)}{\nu(E)} \geq \hat{\psi}\left(\frac{\mu(E)}{\nu(E)}\right).
\end{align}
If $\mu(E)\geq \nu(E)$, we can then invert $\hat{\psi}$  and obtain
\begin{equation}
    \mu(E)\leq \nu(E) \hat{\psi}^{-1}\left(\frac{D_{\hat{\psi}}(\mu\|\nu)+\nu(E^c){\hat{\psi}}^\star(0)}{\nu(E)}\right). \label{eq:probBoundAlphaNorms}
\end{equation}
Moreover, one has that $\hat{\psi}^\star(y) = |y|^q/q + y$, and thus $\hat{\psi}^\star(0)=0$.
Hence, one can rewrite~\Cref{eq:probBoundAlphaNorms} as follows:
\begin{equation}
     \mu(E)\leq \nu(E)\left(\left(p\frac{D_{\hat{\psi}}(\mu\|\nu)}{\nu(E)}\right)^\frac{1}{p}+1\right) = \nu(E)+\nu(E)^\frac1q \left\lVert \frac{d\mu}{d\nu}-1\right\rVert_{L_p(\nu)} . 
\end{equation}
If $\mu(E) \leq \nu(E)$, then the claim holds trivially as $\nu(E)^\frac1q \left\lVert \frac{d\mu}{d\nu}-1\right\rVert_{L_p(\nu)} \geq 0$.
\end{proof}

\subsection{Proof of Theorem~\ref{thm:sdpiConcentrationBound}}
\begin{proof}
    \label{app:proofConcentrationWithoutIndependence}
    From~\cite[Theorem 1]{esposito2024concentration}, we have that 
\begin{align} &\frac{\mathbb{P}\left(\left\lvert f(X_1,\ldots,X_t)-\Pm_{\bigotimes_{i=1}^t X_i}(f)\right\rvert \geq \eta \right)}{2^{\frac1q} \exp\left(\frac{-2t\eta^2}{q }\right)} \leq \prod_{i=2}^t \max_{x_{i-1}} \left\lVert \frac{d\Pm_{X_i|X_{i-1}=x_{i-1}}}{d\Pm_{X_i}}\right\rVert_{L^p(\Pm_{X_i})}\\
    &\hspace{9em}\leq \prod_{i=2}^t \max_{x_{i-1}} \left(\left\lVert \frac{d\Pm_{X_i|X_{i-1}=x_{i-1}}}{d\Pm_{X_i}}-1\right\rVert_{L^p(\Pm_{X_i})}+1\right) \\
    &\hspace{9em}\leq \prod_{i=2}^t \left(\left\lVert K_i^\star \right\rVert_{L_p^0\to L_p^0}\max_{x_{i-1}} \left\lVert \frac{d\delta_{x_{i-1}}}{d\Pm_{X_{i-1}}}-1\right\rVert_{L^p(\Pm_{X_{i-1}})}+1\right) \\
    &\hspace{9em}= \prod_{i=2}^t \left(\left\lVert K_i^\star \right\rVert_{L_p^0\to L_p^0}\omega_p^i+1\right).\label{eq:boundConcentrationViaNorms}
\end{align}
\end{proof}
\bibliographystyle{IEEEtran}
\bibliography{refs}
\end{document}